\documentclass[seceqn,final]{elsart}
\usepackage{natbib}
\usepackage[dvips]{graphicx}
\usepackage{latexsym}
\usepackage[centertags,intlimits]{amsmath}
\usepackage{amssymb}
\usepackage[dvips]{feynmp}

\newcommand{\isk}{T\sum_{\omega_n}\int \frac{dk}{2\pi}}
\newcommand{\isq}{T\sum_{\epsilon_n}\int \frac{dq}{2\pi}}
\newcommand{\isp}{T\sum_{\nu_n}\int \frac{dp}{2\pi}}
\newcommand{\iik}{\int \frac{dk}{2\pi} \int \frac{d\omega}{2\pi}}
\newcommand{\iike}{\int \frac{dk}{2\pi} \int \frac{d\epsilon}{2\pi}}
\newcommand{\iiq}{\int \frac{dq}{2\pi} \int \frac{d\epsilon}{2\pi}}
\newcommand{\iiqO}{\int \frac{dq}{2\pi} \int \frac{d\Omega}{2\pi}}

\newcommand{\wn}{\omega_n}
\newcommand{\On}{\Omega_n}
\newcommand{\en}{\epsilon_n}
\newcommand{\G}[3]{G_0^{#3}({#1},{#2})}
\newcommand{\iskq}{T^2\sum_{\en,\On} \int \frac{dq}{2\pi}\int \frac{dk}{2\pi}}
\newcommand{\iske}{T\sum_{\epsilon_n}\int \frac{dk}{2\pi}}
\newcommand{\ie}{\int \frac{d\epsilon}{2\pi}}
\newcommand{\iO}{\int \frac{d\Omega}{2\pi}}
\newcommand{\csch}[3]{\mathrm{csch}^{#3}\left(\frac{#1}{#2}\right)}
\newcommand{\md}[2]{\sqrt{{#1}^2+{#2}^2}}
\newcommand{\sgn}[1]{\ \mathrm{sgn}(#1)}
\newcommand{\ee}[1]{\mathrm{e}^{#1}}
\newcommand{\nt}[2]{\int\frac{d #1}{#2\pi}}

\newcommand{\kB}{k_\mathrm{B}}
\newcommand{\D}{\tilde{D}}

\renewcommand{\Im}{\mathrm{Im}\,}
\renewcommand{\Re}{\mathrm{Re}\,}
\renewcommand{\vec}[1]{\boldsymbol{#1}}

\begin{document}

\begin{frontmatter}
\title{Theory of the pairbreaking superconductor-metal transition in nanowires}

\author{Adrian Del Maestro},
\ead{adrian@delmaestro.org}
\author{Bernd Rosenow} and
\author{Subir Sachdev}
\address{Department of Physics, Harvard University, Cambridge, MA 02138}

\begin{abstract}
We present a detailed description of a zero temperature phase transition between superconducting and
diffusive metallic states in very thin wires due to a Cooper pair breaking mechanism.  The
dissipative critical theory contains current reducing fluctuations in the guise of both quantum and
thermally activated phase slips.  A full cross-over phase diagram is computed via an expansion in
the inverse number of complex components of the superconducting order parameter (one in the physical
case).  The fluctuation corrections to the electrical ($\sigma$) and thermal ($\kappa$)
conductivities are determined, and we find that $\sigma$ has a non-monotonic temperature dependence
in the metallic phase which may be consistent with recent experimental results on ultra-narrow
wires.  In the quantum critical regime, the ratio of the thermal to electrical conductivity displays
a linear temperature dependence and thus the Wiedemann-Franz law is obeyed, with a new universal
experimentally verifiable Lorenz number.  
\end{abstract} 
\end{frontmatter}

\section{Introduction}
\label{sec:introduction}

At the nanoscale, the basic mechanical, electrical and optical properties of materials that are well
understood at macroscopic length scales can change in interesting and sometimes unexpected ways as
quantization and fluctuation effects manifest themselves. An intriguing question thus arises
regarding the implications of reducing the scale or effective dimensionality of materials, that even
in the bulk, are known to already display interesting quantum mechanical behavior.  Superconductors
are natural physical systems to consider in this context.  Conventional or low temperature
superconductors are well understood in the bulk, unlike their high temperature cousins whose full
description still remains elusive after more than twenty years of intensive research.  A major
obstacle to the study of high temperature superconducting materials is that they are plagued by
their proximity to competing states with both \emph{order} and \emph{disorder} at the atomic scale.
Through a better understanding of the ways in which normal superconductivity is suppressed or
destroyed in different confining geometries and effective dimensions, perhaps progress can be made
towards a mastery of this fascinating emergent phenomena at all length and temperature scales. 

Recent technological advances now allow experimentalists to fabricate and control systems with ever
smaller dimensionality.  In two dimensions, thin films of lead with less than ten atomic layers can
be grown with such precision that mesas and valleys are intentionally engineered to trap vortices,
\emph{hardening} the superconductor in a magnetic field \cite{ozer}.  In one dimension, a technique
known as suspended molecular templating \cite{bezryadin-lau} can produce long metallic wires with
diameters less than 10~nm.  Finally, the fabrication and ultimate measurement of spin excitations in
manganese chains using a scanning tunneling microscope in inelastic tunneling mode is an elegant
example of the observation and control of a system consisting of only a few atoms \cite{stm-atoms}.  

In this paper, we will focus on the transition between a superconductor and a metal in ultra-narrow
wires.  It is well known that the Mermin-Wagner-Hohenberg theorem \cite{mw,hohenberg} precludes the
possibility of long range superconducting order at any non-zero temperature in one dimension.
However, any real wire is three dimensional, and can be approximated as a cylinder with a finite
radius $R$.  In the Landauer picture \cite{landauer} conduction is proportional to the number of
channels in the wire, $N_\perp$, equal to the number of states that can be occupied for a given
energy in all dimensions transverse to transport.  Thus, if we imagine free electron states
propagating down the wire, $N_\perp \propto A/\lambda_\mathrm{F}^2$ where $A$ is the cross-sectional
area and $\lambda_\mathrm{F}$ is the Fermi wavelength.  Any real wire will have  $R \gg
\lambda_{\mathrm{F}}$ implying that $N_\perp \gg 1$ and we can imagine that it will undergo a phase
transition to a superconducting state below some critical temperature.  As the diameter of the wire
decreases, or at suitably low temperatures, it will eventually enter a regime where the
superconducting coherence length, $\xi$ equal to the average separation between Cooper pairs, is
larger than the radius.  This condition defines the \emph{quasi-one dimensional} limit, as paired
electrons necessarily experience the finiteness of the transverse dimension while unpaired electrons
do not.

\subsection{LAMH theory}
\label{subsec:LAMH}
The study of superconducting fluctuations in narrow wires has a long history beginning in 
1968, when Webb and Warburton performed a remarkable transport experiment on thin whisker-crystals
of Sn \cite{whiskers} with diameters between 40 and 400~$\mu$m.  They noticed that for the
thinnest wires, resistive fluctuations leading to a finite voltage persisted below the bulk critical
temperature ($T_c$) for tin and defied any mean field characterization using the maximum
supercurrent \cite{bardeen-mft}.  The understanding of this behavior, which is specific to quasi-one
dimensional superconductors followed rapidly thereafter and is composed of three parts.  The first
was Little's qualitative introduction of thermally activated phase slips \cite{little-qpt}; an
unwinding of the phase of the superconducting order parameter by $\pm2\pi$ in a region of the wire
where the magnitude of the superconducting order parameter has been spontaneously suppressed to zero.
These non-trivial thermal fluctuations are equivalent to a vortex tunneling across the wire.  In an
applied bias current, negative jumps in the phase can exactly balance the linear in time phase
increase needed by the Josephson relation for the system to exhibit a finite voltage below $T_c$
\cite{tinkham}.  Next came the Ginzburg-Landau (GL) theory of Langer and Ambegaokar \cite{la} for
the free energy barrier height of a phase slip event which qualitatively reproduced the most
important features of Webb and Warburton's experiments. The story concluded with the time-dependent
GL theory of McCumber and Halperin \cite{mh} who correctly computed the rate at which these
resistive fluctuations occur, and led to full quantitative agreement.  The contributions of Langer,
Ambegaokar, McCumber and Halperin are now referred to as the \emph{LAMH} theory of thermally
activated phase slips.

Transport near the finite temperature phase boundary between the superconducting and normal state is
controlled by these phase slips, which from the LAMH theory have a free energy barrier height
$\Delta F(T)$ and occur at a rate $\Omega(T)$ at temperature $T$.  When used in conjunction with the
Josephson relation, LAMH showed that their contribution to the resistance of the wire is given by 
\begin{equation}
R_\mathrm{LAMH} = R_q \frac{\hbar \Omega(T)}{\kB T} \ee{-\Delta {F}(T)/\kB T}
\label{eq:RLAMH}
\end{equation}
where $R_q = h/(2e)^2$ is the quantum of resistance.  This result can be applied 
to a nanowire of length $L$ and normal resistance $R_N = (4 L / N_\perp \ell) R_q$ where $N_\perp$
is the number of transverse channels and $\ell$ is the mean free path to obtain the LAMH
contribution to the conductivity \cite{lau-tinkham}
\begin{equation}
\sigma_\mathrm{LAMH} = 3.4 \frac{e^2}{h} N_\perp \ell \left[\frac{T\xi(0)}{T_c N_\perp \ell }
\right]^{3/2} \left(1-\frac{T}{T_c}\right)^{-9/4}
\exp\left[ 0.21 \frac{T_c N_\perp \ell }{T \xi(0)} \left(1-\frac{T}{T_c}\right)^{3/2}\right]
\label{eq:sigmaLAMH}
\end{equation}
or 
\begin{equation}
\sigma_\mathrm{LAMH} = \frac{e^2}{h} N_\perp \ell \
\Phi_\mathrm{LAMH}\left(\frac{T}{T_c}, \frac{N_\perp \ell }{\xi(0)}\right)
\label{eq:PhiLAMH}
\end{equation}
where $\Phi_\mathrm{LAMH}$ is a universal dimensionless function, $T_c$ is the superconducting
transition temperature and  $\xi(0)$ is the zero temperature GL coherence length.  

A multitude of experiments on superconducting nanowires \cite{bezryadin-lau,liu-zadorozhny,
lau-tinkham,rogachev-bezryadin,boogaard, rogachev-bollinger,bollinger-rogachev,chang,rogachev-wei}
have confirmed the accuracy of the LAMH theory by fitting Eq.~(\ref{eq:sigmaLAMH}) to experimental
transport measurements with $T_c$ and $\xi(0)$ as free parameters with great success.  This
description includes the effects of \emph{only} thermally activated phase slips, and neglects the
possibility of quantum phase slips at low temperatures \cite{vanrun,saito} where, if present, one
would expect quantum tunneling to produce deviations from the LAMH theory. There are some
experimental indications that quantum phase slips (QPT) may indeed be present at the lowest
temperatures \cite{giordano-mqt,giordano-physica,shah-pekker} and we attempt to address some of
these issues in Section~\ref{subsec:orderedPhase} by presenting a version of the LAMH theory with
parameters renormalized by quantum fluctuations.  For a recent and comprehensive review describing
the influence of both thermally activated and quantum phase slips on superconductivity in one
dimension see Ref.~\cite{arutyunov}.

\subsection{Ultra-narrow wires}

As the diameter of a wire is reduced, there are two important changes that need to be considered.
The first is the well known volume to surface area ratio, and thus surface effects will begin to
affect bulk behavior.  The second is more subtle and is related to the increased effects of coupling
to an external environment. In the presence of such dissipation, a small system can undergo a
quantum localization transition, as is observed in small Josephson junctions \cite{schmid}.

In the early 1990s, step edge electron beam lithography techniques were used to create narrow indium
strips with diameters between $40$ and $100$~nm.  When transport measurements were performed, there
appeared to be significant deviations from the LAMH resistance at low temperatures resulting in a 
persisting resistance manifest as a ``foot''  raised upwards from the expected exponentially
decreasing resistance \cite{giordano-mqt, giordano-physica}.  It was proposed that this was due to the
onset of quantum phase slips at low temperatures occurring via the macroscopic tunneling mechanism
of Caldeira and Leggett \cite{caldeira-leggett}.  These ideas were vigorously pursued \cite{duan,
zaikin, zaikin-golubev, golubev-zaikin} leading to a host of theories which did not necessarily
agree on the observability of quantum phase slips in experiments.  One of the most interesting
results was an upper bound on the wire diameter of approximately $10$~nm above which quantum
phase slips would be strongly suppressed \cite{zaikin} as their rate $\Omega_{QPS} \sim
\exp(-N_\perp)$ can be exponentially small, where $N_\perp$ is the number of transverse channels in
the wire discussed above.  This upper bound of ten nanometers was far too narrow for step edge
lithography techniques and it would take the invention of new fabrication methods to fully address
these issues.

Wires with truly nanoscale dimensions were not studied until the introduction of a novel and
pioneering nanofabrication technique known as \emph{suspended molecular templating} in early 2000
\cite{bezryadin-lau}.  This remarkable process can be used to manufacture wires with lengths between
$100$ and $200$~nm with diameters less than $10$~nm.  The key feature is the top down approach that
uses a long narrow molecule such as a carbon nanotube or DNA as a backbone on top of which the wire
is deposited.  The fabrication process begins by etching a trench in a substrate formed from a
silicon wafer using electron beam lithography.  The backbone molecules are then placed in solution
and deposited over the substrate.  They are allowed to settle, and at high concentrations some will
end up resting over the trench.  The entire surface of the substrate is then sputter coated with
several nanometers of a metal like Nb or alloy such as MoGe.  The result is that a thin uniform
layer of the deposited material is suspended over the trench by the backbone molecule. It can
be located via scanning electron microscopy (SEM) and then isolated with a mask that is also used to
pattern electrodes that will be used for transport measurements.  

\begin{figure}[t]
\centering
\includegraphics*[width=3.0in]{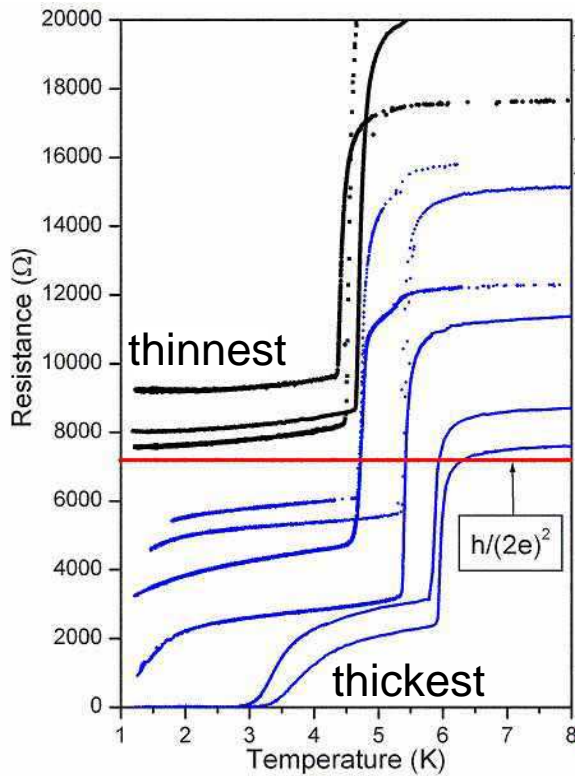}
\caption{\label{fig:nwPD}  Experimental transport measurements on MoGe nanowires reproduced from
Ref.~\cite{nanogallery} showing a distinct difference between thick superconducting and thin
metallic or resistive wires as the temperature is reduced.  At zero temperature, a quantum critical
point would separate the superconducting and metallic phase, with the transition between them being
described by a quantum superconductor-metal transition (SMT).}
\end{figure}

When measuring the resistance of a given wire as a function of temperature, two exponential 
dips are observed. The first at high temperatures, corresponds to the large two dimensional leads
going superconducting while the lower temperature drop is due to the actual wire undergoing a 
transition.  The temperature at which the wire goes superconducting is strongly dependent on its
diameter, with thinner wires being pushed to lower temperatures.  Superb agreement with the LAMH
theory is found for normal state resistances down to $0.5~\Omega$ using Eq.~(\ref{eq:RLAMH}). At
resistances below this value, or for thinner wires, there appears to be a growing experimental
consensus that there is a qualitative change in the resistance including deviations from the theory
of purely thermally activated phase slips.  The wires seem to be entering a regime where resistive
fluctuations coming from other effects, possibly including Coulomb blockade and quantum phase slips
can either postpone or completely destroy the superconducting transition
\cite{bezryadin-lau,lau-tinkham,chang,bollinger-bezryadin}.  This behavior can be seen by measuring
the resistance of thinner and thinner wires as a function of temperature leading to a separation
between superconducting and metallic transport all the way down to the lowest temperatures as seen
in Fig.~\ref{fig:nwPD}.
If superconductivity is indeed being destroyed as the temperature is reduced to zero by quantum
and not thermal fluctuations upon tuning some parameter related to the size of the transverse
dimension, then such a transition is by definition a superconductor-metal \emph{quantum phase
transition} \cite{qpt} (SMT).

\subsection{Pairbreaking quantum phase transition}

From the Cooper instability in BCS theory it is known that the existence of a non-trivial quantum
critical point in a metal implies a finite electron interaction strength.  Naively the existence of
interactions would seem to preclude the possibility of any quantum phase transition between a
superconductor and a metal because the ordered phase will always exist at some strictly non-zero
temperature for arbitrarily weak paring.  Moreover, if the temperature is driven to zero in a pure
BCS superconductor, pair fluctuations will be completely eliminated \cite{ramazashvili}.  The
solution arrives in the form of pairbreaking interactions, or any perturbation that is odd under
time reversal symmetry. Such a term will act differently on the spin and momentum reversed
constituents of a Cooper pair, making pairing more difficult.  The presence of these
interactions effectively cuts off the logarithmic singularity in the pair susceptibility and sets a
critical value for the strength of the pair potential before superconductivity can develop.
Therefore, our proposed SMT must live in the pairbreaking universality class.

The mean-field theory for the SMT goes back to the early work \cite{ag} of Abrikosov and Gor'kov
(AG). In one of the preliminary discussions of a quantum phase transition, they showed that a large
enough concentration of magnetic impurities could induce a SMT at $T=0$ ($T_c$ is protected by
Anderson's theorem \cite{anderson-theorem} for the case of non-magnetic impurities).  The transition
is tuned by a parameter $\alpha$ which is proportional to the impurity concentration and AG derived
an equation for the phase boundary given by
\begin{equation}
\ln\left(\frac{T}{T_{c0}}\right) = \psi\left(\frac{1}{2}\right) - \psi\left( \frac{1}{2} +
\frac{\hbar \alpha}{2\pi \kB T}\right)
\label{eq:AGTheory}
\end{equation}
where $\psi(x)$ is the polygamma function, $\kB T_{c0} = 1.14 \hbar \omega_{\mathrm{D}}
\ee{-1/N(0)V}$ is the BCS transition temperature in the absence of any pairbreaking
perturbations with $\omega_D$ the Debye frequency.  Eq.~(\ref{eq:AGTheory}) shows that by perturbing
a conventional superconductor with a suitably strong interaction that breaks time reversal symmetry,
it is possible to completely destroy the superconducting state at finite temperature at
$\alpha_c(T)$.  Mathematically, this is equivalent to the observation that for large enough
$\alpha$, Eq.~(\ref{eq:AGTheory}) has no non-zero solution for the temperature.

\begin{figure}[t]
\centering
\includegraphics[width=1.6in]{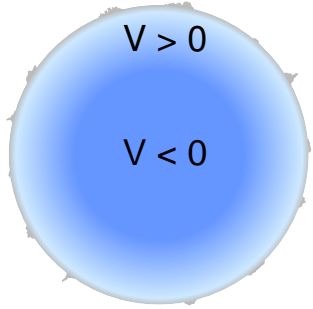}
\caption{\label{fig:wCS}  A schematic cross-section of a metallic wire where magnetic
impurities on the surface are poorly screened leading to a change in sign of the BCS pairing
interaction as one moves from the center, $|\Psi|\ne0$ to the edge, $|\Psi|\simeq0$ where
$\Psi(r)$ is the superconducting order parameter.  Below some temperature, the wire would be
composed of a superconducting core with a normal resistive sheath.}
\end{figure}

It has since been shown that such a theory applies in a large variety of situations with
pairbreaking perturbations \cite{deGennes}.  For the general case, $\hbar \alpha$ can be
interpreted as the depairing energy or splitting between the two time-reversed electrons of a Cooper
pair, averaged over the time required to completely uncorrelated their phases.  Relevant examples
include anisotropic gap superconductors with non-magnetic impurities
\cite{herbut,galitski-rc,galitski-prl}, lower-dimensional superconductors with magnetic fields
oriented in a direction parallel to the Cooper pair motion \cite{lopatin-vinokur,shah-lopatin}, and
$s$-wave superconductors with inhomogeneity in the strength of the attractive BCS interaction
\cite{spivak}.  Indeed, it is expected that pairbreaking is present in any experimentally
realizable SMT at $T=0$. In the nanowire experiments, explicit evidence for pairbreaking magnetic
moments on the wire surface was presented recently by Rogachev {\em et al.} \cite{rogachev-wei}.
Any impurity at the surface would be much less effectively screened, and one could imagine a BCS
coupling $V(r)$ which depends on the radial coordinate of the wire $r$. In this picture, $V(r)$
would change sign from negative (attractive) to positive (repulsive) as $r$ changes from $r= 0$ to
$r= R$ where $R$ is the diameter of the wire.  This behavior is schematically outlined in
Fig.~\ref{fig:wCS}.
For the thickest wires, $R > \xi$ and the mean field solution to the BCS equations will lead to the
wire being described by a superconducting core surrounded by a cylindrical metallic envelope.  A
similar picture was recently put forward to describe the SMT in two dimensions by imagining
superconducting grains embedded into a film with a pairing interaction which depended on the
distance from the center of the islands \cite{spivak}.  In the nanowire case, as the transverse
dimension is reduced, there will be a critical radius, $R \lesssim \xi$, where the superconducting
core will vanish and the wire will enter a metallic state.  This is a rather physically appealing
picture as it is suitable for the destruction of superconductivity in a wire that is only weakly
disordered in the bulk and is well suited to theoretical models.

It is clearly impossible to continually reduce the diameter of a single wire while measuring its
transport properties and thus other more systematic examples of a SMT would be beneficial.  The most
obvious candidate is a transition that can be observed in a single wire by increasing the strength
of a magnetic field oriented parallel to its long axis.  Such an experiment has been performed by
Rogachev, Bollinger and Bezryadin \cite{rogachev-bollinger} on individual Nb nanowires.  As the
strength of the parallel magnetic field is increased, the superconducting transition appears at
lower and lower temperatures with the expectation that for suitably strong fields it will vanish all
together and the wire will exhibit metallic behavior;  a quantum SMT.  To completely destroy
superconductivity, pairbreaking events must be uncorrelated over long time scales and a theoretical
description would require diffusive electrons or suitably strong boundary scattering.

\subsection{Microscopic approach}

Fluctuations about the AG theory have been considered \cite{lopatin-vinokur,shah-lopatin} in the
metallic state, and lead to the well-known Aslamazov-Larkin (AL) \cite{al}, Maki-Thompson (MT)
\cite{maki,thompson} and Density of States (DoS) \cite{altshuler-aronov} corrections to the
conductivity.  The form of these corrections is usually introduced in terms of the structure of
their diagrammatic representation within the finite temperature disordered electron perturbation
theory but they all have the same physical origin: in the presence of strong pairbreaking, the
normal metal still experiences pairing fluctuations near the Fermi surface as a result of its
proximity to the superconducting state.   Specifically, the AL effect comes from the direct charge
transfer from fluctuating Cooper pairs, the MT correction results from coherent Andreev scattering
off the fluctuating pairs and the Density of States (DoS) correction is due to the reduction of the
normal electron density of states near the Fermi surface accounting for the paired electrons.  

The exact form of these contributions are known from recent microscopic finite temperature
perturbative computations in BCS theory \cite{lopatin-vinokur,shah-lopatin}.  These results are
valid at low temperatures, with the pairbreaking parameter $\alpha$ larger than critical $\alpha_c$
of the SMT.  Denoting the BCS coherence length by $\xi_0$, one can define a clean limit by $\xi_0
\ll \ell$ and a dirty limit by $\ell \ll \xi_0$.  The total conductivity was obtained in the dirty
limit and it was found that
\begin{equation}\begin{split}
\sigma &= \sigma_0 + \frac{e^2}{\hbar} \left(\frac{k_B T}{\hbar D} \right)^{-1/2} \left[ \frac{ \pi }
{12 \sqrt{2}} \left( \frac{k_B T}{\hbar(\alpha - \alpha_c)} \right)^{5/2} \right]  
+\frac{e^2}{\hbar} \left(\frac{k_B T}{\hbar D} \right) \left[ c \frac{\hbar(\alpha - 
\alpha_c)}{k_B T} \right] 
\label{eq:shahSigma}
\end{split}\end{equation}
where $\sigma_0$ is a background metallic conductivity, $c$ is a non-universal constant, $D$ is the
diffusion constant in the metal, and the remaining corrections from pairing fluctuations have been
written in the form of a power of $T$ times a factor within the square brackets which depends only
upon the ratio $\hbar(\alpha-\alpha_c)/k_B T$. Writing the conductivity in this way allows us to
determine the relative importance of the fluctuations corrections, in the renormalization group
sense, to the SMT. The first square bracket represents the usual Aslamazov-Larkin (AL) correction
\cite{al} and has a prefactor of a negative power of $T$ and so is a relevant perturbation. This is
a result of the large inverse power of $\alpha-\alpha_c$ which will be dominant near criticality 
as the critical point is approached.  The second term arises from the additional AL, MT and DoS
corrections: the prefactor has no divergence as a power of $T$, and so this correction is
formally irrelevant at the SMT. The complete second term has a finite limit as $T \rightarrow 0$,
and so becomes larger than the formally relevant AL term at sufficiently low $T$ in the metal.  The
second term is therefore identified as being \emph{dangerously irrelevant} in critical phenomena
parlance: it is important for the properties of the low $T$ metallic region, but can be safely ignored
at finite temperatures near the critical coupling.

\subsection{Field-theoretic approach}

At the SMT, field-theoretic analyses \cite{sachdev-troyer,podolsky-sachdev} show that the AG theory,
along with the AL, MT and DoS corrections, is inadequate in spatial dimension $d \leq 2$, and
additional repulsive self-interactions among Cooper pairs have to be included. Here, $d$ defines the
dimensionality of the Cooper pair motion. The confining dimension, or radius of the wire, $R$, is
larger than the inverse Fermi wavevector, but smaller than the superconducting coherence length or
Cooper pair size, $\xi$. This is the exact condition discussed previously for the quasi-one
dimensional limit.  While the Cooper pairs are effectively one dimensional, any unpaired electrons
have a three dimensional Fermi surface and thus strictly 1d Luttinger liquid physics do not apply.

The goal of this work is to understand the aforementioned experiments on nanoscale metallic wires
and we are interested in the fluctuation corrections to the thermal and electrical conductivity
across the SMT as well as the nature of the crossovers from this universal quantum critical physics
to previously studied regimes at low $T$ about the superconducting and metallic phases.  
In the remaining sections, we will examine the $d=1$ SMT in great detail and begin by introducing a 
theory for an ultra-narrow superconducting wire in terms of a strongly-coupled field theory of bosonic
Cooper pairs, overdamped by their coupling to fermionic quasiparticles (normal electrons). Both the
zero frequency thermal and electrical transport coefficients are computed, first in the large-$N$
limit, where the number of components of the superconducting order parameter ($N$) is assumed to be
infinite, and then as an expansion in $1/N$.  Along the way we make contact with the microscopic BCS
theory, touch upon an extension of the LAMH theory near $T_c$ and investigate the ratio of the
thermal to electrical conductivity, known as the Wiedemann-Franz ratio.

\begin{figure}[t]
\centering 
\includegraphics*[width=3.2in]{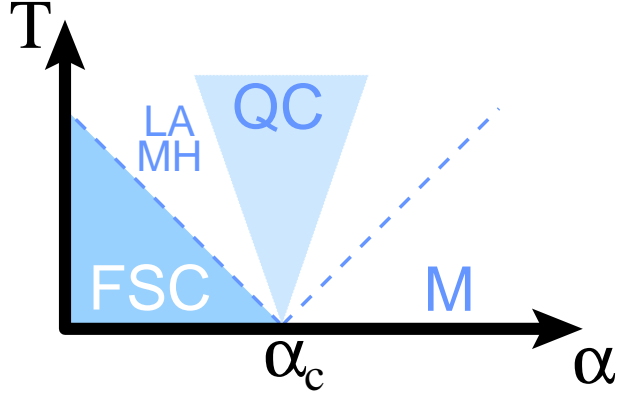} 
\caption{ Crossover phase diagram of the superconductor-metal transition in a 
quasi-one dimensional wire. The \emph{metal} (M) is described by the perturbative theory of 
Ref.~\cite{lopatin-vinokur}. The \emph{quantum critical} (QC) region is described by our effective
theory $\mathcal{S}_\alpha$ for temperatures above $T_{\rm dis}$ where the effects of disorder may
be neglected.  The Mooij-Sch\"on mode is present everywhere, but couples strongly to superconducting
fluctuations only in the \emph{fluctuating superconductor} (FSC) regime, where it is described by
Eq.~(\ref{eq:Sphi}); note that $\mathcal{S}_\alpha$ does \emph{not} apply here. The dashed lines are
crossover boundaries which occur at $T \sim |\alpha-\alpha_c|$ (from Eq.~(\ref{eq:PhiSigScaleG}))
and the intermediate LAMH region should be described by the theory of thermally activated phase
slips discussed in Section~\ref{subsec:LAMH} with the possible modifications of
Chapter~\ref{chap:largeN}.  Note that $\mathcal{S}_\alpha$ does not include the MT and DoS
corrections in the metallic regime where they are actually larger than the AL contribution to the
conductivity.}
\label{fig:phaseDiagram} 
\end{figure}

To frame our discussion, we conclude the introduction with a summary of the full crossover phase
diagram that is constructed throughout this paper in Fig.~\ref{fig:phaseDiagram}.
The important features of this global pairbreaking phase diagram include a strongly fluctuating
quantum critical regime, the main focus of this study, where phase and amplitude fluctuations must
be treated on equal footing.  For a pairbreaking strength greater than the critical one, as the
temperature is reduced, there is a crossover to a low-$T$ metallic regime described by the theory
\cite{lopatin-vinokur,shah-lopatin} of AL+MT+DoS corrections in $d=1$ (dashed line in
Fig.~\ref{fig:phaseDiagram}). On the superconducting side, there is a regime of intermediate
temperatures where the classical LAMH phase slip theory applies \cite{la,mh}, with possible
modifications coming from quantum renormalized coefficients and eventually another crossover at
still lower temperatures to a phase fluctuating regime whose description requires a non-linear
$\sigma$-model of fermion pair fluctuations coupled to the superconducting order \cite{andreev}.  

Models similar to the one we analyze, but lacking amplitude fluctuations (so called phase-only
theories) have been previously applied to the physical situation considered here
\cite{zaikin,buchler,rdof,meidan,rdo}. In these continuum theories, the destruction of
superconductivity is due to the proliferation of quantum phase slips resulting in a normal phase
which we maintain is \emph{insulating} and not metallic at $T=0$.  Within the phase only
approach, the superfluid's conductivity would be controlled by irrelevant phase-slip operators
as the resulting $d=1$ superconductor-insulator transition (SIT) lives in the Kosterlitz-Thouless
universality class \cite{giamarchi}.  Such theories would thus be applicable to a quantum SIT which
could be appropriate in short inhomogeneous wires.  

\section{Dissipative model}
\label{sec:dissipativeModel}

The approach to the SMT taken here is akin to previously studied theories
\cite{fl,feigelman-skvortsov,galitski-larkin} of disordered superconducting films with
unconventional pairing symmetry.  Such films are assumed to be composed of a network of small
Josephson coupled superconducting islands.  The transition between the normal and superconducting
state can be tuned by altering the distance between the grains; if the separation becomes large
enough, quantum fluctuations can destroy superconducting order even at zero temperature.
Physically, the transition occurs when the Josephson coupling becomes less than the Coulomb energy
due to the transfer of a single Cooper pair between islands. These arguments can be made more
rigorous by starting from microscopic BCS theory and deriving an effective model for the
fluctuations of the Cooper pair order parameter.  The order parameter is strongly overdamped by decay
into quasiparticle excitations manifest as an interaction that is long range in imaginary time
\cite{herbut,ramazashvili, galitski-prl, lopatin-vinokur,shah-lopatin,spivak}.  

These ideas can be directly applied to the $d=1$ quantum superconductor-metal transition in the
pairbreaking universality class which is relevant for ultra-narrow quasi one-dimensional metallic
wires.  The physical ingredients include repulsive paired states and a lack of charge conservation
of the condensate requiring the existence of a bath (the large number of transverse
conduction channels) into which the former member electrons of a disassociated Cooper pair can flow
into upon breaking.  The final description will be in terms of a strongly-coupled field theory of
bosonic Cooper pairs, overdamped by their coupling to the unpaired fermionic states of the
metal.  

The fluctuation corrections to transport associated with the AL correction discussed above are
naturally captured in this picture.  These have \cite{lopatin-vinokur,shah-lopatin} a Cooper pair
propagator $(\D k^2 + |\wn| + \alpha)^{-1}$ at wavevector $k$ and imaginary bosonic Matsubara
frequency $i\omega_n$ in the metal in both the clean and dirty limits.  Here, the ``mass'' or
bare pairbreaking frequency $\alpha$ measures the strength of the pairbreaking interaction which
could come from a variety of sources as mentioned in the introduction.  $\D$ is equal to the usual
diffusion constant $\D = D = v_\mathrm{F} \ell/3$ ($v_\mathrm{F}$ is the Fermi velocity) in the
dirty limit where the mean free path is much smaller than the superconducting coherence length
($\ell \ll \xi_0$). In the clean limit, where $\ell \gg \xi_0$, $\D$ will be in general some
non-universal number that depends on the specific microscopic details (such as the lattice constant)
of the system in question.  This motivates the quantum critical theory of
Ref.~\cite{sachdev-troyer,wireslett} for a field $\Psi(x,\tau)$ which represents the local
Cooper pair operator
\begin{equation}\begin{split}
\mathcal{S}_\alpha &= \int_0^L dx \int_0^{1/T} d\tau \left[ \D |\partial_x \Psi(x,\tau)|^2 + \alpha 
|\Psi(x,\tau)|^2 + \frac{u}{2} |\Psi(x,\tau)|^4 \right] 
\\
& \qquad +\; T \sum_{\wn} \int_0^L dx\,  \gamma |\wn| |\Psi(x,\omega_n)|^2 ,
\label{eq:Salpha}
\end{split}\end{equation}
where we have used the temporal Fourier transform of $\Psi(x,\tau)$
\begin{equation}
\Psi(x,\wn) = \int_0^{1/T} d \tau\ \Psi(x,\tau) \ee{i\wn\tau}.
\end{equation}
with $\omega_n = 2\pi n T$ to more compactly express the non-locality of the dissipative term in
imaginary time and have chosen units where $\hbar=\kB = 1$ for convenience.  From this point
forward, we will suppress the limits of integration for the sake of compactness unless their
inclusion is required for clarity.  The quartic coupling $u$ must be positive to ensure stability,
and describes the repulsion between Cooper pairs. The pairs are strongly overdamped, and the rate of
their decay into the metallic bath is characterized by the coupling constant multiplying the $|\wn|$
term, $\gamma$,  which is required to be positive by causality.  It will be convenient to rescale
the field $\Psi$ such that the coefficient of the Landau damping term is equal to unity. In
addition, we rescale all couplings according to
\begin{equation}
\Psi \to \frac{\Psi}{\sqrt{\gamma}}; \quad \D \to \gamma \D; \quad
\alpha \to \gamma \alpha; \quad u \to \gamma^2 u .
\label{eq:rescale}
\end{equation}
This theory describes the vicinity of a superconductor-metal quantum critical point, corresponding
to the (bare) value $z=2$ for the dynamic critical exponent.  A different description ($z\ne2$)
cannot be completely ruled out, but it would most likely require additional tuning parameters as
well as the inclusion of unusual pairing phenomena \cite{ramazashvili}.  

The quantum phase transition is driven by altering the strength of the pairbreaking frequency
$\alpha$ as was shown schematically in Fig.~\ref{fig:phaseDiagram}.  While $\alpha \gg  \alpha_c$
there is normal metallic conduction and for $\alpha \ll \alpha_c$ the system is fully
superconducting.  For $\alpha \gtrsim \alpha_c$ pairing fluctuations enhance the conductivity while
for $\alpha \lesssim \alpha_c$, both thermal and quantum phase slips, included as amplitude
fluctuations of $\Psi$ that destroy the superflow.

The field theory in Eq.~(\ref{eq:Salpha}) is identical in form to the Hertz-Millis-Moriya theory
\cite{hertz,millis,moriya1} describing the Fermi liquid to spin-density wave (SDW) transition, with
the Cooper pair operator $\Psi$ replaced by an $O(3)$ order parameter representing diffusive
paramagnons.  In the neighborhood of this transition, $k$ measures the magnitude of the deviation
from the SDW ordering wave vector $\vec{K}$ and the dissipative $|\wn|$ term arises from the damping
of order parameter fluctuations resulting from coupling to gapless fermionic excitations of the
metal near points on the Fermi surface connected by $\vec{K}$. A more careful analysis leads to the
realization that at $T=0$ on the ordered (SDW) side of the transition, a gap appears in the fermion
spectrum for small $k$.  Thus, this description is only fully accurate at $T=0$ on the disordered
(metallic) side of the transition or at finite temperature in the quantum critical regime above the
SDW state.  The same logic applies to the role of phase fluctuations at low temperatures in the
superconducting phase near the SMT.  We will return to this point later with a thorough discussion
of the Mooij-Sch\"on normal mode, but for now we begin with a detailed scaling analysis of
$\mathcal{S}_\alpha$.

\subsection{Scaling analysis}
\label{subsec:scalingAnalysis}
Simple power counting for the rescaled self-interaction term in Eq.~(\ref{eq:Salpha}) in $d$ spatial
dimensions fixes the upper critical dimension at $d=2$ and necessitates a non-perturbative treatment
for the quasi-one dimensional limit considered here.  In $d=1$, $u$ is a relevant perturbation but
in the strong coupling regime ($u \to \infty$) we expect all results to be universal ($u$
independent).  

Pankov \emph{et al.} \cite{pankov} studied a theory similar to $\mathcal{S}_\alpha$ with $\Psi$ 
replaced by a $N$-component field via the renormalization group (RG) in an $\epsilon =
2-d$ expansion in one and two dimensions at zero temperature.  The most important result obtained
from their RG analysis is that the damping term, $|\wn|$, generated from the long-range $1/\tau^2$
interaction between order parameter fluctuations does not require an independent renormalization,
and thus the frequency dependence of the propagator only involves wavefunction renormalization.  At
$T=0$ and $\alpha = \alpha_c$ in one dimension, they find a non-trivial fixed point and an analysis
of the RG equations leads to an expression for the dynamical susceptibility at small frequencies and
momenta
\begin{align}
\chi(k,\omega) &= \int d\tau \int dx \langle \Psi^{\ast}_a (x,\tau)\Psi^{\phantom{\ast}}_a(0,0) \rangle
\ee{-i(k x-\omega \tau)} \nonumber \\ 
&= k^{-2+\eta} \Phi_{\chi,0}\left(\frac{\omega}{c_0 k^{2-\eta}}\right)
\label{eq:ChiTeq0}
\end{align}
where $a=1,\ldots,N$, $\Phi_{\chi,0}$ is a universal scaling function and $c_0$ is a non-universal
constant that will depend on $\D$. The dynamical critical exponent can thus be read off as
\begin{equation}
z = 2 - \eta
\end{equation}
where its bare value has been corrected by an anomalous dimension $\eta \sim \epsilon^2$.  This 
result holds to all orders due to the existence of only wavefunction renormalization
\cite{pankov,gamba}.  Eq.~(\ref{eq:ChiTeq0}) can be generalized to finite temperatures 
where
\begin{equation}
\chi(k,\wn,T) = \frac{1}{T} \Phi_\chi \left(\frac{\wn}{T},\frac{c_1 k}{T^{1/z}}\right),
\end{equation}
with $\Phi_\chi$ another universal scaling function and $c_1$ a non-universal constant.  Most
interestingly, at $k=0$ and $\wn=0$ the value of the inverse susceptibility in the quantum
critical region will be fixed by temperature alone,
\begin{equation}
\chi^{-1}(0,0) = \mathcal{A} T
\end{equation}
and the highly non-trivial universal constant $\mathcal{A}$ will be computed in a $1/N$ expansion in
Section~\ref{sec:Nfluc}. 

Scaling functions for the most singular parts of the dc electrical $\sigma$ and thermal $\kappa$
conductivities can also be derived with a knowledge of their scaling dimensions alone.  
In one dimension, the longitudinal conductivity of a wire is equal to $e^2/h$ times a length, and
for $k=\wn=0$ but finite temperature there is only one length scale available, the $\emph{thermal
length}$ $L_T \sim T^{-1/z}$, and the energy scale is set by the distance from the critical point.
Similar arguments apply for the thermal conductivity leading to (returning to physical units)
\begin{align}
\label{eq:PhiSigScaleG}
\sigma &=  \frac{e^2}{\hbar} \left(\frac{k_B T}{\hbar \widetilde{D}} \right)^{-1/z}
\Phi_\sigma \left ( \frac{[\hbar(\alpha-\alpha_c)]^\nu}{(k_B T)^{1/z }} \right) 
\\
\label{eq:PhiKapScaleG}
\kappa &= \frac{k_B^2 T}{\hbar} \left(\frac{k_B T}{\hbar \widetilde{D}} \right)^{-1/z}
\Phi_\kappa \left( \frac{[\hbar (\alpha-\alpha_c)]^\nu}{(k_B T)^{1/z}} \right) 
\end{align}
where $\nu$ is the usual correlation length exponent defined by $\xi \sim |\alpha-\alpha_c|^{-\nu}$
and $\Phi_\sigma$ and $\Phi_\kappa$ are two dimensionless universal scaling functions.

In the Gaussian (non-interacting) limit, $z=2$, $\eta=0$ and $\nu=1/2$ and corrections can be
computed in the $\epsilon = 2-d$ expansion \cite{pankov,sachdev-troyer} 
\begin{align}
\label{eq:etaeps}
\eta &= \frac{(N+2)(12-\pi^2)}{4(N+8)^2} \epsilon^2 + O(\epsilon^3) 
\\
\label{eq:nueps}
\nu &= \frac{1}{2} + \frac{(N+2)}{4(N+8)}\epsilon 
+ \frac{(N+2)[6N^2 + (228-7\pi^2)N + 792 - 38 \pi^2]}{48(N+8)^3} \epsilon^2 + O(\epsilon^3).
\end{align}
These results are in agreement with Monte Carlo simulations \cite{werner-troyer} which found $z =
1.97(3)$, $z+\eta = 1.985(20)$ and $\nu = 0.689(6)$.  

If interactions are included but calculations are performed in the limit where the number of order
parameter components is large, the \emph{large-$N$ limit}, $z$ and $\eta$ are unchanged from their
Gaussian values, but $\nu=1$ and the replacement
\begin{equation}
\frac{[\hbar (\alpha-\alpha_c)]^\nu}{(k_B T)^{1/z}} \to
\frac{\hbar (\alpha-\alpha_c)}{(k_B T)^{2/z\nu}} 
\label{eq:largeNScaleVar}
\end{equation}
is required in the scaling functions above.  Corrections from the $N=\infty$ values of the exponents
$\nu$ and $\eta$ can be computed in the $1/N$ expansion and the results are detailed in
Section~\ref{sec:Nfluc}.

The RG analysis of Ref.~\cite{pankov} confirms that  $\mathcal{S}_\alpha$ satisfies
conventional hyperscaling relations at the $T=0$ SMT in the absence of disorder.  This implies that
all irrelevant operators can be neglected, and transport should be fully described by
Eq.~(\ref{eq:Salpha}).  Moreover, the most singular part of the dc conductivity of a given
wire will be described by Eq.~(\ref{eq:PhiSigScaleG}) which is independent of its length $L$. 

\subsection{Particle-hole asymmetry}
\label{subsec:particleHoldAsymmetry}
In the scaling analysis of the previous section, we neglected an important detail; for $z=2$, when
the energy dependence of the electronic density of states near the Fermi level is included, 
a propagating term in the action can arise from the weak particle-hole asymmetry of the electronic 
spectrum
\begin{equation}
\mathcal{S}_{\rho} = \int dx \int d\tau \rho \, \Psi^\ast(x,\tau)\frac{\partial}{\partial \tau}
\Psi(x,\tau).
\label{eq:Seta}
\end{equation}
The magnitude of particle-hole symmetry breaking, $\rho$ is proportional to the energy derivative
of the density of states at the Fermi energy, and we therefore expect its bare value to be small.
This is supported by the fact that our quasi-one dimensional treatment of Cooper pairs coupled to a 
bath composed of three dimensional electrons, required the ratio of the pairing to Fermi energy to be
small. Specifically, if $\lambda_\mathrm{F} \ll R \lesssim \xi$, then using the BCS result for the
zero temperature energy gap $\Delta = \epsilon_\mathrm{F} / (2\pi \xi k_\mathrm{F})$ we find that
$1/k_\mathrm{F} \xi \sim \Delta / \epsilon_\mathrm{F} \ll 1$.

In fact, a microscopic weak coupling derivation of Eq.(\ref{eq:Salpha}) for a dirty two 
dimensional superconductor with d-wave pairing symmetry \cite{herbut,galitski-prl} finds that the
ratio of the dissipative to propagative terms is proportional to the dimensionless conductivity of
the normal phase, $\rho \propto 1/\epsilon_\mathrm{F}\tau$ where $\tau$ is the scattering time in
the self-consistent Born approximation.  For a good metal, the product of the scattering time and
Fermi energy is large, and thus $\rho \ll 1$.  This is consistent with results at finite temperature
for a weak-coupling short coherence length superconductor \cite{stintzing}.

At tree level, $\rho$ is marginal, and as just argued, we expect its bare value to be small.
However, we can examine the renormalization group fate of $\mathcal{S}_{\rho}$ near the fixed point
of $\mathcal{S}_\alpha$.  The scaling dimension of $\rho$ can be computed in a $d = 2-\epsilon$
expansion for a massless ($\alpha=0$) quantum critical theory through the conventional method of
isolating any logarithmic singularities (or $1/\epsilon$ poles) in the Feynman diagrams
corresponding to all possible insertions of the perturbing term, $i\rho\omega$ in
Eq.~(\ref{eq:Seta}).  For the case considered here, there are two unique graphs 
\begin{equation}
2\;
\parbox{24mm}{%
  \begin{fmffile}{phas1}
    \begin{fmfgraph*}(60,60)
      \fmfleft{l} 
      \fmfright{r}
      \fmf{plain}{l,vl}
      \fmf{plain}{vr,r}
      \fmf{plain,left,tension=0.2}{vl,vr,vl}
      \fmfdot{vl,vr}
	  \fmffreeze
	  \fmf{plain}{vl,vr}
	  \fmfforce{(.5w,0.78h)}{vt}
	  \fmfv{decor.shape=cross,decor.size=3mm,label=$i\omega_1$}{vt}
    \end{fmfgraph*} 
  \end{fmffile}}
+\;\;
\parbox{24mm}{%
  \begin{fmffile}{phas2}
    \begin{fmfgraph*}(60,60)
      \fmfleft{l} 
      \fmfright{r}
      \fmf{plain}{l,vl}
      \fmf{plain}{vr,r}
      \fmf{plain,left,tension=0.2}{vl,vr,vl}
      \fmfdot{vl,vr}
	  \fmffreeze
	  \fmf{plain}{vl,vr}
	  \fmfforce{(.5w,0.5h)}{vt}
	  \fmfv{decor.shape=cross,decor.size=3mm,label=$i\omega_1$,label.angle=90,label.dist=0.7mm}{vt}
    \end{fmfgraph*} 
  \end{fmffile}}
\label{eq:phasGraphs}
\end{equation}
where a solid line is equal to the bare propagator $(k^2 + |\omega|)^{-1}$, a dot represents the
quartic interaction $u$ and a cross is an insertion coming from Eq.~(\ref{eq:Seta}).

If the external lines have frequency $\Omega$ and zero momentum, then the combination of these
graphs leads to the integral
\begin{equation}\begin{split}
I(\Omega) &= -i 2 \rho u^2 \int \frac{d \omega_1}{2\pi} \int \frac{d \omega_2}{2\pi}
\int \frac{d^d k}{(2\pi)^d} \int \frac{d^d q}{(2\pi)^d} \\
& \quad \times\; \frac{\omega_1}
{(k^2 + |\omega_1|)^2(q^2+|\omega_2|)[(k+q)^2+|\omega_1+\omega_2+\Omega|]}.
\end{split}\end{equation}
A simple power counting analysis of the integrand in $d = 2-\epsilon$ dimensions leads to the
appearance of the predicted pole
\begin{equation}
I(\Omega) = i\Omega \rho \frac{A}{2\epsilon},
\label{eq:IOmegaA}
\end{equation}
where as usual, the flow equation for $\rho$ is related to the residue $A$ via
\begin{equation}
\frac{d \rho}{d \ell} = A \rho.
\label{eq:dleta}
\end{equation}
$I(\Omega)$ is computed in Appendix~\ref{app:phas} leading to the flow equation
\begin{equation}
\frac{d \rho}{d \ell} =  \frac{u^2}{16 \pi^2}\left(1-\frac{8}{\pi^2}\right)\rho .
\end{equation}
The fixed point value of $u$ is given in Ref.~\cite{pankov} for the equivalent $z=2$ $O(N)$ model
with the change of notation $u_0 = 3 u$.  Equivalently it can be easily computed to one loop order
as
\begin{equation}
u^{*} = \frac{2 \pi^2}{5} \epsilon
\end{equation}
leading to
\begin{equation}
\frac{d \rho}{d \ell} =  \frac{\epsilon^2}{100}(\pi^2-8)\rho
\label{eq:detadl}
\end{equation}
or $\rho(\ell) \sim \mathrm{e}^{0.02 \epsilon^2 \ell}$ at RG scale $\ell$. Thus we conclude that
although $\rho$ is relevant, its scaling dimension is extremely small.  In conjunction with a bare
value that we have argued should be diminutive, we will neglect $\mathcal{S}_\rho$ in future
calculations. 

There is still one piece missing in our analysis of $\mathcal{S}_\alpha$ as alluded to in the
previous section; the role of charge conservation (which $\mathcal{S}_\alpha$ breaks) and the
associated normal modes.

\subsection{Phase fluctuations}
\label{subsec:phaseFluctuations}

From hydrodynamic arguments, it is known that a one-dimensional metal or superconductor should
support a gapless plasmon, or a Mooij-Sch\"on normal mode \cite{ms}, which disperses as $\omega \sim
k \ln^{1/2} (1/(kR))$.  Our discussion of this issue parallels that in Refs.~\cite{ioffe-millis,scs}
on the role of conservation laws in the critical fluctuations of quantum transitions in metallic
systems for which the order parameter is overdamped (as is the case here).  To make this explicit,
couple $\Psi$ to a fluctuating scalar potential $A_\tau$ with bare action
\begin{equation}
\mathcal{S}_A = \int \frac{dk}{2\pi} \int \frac{d \omega}{2 \pi} \frac{|A_\tau (k, \omega)|^2 }
{4 \ln (1/(k R))}.  
\label{eq:SA}
\end{equation}
However, the nature of the $A_\tau$-$\Psi$ coupling differs  between the ``quantum 
critical'' and ``fluctuating superconductor'' regimes of Fig.~\ref{fig:phaseDiagram}. For the main
results of this study, we need only the coupling in the quantum critical region, where the physics of 
the plasmon mode is unchanged from that in the ``Metal'' region of Fig.~\ref{fig:phaseDiagram}. 
After integrating out the fermions, we obtain the $A_\tau$ action
\begin{equation}
\mathcal{S}_{\Pi} = \int \frac{dk}{2\pi} \int \frac{d \omega}{2 \pi} \frac{1}{2} \Pi (k, 
\omega) |A_\tau (k, \omega)|^2  ,
\label{eq:SPi}
\end{equation}
where $\Pi$ is the irreducible density correlation function (the ``polarizability'') of the 
metal. For $\omega \gg k$, we have $\Pi(k,\omega) \sim k^2/\omega^2$, and then 
$\mathcal{S}_A + \mathcal{S}_\Pi$ has a pole at the plasmon frequency noted above. We also observe that
the coupling between these $A_\tau$ fluctuations and $\Psi$ is negligible: a slowly varying 
$A_\tau$ is a shift in the local chemical potential, and to the extent we can ignore the variation 
in the electronic density of states at these energy scales, the effective couplings in
$\mathcal{S}_{\alpha}$ do not change with $A_\tau$, and there is no $A_\tau - \Psi$ coupling.  
The existence of an $A_\tau - \Psi$ coupling that is non-analytic in frequency has been discussed by
Ioffe and Millis \cite{ioffe-millis}.  However, they show by Ward identities, which also apply here,
that these couplings do not contribute to the physical charge correlations.  So, in the quantum
critical region, $\mathcal{S}_{\alpha}$ and $\mathcal{S}_A + \mathcal{S}_{\Pi}$ are independent
theories describing the pairing and charge fluctuations respectively. 

As described in the previous section, the coupling constant which measures particle-hole asymmetry
is formally, albeit weakly relevant (see Eq.~(\ref{eq:detadl})) and thus a small $A_\tau - \Psi$
coupling appears by making the $\tau$-derivative in Eq.~(\ref{eq:Seta}) Gauge covariant
\begin{equation}
\mathcal{S}_{\rho} = \int dx d \tau \left[ \rho\, \Psi^\ast(x,\tau)\left( \frac{\partial}{\partial 
\tau} - 2ei A_\tau \right) \Psi(x,\tau) \right].
\label{eq:SetaA}
\end{equation}
However, the combination of the small scaling dimension and the small bare value of $\eta$ implies 
that such particle-hole asymmetric effects, and the consequent coupling between pairing and charge 
fluctuations, can justifiably be ignored in theories hoping to describe realistic experiments.

We conclude by addressing the physics in the ``Fluctuating superconductor'' regime of
Fig.~\ref{fig:phaseDiagram} for $\alpha < \alpha_c$. 
In this discussion, we neglect the possibility of a narrow region 
$\alpha \sim \alpha_c$ with gapless superconductivity, which is known to occur in the vicinity of a
pair breaking transition \cite{deGennes}.  Now coupling between the pairing and charge fluctuations
is much stronger.  When $T  < (\alpha_c - \alpha)$ the action $\mathcal{S}_{\alpha}$ does \emph{not}
apply for the smallest frequencies. The reasons for this are again analogous to
arguments made for the spin-density-wave ordering transition in metals, as discussed in
Section~\ref{sec:dissipativeModel} and Ref.~\cite{qpt}.  For the latter case, it was argued that
with the emergence of long-range spin density wave order, the low energy fermionic particle-hole
excitations at the ordering wavevector were gapped out, and so the diffusive paramagnon action
applied only for energies larger than this gap. At energies smaller than the gap, spin-waves with
dispersion $\omega \sim k$ emerge.  In the superconducting case, there is no true long-range  order
at any $T>0$, but the order is disrupted primarily by `renormalized classical'  thermal fluctuations
of the phase, $\phi$ of the complex $\Psi$ field.  We assume that there is a local pairing amplitude
in the fermion spectrum, analogous to the spin-density wave order. The low energy effective action
for $\phi$ obtained by integrating the fermions in the presence of a local pairing, is
\begin{equation}
\mathcal{S}_\phi = \int dx d\tau \left\{[K_1 \left[ \partial_\tau \phi(x,\tau) - 2 e A_\tau \right]^2 
+ K_2 \left[\partial_x \phi(x,\tau)\right]^2 \right\} 
\label{eq:Sphi}
\end{equation}
where $K_{1,2}$ vanish as power of $(\alpha_c - \alpha)$ \cite{scs}.  The strongly coupled pairing
and charge fluctuations in the ``Fluctuating superconductor'' regime of Fig.~\ref{fig:phaseDiagram} are 
described by $\mathcal{S}_A+\mathcal{S}_\phi$, and this theory contains the Mooij-Sch\"on 
mode, which is the analog of the `spin-wave' mode. We do \emph{not} claim that
$\mathcal{S}_A+\mathcal{S}_\phi$ can extended across quantum criticality into the normal 
phase, in contrast to other works \cite{zaikin} which consider vortex unbinding in such a theory.

The arguments presented in this section strive to justify the use of $\mathcal{S}_{\alpha}$
throughout the quantum critical regime where the presence of repulsive interactions between Cooper
pairs will be essential in obtaining the correct form of the full crossover transport scaling
functions $\Phi_\sigma$ and $\Phi_\kappa$.  This is most clearly seen by examining the energy and
length scales at which various terms in $\mathcal{S}_\alpha$ become dominant in both the clean limit
where the mean free path is longer than the superconducting coherence length, and the dirty limit,
where the opposite is true.  

\subsection{Universality and interactions in the quantum critical regime}

The microscopic theory of the superconductor-metal transition was considered in great detail 
by Shah and Lopatin (SL) \cite{shah-lopatin} in a Gaussian theory of superconducting fluctuations that
corresponds to the effective field theory presented here with $u$ set to zero. In 
Section~\ref{chap:largeN} we will compare the transport properties computed from the effective action
$\mathcal{S}_{\alpha}$ with Shah's results in the metallic regime where $\alpha \gg \alpha_c$.
Before doing so we will need to connect the (until now) phenomenological coupling constants of
Eq.~(\ref{eq:Salpha}) to those of microscopic BCS theory.  The details are given in
Appendix~\ref{app:microBCS} with the main result being that in the dirty ($d$) limit, the diffusion
($\D$), dissipation ($\gamma$) and interaction ($u$) constants are given by 
\begin{align}
\D_d &= D = \frac{1}{3} v_\mathrm{F} \ell 
\\
\gamma_d &\simeq \frac{1.5}{k_\mathrm{F} \ell} 
\\
u_d &\simeq 2.9\frac{v_F}{\hbar N_{\perp}}
\end{align}
where the number of transverse conduction channels in the wire $N_\perp$ is assumed to be large
and in the clean ($c$) limit
\begin{align}
\D_c &= \frac{1}{4} v_F \xi_0  
\\
\gamma_c &\simeq \frac{2.0}{k_\mathrm{F} \xi_0} 
\\
u_c &= u_d \simeq 2.9\frac{v_F}{\hbar N_{\perp}}.
\end{align}
Note that the microscopic value of the quartic coupling constant $u$ is identical in both the clean
and dirty limit.

Armed with the microscopic values of the model parameters we can make a number of observations
regarding the validity of the non-interacting theory in the strongly fluctuating quantum critical 
regime (which is referred to as the \emph{classical regime} in SL).  In the non-interacting theory, 
the temperature dependence of the superconductor-metal phase boundary $\alpha_c(T)$ is 
computed from an expansion of the mean-field result of Abrikosov and Gor'kov given in
Eq.~(\ref{eq:AGTheory}) for $T \ll \alpha_c$. In SL, the deviation from criticality is measured
with respect to the finite temperature mean field phase boundary 
\begin{equation}
\alpha_c(T) = \alpha_c - \frac{\pi \gamma \kB  T^2}{3\hbar T_{c0}},
\label{eq:alphaT}
\end{equation}
where $\gamma \approx 0.577$ is the Euler-Mascheroni constant and $T_{c0}$ is the classical 
BCS transition temperature in the absence of pairbreaking.  This notation (which
we temporarily adopt) requires comment.  The location of the quantum critical point is as usual
defined as $\alpha_c$ at $T=0$, but in Eq.~(\ref{eq:alphaT}) $\alpha_c(T)$ is the function which
locates the value of the pair breaking frequency at which superconducting order is lost for a given
temperature.  It is the approximate functional inverse of Eq.~(\ref{eq:AGTheory})
in the low temperature limit.  To summarize, SL are interested in large positive values of $\alpha$
far into the metallic phase and  have chosen to define a coupling constant that measures the
distance from classical criticality defined by the temperature dependent mean field phase boundary
and \emph{not} the distance from quantum criticality.

To successfully compare the approach considered here with that of SL, we shift the definition of
$\alpha$ accordingly and write (returning to physical units)
\begin{equation}\begin{split}
\mathcal{S} &= \int d x \int d\tau \left[ \D |\partial_x \Psi(x,\tau)|^2 +
\frac{\pi \gamma \kB }{3\hbar T_{c0}}T^2 |\Psi(x,\tau)|^2 
+ \frac{u}{2} |\Psi(x,\tau)|^4 \right] 
\\
& \qquad +\; \frac{\kB T}{\hbar} \sum_{\wn} \int d x\,  |\wn| |\Psi(x,\omega_n)|^2 .
\label{eq:SalphaT}
\end{split}\end{equation}
In this form, it is clear that the theory only goes quantum critical at $T=0$.  The coupling
constants $\D$ and $u$ can take on the values computed in Appendix~\ref{app:microBCS} for the clean
and dirty limits, but we are primarily concerned with the role of the quartic coupling $u$,
characterizing the strength of the Cooper pair self interaction.  For $d=1$,  $u$ has scaling
dimension one and from Eq.~(\ref{eq:uDirty}) it has engineering dimensions of inverse mass times
inverse length or frequency squared times length over energy.  There are three distinct regions of
the phase diagram as the temperature is reduced in the quantum critical regime set by the size of
the bare value of $u$ defined by the conditions:
\begin{enumerate}
\renewcommand{\labelenumi}{\Roman{enumi}}
\item The quartic coupling can be ignored and the Gaussian theory of SL fully describes transport.
\item Interactions are important, and Hartree corrections to the mass must be included
leading to non-universal results.
\item The quartic coupling is relevant, its bare value is large, and all results are universal.
\end{enumerate}

Cases I and II can be distinguished by examining the lowest order correction to the
perturbatively renormalized or Hartree corrected mass coming from Eq.~(\ref{eq:SalphaT}) at one loop
order
\begin{equation} \begin{split}
R &= \frac{\pi \gamma \kB T^2}{3 \hbar T_{c0}} 
\\
& \quad -\; \hbar u \int \frac{dk}{2\pi} 
\left( \int \frac{d\omega}{2\pi} \frac{1}{\D k^2 + |\omega|} 
- \frac{\kB T}{\hbar} \sum_{\omega_n} \frac{1}{\D k^2 + |\omega_n| 
+ \pi \gamma  \kB T^2/ 3\hbar T_{c0}}  \right)
\end{split}\end{equation}
where we have applied the usual shift to subtract off a zero temperature contribution so that our
renormalized mass $R=0$ at quantum criticality.  The Hartree correction can be separated into two
contributions, one coming from the integral, and one coming from the most dominant contribution to
the sum, the $\omega_n = 0$ term.  These are given by
\begin{equation}\begin{split}
& \int \frac{d\omega}{2\pi}\int \frac{dk}{2\pi} \left( \frac{1}{\D k^2 +
|\omega_n| + \pi \gamma \kB  T^2/3\hbar T_{c0}} 
- \frac{1}{\D k^2 + |\omega|} \right) = -\frac{1}{\pi} 
\sqrt{\frac{\pi \gamma \kB  T^2}{3 \hbar \D T_{c0}}}
\end{split}\end{equation}
and 
\begin{equation}
\frac{\kB T}{\hbar} \int \frac{dk}{2\pi} \frac{1}{\D k^2 + \pi \gamma  T^2/3T_{c0}} 
= \frac{1}{2}\sqrt{\frac{3 \kB T_{c0}}{\pi \gamma \hbar \D}} 
\end{equation}
respectively.  Provided that $T<T_{c0}$, the second is the most dominant contribution, and thus the
renormalized mass is significant when it is greater than the bare mass of Eq.~(\ref{eq:alphaT}),
\begin{equation}
\frac{\hbar u}{2} \sqrt{\frac{3 \kB  T_{c0}}{\pi \gamma \hbar \D}} > 
\frac{\pi \gamma \kB T^2}{3\hbar T_{c0}}
\end{equation}
which defines the Hartree temperature
\begin{equation}
\kB T_\mathrm{H} = \left(\frac{3 \hbar\kB T_{c0}}{\pi\gamma}\right)^{3/4}
\left(\frac{u}{2\sqrt{\D}}\right)^{1/2}.
\label{eq:TH}
\end{equation}
For temperatures above $T_\mathrm{H}$ one can ignore the presence of a repulsive interaction between
the Cooper pairs, and the non-interacting results of Ref.~\cite{shah-lopatin} will be accurate.

The temperature below which all results scale to universal values can be obtained by considering the
thermal length which follows naturally from our previous scaling analysis $L_T \sim T^{-1/z}$ or
more precisely for $z=2$ 
\begin{equation}
L_T = \sqrt{\frac{\hbar \D}{\kB T}}.
\label{eq:LT}
\end{equation}
The bare quartic coupling can be assumed to be large with respect to all other parameters and thus 
flow to infinite strength when the potential energy is greater than the kinetic energy, i.e., 
\begin{equation}
\frac{\hbar^2 u}{L_T} > \frac{\hbar \D}{L_T^2}.
\end{equation}
This relation sets the temperature $T_\mathrm{U}$ below which one can safely take $u\to\infty$ and
obtain universal results to be
\begin{equation}
\kB T_\mathrm{U} = \frac{\hbar^3 u^2}{\D}.
\label{eq:TU}
\end{equation}
The values of the microscopic parameters computed in Appendix~\ref{app:microBCS} and repeated above
can be used to evaluate the temperatures defined in Eqs.~(\ref{eq:TH}) and (\ref{eq:TU}) which
separate the regions I-II and II-III.  As before, the results depend on whether the system is taken
to be in the clean or dirty limit.

\subsubsection{Dirty Limit $(\xi_0 \gg \ell)$}
The Hartree temperature in the dirty limit can be found by substituting Eqs.~(\ref{eq:Tc0}),
(\ref{eq:DDirty}) and (\ref{eq:uDirty}) in Eq.~(\ref{eq:TH})
\begin{equation}
T_{\mathrm{H},d} = 0.83\frac{\hbar v_\mathrm{F}}{(\xi_{\mathrm{loc}} N_\perp \xi_0^3)^{1/4}},
\label{eq:THd}
\end{equation}
where the single electron localization length is defined to be
\begin{equation}
\xi_\mathrm{loc} = N_\perp \ell .
\end{equation}
This temperature can be converted into a length scale, which gives a lower bound on lengths over
which one must explicitly include Hartree corrections
\begin{equation}
L_{\mathrm{H},d} = 0.63 \xi_{\mathrm{loc}}^{1/4} (\ell \xi_0)^{3/8}.
\label{eq:LHd}
\end{equation}
The universal temperature scale is found from Eq.~(\ref{eq:TU}) to be
\begin{equation}
T_{\mathrm{U},d} = \frac{25}{N_\perp} \frac{ \hbar v_\mathrm{F}}{\xi_{\mathrm{loc}}},
\label{eq:TUd}
\end{equation}
corresponding to length scales longer than 
\begin{equation}
L_{\mathrm{U},d} = 0.12\xi_{\mathrm{loc}}.
\label{eq:LUd}
\end{equation}

\subsubsection{Clean Limit $(\xi_0 \ll \ell)$}
The same analysis can be repeated using Eqs.~(\ref{eq:DClean}) and (\ref{eq:uClean}) for the clean
limit.  The Hartree temperature is given by
\begin{equation}
T_{\mathrm{H},c} = 0.96 \frac{\hbar v_\mathrm{F}}{\sqrt{N_\perp} \xi_0}
\label{eq:THc}
\end{equation}
with associated length scale
\begin{equation}
L_{\mathrm{H},c} = 0.51 N_\perp^{1/4} \xi_0.
\label{eq:LHc}
\end{equation}
For universal results we find
\begin{equation}
T_{\mathrm{U},c} = 33\frac{\hbar v_\mathrm{F}}{N_\perp^2 \xi_0}
\label{eq:TUc}
\end{equation}
with 
\begin{equation}
L_{\mathrm{U},c} = 0.090N_\perp \xi_0.
\label{eq:LUc}
\end{equation}
The results are summarized in Table~\ref{tab:TLScales}, but it is immediately clear that 
when in the clean limit, $L_T > L_{\mathrm{U},c}$ can be easily satisfied, whereas in the dirty
limit $L_T > L_{\mathrm{U},d}$ would require lengths on the order of $\xi_\mathrm{loc}$, and
thus weak localization effects could become important.  
\begin{center}
\begin{table}
\renewcommand{\tabcolsep}{8pt}
\renewcommand{\arraystretch}{2}
\begin{tabular}{|c|c|c|} \hline
\textbf{Gaussian} & \textbf{non-Gaussian} & \textbf{non-Gaussian} \\ 
& \textbf{non-universal} & \textbf{universal} \\ \hline \hline
\multicolumn{3}{|c|}{\emph{Dirty Limit} $(\xi_0 \gg \ell)$} \\ \hline
$\kB T > 0.83 \frac{\hbar v_\mathrm{F}}{(\xi_{\mathrm{loc}} N_\perp \xi_0^3)^{1/4}}$ &
$\frac{25}{N_\perp} \frac{\hbar v_\mathrm{F}}{\xi_{\mathrm{loc}}} < 
\kB T < 0.83\frac{\hbar v_\mathrm{F}}{(\xi_{\mathrm{loc}} N_\perp \xi_0^3)^{1/4}}$  &
$ \kB T < \frac{25}{N_\perp} \frac{\hbar v_\mathrm{F}}{\xi_{\mathrm{loc}}} $ \\ \hline 
$ L < 0.63 \xi_{\mathrm{loc}}^{1/4} (\ell \xi_0)^{3/8}$ & 
$ 0.63 \xi_{\mathrm{loc}}^{1/4} (\ell \xi_0)^{3/8} < L <  0.12\xi_{\mathrm{loc}}$ &
$ L > 0.12\xi_{\mathrm{loc}}$ \\ \hline
\multicolumn{3}{|c|}{\emph{Clean Limit} $(\xi_0 \ll \ell)$} \\ \hline
$ \kB T > 0.96\frac{\hbar v_\mathrm{F}}{\sqrt{N_\perp} \xi_0} $ &
$ 33\frac{\hbar v_\mathrm{F}}{N_\perp^2 \xi_0} < \kB T < 
0.96\frac{\hbar v_\mathrm{F}}{\sqrt{N_\perp} \xi_0}$ &
$ \kB T < 33\frac{\hbar v_\mathrm{F}}{N_\perp^2 \xi_0}$ \\ \hline
$L < 0.51 N_\perp^{1/4} \xi_0$ & 
$ 0.51 N_\perp^{1/4} \xi_0 < L < 0.090N_\perp \xi_0 $ & 
$L > 0.090N_\perp \xi_0$ \\ \hline
\end{tabular}
\caption{The temperature and length scales in the clean and dirty limits corresponding to the
regions of applicability described in I-III for the effective action $\mathcal{S}_\alpha$.}
\label{tab:TLScales}
\end{table}
\end{center}

We must therefore restrict our analysis to temperatures greater than those where disorder 
effects set in, which we now investigate.

\subsection{The role of disorder}
\label{subsec:roleDisorder}
Until now, the presence of disorder in the wire, manifest as spatially dependent
coefficients in $\mathcal{S}_\alpha$ has been neglected.  This topic has already been studied in
great detail by the authors \cite{diswireslett} and will not be focused upon here.  Instead, 
we choose to consider only those temperatures above which weak localization effects can be safely
neglected and limit the scope of our results to the non-random universality class.  An
estimate of the temperature scale $T_\mathrm{dis}$ where disorder effects must be included is found
by equating the thermal length with the localization length. This yields
\begin{equation}
L_T = \xi_\mathrm{loc} = N_\perp \ell,
\end{equation}
and using Eq.~(\ref{eq:LT}) and Eq.~(\ref{eq:DDirty}) we find
\begin{equation}
\kB T_\mathrm{dis} = \frac{\hbar}{3 N_\perp^2 \tau}
\label{eq:Tdis}
\end{equation}
where $\tau = \ell/v_\mathrm{F}$ is the elastic scattering time.  $T_\mathrm{dis}$ can
therefore be made arbitrarily small by considering thicker or cleaner wires.

The analysis performed in this section has provided a firm foundation for the applicability of the
effective action $\mathcal{S}_\alpha$ to the SMT in ultra-narrow wires.  In the next section 
transport results are computed near this quantum phase transition in the limit where the number of
complex components of $\Psi$ is large.

\section{Transport in the Large-$N$ limit}
\label{chap:largeN}
In order to incorporate the repulsive interaction between pairing fluctuations, the effective
dissipative action must be generalized from the physical case of describing a $1$-component complex
field $\Psi$ corresponding to the Cooper pair operator, to an $N$-component complex field $\Psi_a$
with $a=1,\ldots, N$. This section presents a calculation of the thermal $\kappa$ and electrical
$\sigma$ dc transport coefficients in the ``LAMH'', ``quantum critical'' and ``metal'' regimes
described in Fig.~\ref{fig:phaseDiagram} through the application of both analytical and numerical
methods.  It is always assumed, unless otherwise specified, that these fluctuation corrections are
the most singular terms at finite temperature resulting from the direct contribution to transport
due to Cooper pairs, (i.e. any subscripts on transport coefficients are suppressed).  It is
understood that making contact with real experimental measurements would require a subtraction of
normal state transport results.

\subsection{Effective classical theory}
\label{subsec:effClassTheory}
Generalizing the effective action of Eq.~(\ref{eq:Salpha}) to describe the fluctuations of a 
$N$-component pairing field at finite temperature
\begin{equation}\begin{split}
\mathcal{S}_\alpha &= \int d x \int d\tau \left[ \D |\partial_x \Psi_a(x,\tau)|^2 
+ \alpha |\Psi_a(x,\tau)|^2 
+ \frac{u}{2} |\Psi_a(x,\tau)|^4 \right]  
\\
& \qquad +\; T \sum_{\wn} \int d x\,  |\wn| |\Psi_a(x,\wn)|^2 ,
\label{eq:SalphaN}
\end{split}\end{equation}
where we have used the short hand notation
\begin{align}
|\Psi_a(x,\tau)|^2 &\equiv |\Psi_1(x,\tau)|^2 + \cdots + |\Psi_N(x,\tau)|^2 \\
|\Psi_a(x,\tau)|^4 & \equiv \left[|\Psi_1(x,\tau)|^2 + \cdots + |\Psi_N(x,\tau)|^2\right]^2
\end{align}
for the sake of compactness.  

The aforementioned goal of calculating uniform electrical and thermal transport properties in
the dc limit will require special care to ensure that the zero frequency limit is taken while the
temperature remains finite.  The universal properties of the SMT can be most easily accessed in the
strong coupling regime, $u\to\infty$, while the ratio of $u$ to $\alpha$ is held fixed.  The result
is a much simpler \emph{hard spin} quadratic action which is written in Fourier space as
\begin{equation}
\mathcal{S}_g = \frac{T}{g} \sum_{\wn} \int \frac{dk}{2\pi}
|\Psi_a (k, \wn)|^2 (\D k^2 + |\wn|). 
\label{eq:Sg}
\end{equation}
along with the constraint $|\Psi_a (x, \tau)|^2 = 1$ where $\Psi_a(k,\wn)$ is the Fourier transform
of $\Psi_a(x,\tau)$ defined by
\begin{equation}
\Psi_a(k,\wn) = \int dx \int d\tau \Psi_a(x,\tau) \ee{-i(kx - \wn \tau)}.
\end{equation}
The parameter $g$ now tunes across the quantum critical point and  the quantum partition function is
given by 
\begin{equation}
\mathcal{Z} = \int \mathcal{D}\Psi^{\phantom{\ast}}_a 
\mathcal{D}\Psi_a^\ast\ \delta(|\Psi^{\phantom{\ast}}_a|^2-1)\ \ee{-\mathcal{S}_g}.
\label{eq:Zquantum}
\end{equation}

We will first investigate observables at frequencies much smaller than the temperature 
in the continuum quantum critical regime ($\hbar \omega \ll \kB T$).  Here, temperature plays the
role of an infrared cutoff and the effects of quantum fluctuations can be integrated out, producing
an effective classical theory with renormalized parameters.  This can be done in the large-$N$ limit
by first imposing the constraint $|\Psi_a(x,\tau)|^2 = 1$ via a Lagrange multiplier $\mu$ and then
integrating out all non-zero Matsubara frequencies from $\mathcal{Z}$ over their Gaussian action.
The resulting effective action has an overall factor of $N$, and as $N \to \infty$ we can perform
the functional integral over $\mu$ using the saddle point approximation where we replace $r = i\mu$.
This yields the classical partition function 
\begin{equation}
\mathcal{Z}_c = \int \mathcal{D} \Psi^{\phantom{\ast}}_a\mathcal{D}
\Psi_a^\ast \exp \left\{ - \frac{N}{T} \int dx
\left[ \D |\partial_x \Psi_a(x) |^2 + V (|\Psi_a (x)|^2) \right]\right\}
\label{eq:Zc}
\end{equation}
where 
\begin{equation}
\Psi_a (x) = \frac{T}{\sqrt{g}} \int_0^{1/T} d \tau \Psi_a (x, \tau)
\end{equation}
is an imaginary time independent classical field governed by the sombrero shaped effective potential
$V(z)$ ($z \equiv |\Psi_a|^2$) given by 
\begin{equation}
V(z) = z r(z)  + T \sum_{\wn \neq 0} \int \frac{dk}{2 \pi} \ln
[ \D k^2 + |\wn| + r(z)] - \frac{r(z)}{g}  .
\label{eq:Vz}
\end{equation}
The function $r = i\mu$ is to be determined by solving the saddle point constraint equation 
$\partial V / \partial r = 0$,
\begin{equation}
z + T \sum_{\wn \neq 0} \int \frac{dk}{2 \pi} \frac{1}{\D k^2 +
|\wn| + r(z)} = \frac{1}{g}. 
\label{eq:rz}
\end{equation}

\begin{figure}[t]
\centering
\includegraphics*[width=3.2in]{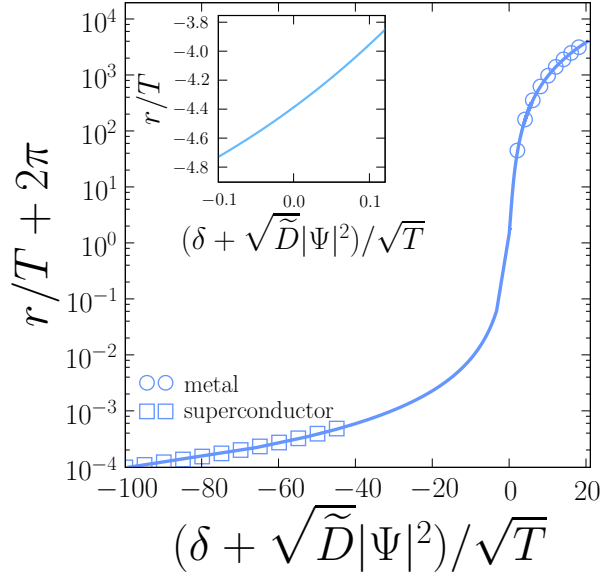}
\caption{\label{fig:rzT}  The numerical solution of the transcendental saddle point
equation (\ref{eq:deltaz}) which will be used in all computations of the effective classical
potential $V(z)$. The symbols were calculated using the approximate solution to $r/T$ found in the
metallic (Eq.~(\ref{eq:szM})) and superconducting (Eq.~(\ref{eq:szSC})) limits.}
\end{figure}

The scaling limit of equations (\ref{eq:Vz}) and (\ref{eq:rz}) can be reached leading to a 
universal, cutoff-independent expression for $V(z)$.  First consider Eq.~(\ref{eq:rz}) and note that
the $T=0$ quantum critical point is at $g=g_c$, where $g_c$ is determined in the large-$N$ limit by
\begin{equation}
\int \frac{d\omega}{2\pi} \int \frac{dk}{2\pi} \frac{1}{\D k^2 +
|\omega|} = \frac{1}{g_c} 
\label{eq:gc}
\end{equation}
and an ultra-violet (UV) cutoff must be included for finiteness. Defining
\begin{equation}
\delta \equiv \sqrt{\D}\left( \frac{1}{g_c} - \frac{1}{g} \right),
\label{eq:delta}
\end{equation}
as a renormalized tuning parameter, the quantum critical point now resides at $\delta=0$,
$T=0$. Subtracting Eq.~(\ref{eq:gc}) from (\ref{eq:rz}) and performing the sum and integral up to a
UV frequency cutoff $\Lambda_\omega$
\begin{equation}\begin{split}
&\delta + z \sqrt{\D}  
= \sqrt{\D} \int \frac{dk}{2\pi^2} \left[ \ln \left(
\frac{\Lambda_\omega + \D k^2}{\D k^2} \right) \right. 
\\
&\left. \qquad -\; \psi \left( 1 + \frac{
\Lambda_\omega + \D k^2 + r(z)}{2 \pi T} \right) 
+ \psi\left(1 + \frac{\D k^2 + r(z)}{2\pi T} \right) \right]
\end{split}\end{equation}
where $\psi$ is the digamma function. In this form, the limit
$\Lambda_\omega \rightarrow \infty$ can now be safely be taken, and after rescaling to a
dimensionless momentum
\begin{equation}
\frac{\delta}{\sqrt{T}} + z \sqrt{\frac{\D}{T}} = 
\int \frac{dk}{2\pi^2} \left[ \ln \left( \frac{2 \pi}{ k^2} \right) +
\psi \left( 1 + \frac{ k^2 + r(z)/T}{2 \pi} \right) \right] .
\label{eq:deltaz}
\end{equation}
This is one of the most important results in the scaling limit, and determines $r(z)/T$ as a
universal function of $\delta/\sqrt{T}$ and $z \sqrt{\D/T}$. A numerical solution of
Eq.~(\ref{eq:deltaz}) is shown in Fig.~\ref{fig:rzT}, and we note that it has a minimum possible
value of $-2\pi$ due to the argument of the polygamma function.
The effective potential $V=V(z,\delta,T)$ and renormalized mass $r=(z,\delta,T)$ are actually
functions of three variables $z=|\Psi_a|^2$, $\delta$ and $T$.  For the sake of brevity, we will
usually just explicitly indicate their $z = |\Psi_a|^2$ dependence whenever possible.

\begin{figure}[t]
\centering
\includegraphics*[width=3.2in]{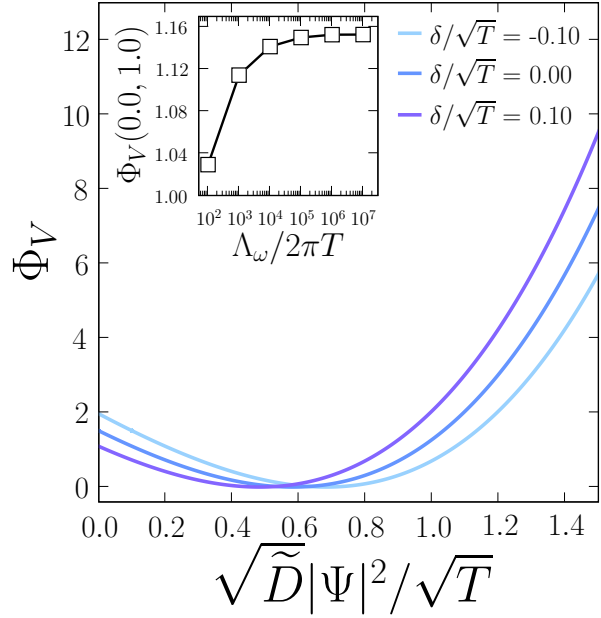}
\caption{\label{fig:vz} The scaling form of the effective potential calculated by
including $10^7$ terms in the frequency summation for $\delta/\sqrt{T} = -0.1,0.0,0.10$ (right 
to left curve).  The inset shows the convergence properties of
the sum for fixed $\delta/\sqrt{T}$ and $z = |\Psi_a|^2$.}
\end{figure}

Substituting the expression for $1/g$ in Eq.~(\ref{eq:rz}) into Eq.~(\ref{eq:Vz}), and subtracting a
constant which is independent of $z$, we obtain using $\wn = 2\pi n T$
\begin{equation}
V(z) =  \sqrt{\frac{2\pi}{\D}} T^{3/2}\sum_{n=1}^{\infty} \left[
\frac{2n + r(z)/2\pi T}{\sqrt{n + r(z)/2\pi T}} - 2\sqrt{n}\right].
\label{eq:Vzr}
\end{equation}
The structure of the effective classical potential indicates that it can be written in the scaling
form
\begin{equation}
V(z,T,\delta) = \frac{T^{3/2}}{\sqrt{\D}} \Phi_{V} \left(
\frac{\delta}{\sqrt{T}}, z \sqrt{\frac{\D}{T}} \right) 
\label{eq:VzScaling}
\end{equation}
where $\Phi_{V}$ is a universal dimensionless function.  By truncating the sum in Eq.~(\ref{eq:Vzr})
at some large, but finite value, the scaling function $\Phi_{V}(z)$ can be evaluated at fixed
$\delta/\sqrt{T}$ as seen in Fig.~\ref{fig:vz}.  For $\delta/\sqrt{T} = z \sqrt{\D/T} = 0$ we find
$r = -0.69728$ leading to $V(0) = 1.5100$.

\subsection{Limiting forms of $V(z)$}
An analytic form for the effective potential can be determined for the limiting cases
$\delta/\sqrt{T} \to \infty$ and $\delta/\sqrt{T} \to -\infty$ corresponding to the metallic and
superconducting phases respectively.  

\subsubsection{Metallic phase}
In the metallic phase, a large positive value of $\delta$ is reached by investigating the
${r(z)} \to \infty$ behavior of Eq.~(\ref{eq:deltaz}). In this limit, the asymptotic expansion of
the gamma function can be used, leading to
\begin{equation}
r(z) \simeq \pi^2 \left(\delta + z\sqrt{\D}\right)^2
\label{eq:szM}
\end{equation}
with the the result plotted using circular symbols in Fig.~\ref{fig:rzT}.  For $r(z) \gg 1$, the
effective potential in Eq.~(\ref{eq:Vzr}) can be rewritten in terms of a sum over Matsubara
frequencies.  After adding and subtracting the $n=0$ term 
\begin{equation}
V(z) = -\frac{T}{\sqrt{\D}} \sqrt{r(z)} 
+ \frac{T}{\sqrt{\D}} \sum_{n=0}^{\infty}\left( \frac{2\wn + r(z)}{\sqrt{\wn + r(z)}} 
- 2\sqrt{\wn}\right).
\end{equation}
In the low temperature limit, the Matsubara summation can be converted into
an integral
\begin{equation}
V(z) \simeq -\frac{T}{\sqrt{\D}} \sqrt{r(z)} + \frac{1}{3\pi\sqrt{\D}}[r(z)]^{3/2}
\end{equation}
and using Eq.~(\ref{eq:szM}) we find the temperature independent result
\begin{equation}
V(z) \sim \left(\delta + z\sqrt{\D}\right)^{3}.
\label{eq:VzMT}
\end{equation}

\subsubsection{Superconducting phase}
For the strongly ordered phase, ${\delta}/{\sqrt{T}} \to -\infty$ and the analysis is less
straightforward.  As noted previously, $r(z)$ is bounded from below by the first negative Matsubara
frequency $-\omega_1 = -2\pi T$, and near this value, the most divergent term in
Eq.~(\ref{eq:deltaz}) produces
\begin{equation}
r(z) \simeq -\omega_1 + T^2 \left( \delta + z\sqrt{\D}\right)^{-2},
\label{eq:szSC}
\end{equation}
plotted as open squares in Fig.~\ref{fig:rzT}.  With $r(z)$ taking this extreme value,
all terms in Eq.~(\ref{eq:Vzr}) are well behaved, \emph{except} the $n = 1$ term. Extracting 
the culprit
\begin{equation}
V(z) = \frac{T}{\sqrt{\D}}\left[  \left( \frac{2\omega_1 + r(z)}{\sqrt{r(z) + 
\omega_1}} - 2\sqrt{\omega_1} \right) 
+ \sum_{n=1}^{\infty} \left( \frac{ 2\omega_{n+1} + r(z)}{\sqrt{r(z) +
\omega_{n+1}}} - 2\sqrt{\omega_{n+1}} \right) \right]
\end{equation}
where the sum has been shifted by one.  Substituting Eq.~(\ref{eq:szSC}) in the first term to 
investigate the divergence, and setting $r(z) = -\omega_1$ in the second, the low temperature form
of the interaction potential is given by  
\begin{equation}
V(z) \sim -2\pi \left(\delta + z\sqrt{\D}\right)T .
\label{eq:VzSCT0}
\end{equation}

It seems somewhat surprising that in the superconducting phase, the effective potential vanishes
linearly with temperature, as $T\to0$.  However, at this point, we will endeavor to calculate
transport only in the quantum critical regime where $\delta/\sqrt{T} \ll 1$ and can thus safely
ignore the irregularity. We will return to the issue in Section~\ref{subsec:orderedPhase} by
investigating the low temperature ordered phase through a calculation of the Coleman-Weinberg
effective potential at $T=0$, as well as constructing an effective Ginzburg-Landau potential, near
$T_c$.

\subsection{Renormalized classical conductivity}
\label{subsec:classicalConductivity}
Now the fruits of our labor become apparent as the dc electrical conductivity can be calculated for
the effective classical action described by Eq.~(\ref{eq:Zc}) via the Kubo formula \cite{mahan} by
reintroducing a \emph{real} time dependence to the classical order parameter and approximating its
low frequency dynamics by a Langevin equation \cite{podolsky-sachdev}.  The corresponding equation
of motion is 
\begin{equation}
\frac{\partial \Psi(x,t)}{\partial t} = \D \frac{\partial^2
\Psi(x,t)}{\partial x^2} - V'(|\Psi(x,t)|^2) \Psi(x,t)  + f(x,t) 
\label{eq:eom}
\end{equation}
where $f$ is a complex Gaussian correlated random noise obeying
\begin{equation}
\langle f(x,t) f^\ast (x',t') \rangle = 2 T \delta(x-x') \delta (t-t').
\end{equation}
By taking a derivative of Eq.~(\ref{eq:Vz}) and using the saddle point equation (\ref{eq:rz}),
$V'(z) = r(z)$.  Eq.~(\ref{eq:eom}) represents the simple Model A dynamics of
Ref.~\cite{hh} and should capture the correct quantum critical dynamics whenever the
renormalized mass $r$ takes a value such that the $\wn\ne0$ modes are sufficiently gapped.

The electrical current is defined to be
\begin{equation}
J = i e^\ast \D \left( \Psi^\ast \partial_x \Psi - \partial_x
\Psi^\ast \Psi \right)
\end{equation}
and thus the uniform dc conductivity can be found from an integral over all space and time of the
current-current correlation function over the partition function $\mathcal{Z}_c$
\begin{equation}
\sigma = \frac{1}{T} \int_0^L d x \int_0^\infty dt \langle J(x,t) J(0,0)
\rangle 
\label{eq:sigmaDef}
\end{equation}

Dimensional analysis of the equation of motion above in conjunction with Eq.~(\ref{eq:VzScaling}),
implies that the classical conductivity obeys the scaling form
\begin{equation}
\sigma = \frac{e^{\ast 2}}{\hbar} \sqrt{\frac{\hbar \D}{\kB T}} \Phi_\sigma
\left( \frac{\delta}{\sqrt{\hbar \kB T}} \right)
\label{eq:PhiSigma}
\end{equation}
where we have inserted the dimensionally correct powers of $\hbar$ and $\kB$ in the final result.
This is simply Eq.~(\ref{eq:PhiSigScaleG}) with the replacement given in Eq.~(\ref{eq:largeNScaleVar})
written in terms of our new measure of the distance from criticality $\delta$.

The scaling function $\Phi_\sigma (x)$ is a smooth function of $x$ through $x=0$ and it can be
determined by finding a numerical solution to the classical equation of motion (\ref{eq:eom}) 
for a one-component classical complex field $\Psi(x,t)$.  By employing both classical
Monte Carlo simulations and a stochastic partial differential equation solver, its full
time-evolution can be determined.  

\begin{figure}[t]
\centering
\includegraphics*[width=3.2in]{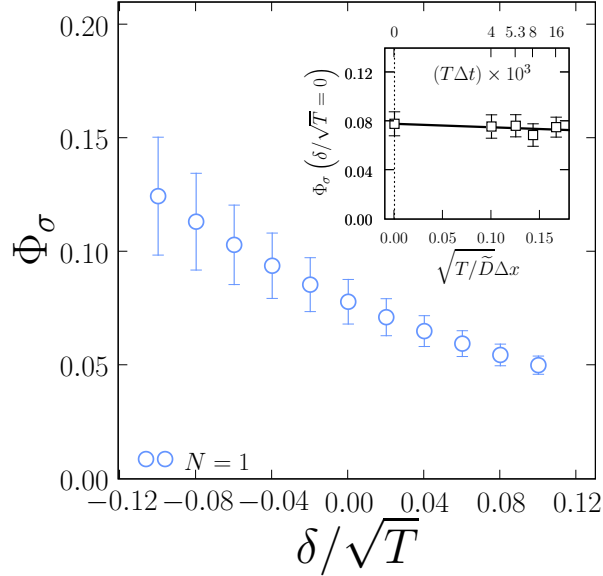}
\caption{\label{fig:PhiSN1}  The dc conductivity scaling function $\Phi_\sigma$ near
the quantum critical point calculated by brute force integrating the current-current correlator
measured using the Langevin dynamics formalism for a one-component complex field $(N=1)$.  The inset
shows the spatial and temporal finite size scaling for a single data point corresponding to
$\delta/\sqrt{T} = 0$. Note the non-linear relationship between the spatial and temporal mesh
sizes.} 
\end{figure}

We begin by fixing the value of $\delta/\sqrt{T}$ and discretize the Hamiltonian described by the
classical partition function Eq.~(\ref{eq:Zc}) to a unit spaced lattice of $L$ sites.  In order to
solve the full equation of motion, the set of initial conditions, given by the
equilibrium configurations of the order parameter field $\Psi$ must be determined. These can be
obtained by classical Monte Carlo methods using a large number of different seed configurations
where the equilibrium order parameters are stored only after a suitable number of Monte Carlo time
steps (large enough to eliminate any possible autocorrelations) have been performed.  The set of
configurations are then used as the initial $(t=0)$ states of the stochastic equation of motion,
Eq.~(\ref{eq:eom}).  At each time step, the noise function $f(x,t)$ is drawn from a Gaussian
distribution and by using the second order stochastic Runge Kutta (or Heun) algorithm
\cite{noisebook} the time dependence of $\Psi(x,t)$ is determined.  The current-current correlator
in Eq.~(\ref{eq:sigmaDef}) is computed as an average over all temporal trajectories of $\Psi$ and
the dc conductivity can be found after integrating over all space and time.  The results are
necessarily dependent on the size of the spatial and temporal discretization and the final value for
the scaling function $\Phi_\sigma$ must be finite size scaled in both space and time.  The resulting
value of $\Phi_\sigma$ directly above the quantum critical point, ($\delta =0$ $T>0$) was found to be
$\Phi_\sigma(0) = 0.07801 \pm 0.01$.  This fully universal number is independent of any of the
specific details of the particular quasi-one dimensional system under consideration.  We have also
computed the value of $\Phi_\sigma$ for a range of $\delta$ near the critical coupling as seen in
Fig.~\ref{fig:PhiSN1}. 
The semi-classical result for the physical one-component complex order parameter determined here
should quite accurately reproduce the real electrical transport in the quantum critical regime, and
we will use it to benchmark our $N=\infty$ results in Section~\ref{subsec:largeNExpansion}.  

\subsection{The ordered phase}
\label{subsec:orderedPhase}

To address the physics of the ordered phase we again consider the action for an $N$ component 
complex field $\Psi_a(x,\tau)$ with magnitude $|\Psi_a(x,\tau)|=1$ but instead of treating the
fluctuations semi-classically, we take a $\sigma$-model approach \cite{zinn-justin}. 
After enforcing the fixed magnitude constraint on $\Psi_a$ with a Lagrange multiplier field 
$\mu(x,\tau)$, its action in the presence of a finite conjugate field $h_a$ reads
\begin{equation}\begin{split}
\mathcal{S} &= \frac{1}{g} \int dx \int d\tau \bigg\{ \Psi_a^\ast(x,\tau) \left[-\D \partial_x^2 
+ |\partial_\tau| + i\mu(x,\tau)\right]\Psi_a(x,\tau) - i\mu(x,\tau)
\\
& \qquad -\; g\left[h^{\phantom{\ast}}_a \Psi_a^\ast(x, \tau) + 
h^\ast_a\Psi^{\phantom{\ast}}_a(x,\tau) \right] \bigg\}.
\label{eq:psiaction}
\end{split}\end{equation}
where we have integrated the kinetic term by parts, and used the abuse of notation 
$|\partial_\tau|$ to infer the dissipative $|\wn|$ term in frequency space.  To derive saddle point 
equations in the large-$N$ limit, we explicitly break the $O(N)$ symmetry by choosing only one
component of the conjugate field $h_a$ to be non-zero and equal to $h$ in the $N^{th}$ direction.
Integrating over $N-1$ components of $\Psi$ and keeping only one component $\sigma$ (not to be
confused with the conductivity and dropping the space and imaginary time dependence of fields)
\begin{equation}
\mathcal{S}_\mathrm{eff} = (N-1) \int dx \int d\tau 
\left[ \frac{1}{g}\sigma^\ast\left(-\D \partial_x^2 + |\partial_\tau|\right)\sigma 
+  V\left(|\sigma|^2,i\mu\right) - \left(h \sigma^\ast + h^\ast\sigma \right) \right] 
\label{eq:SigmaAction}
\end{equation}
where we have rescaled the coupling $g$ by a factor of $N-1$. In the limit of large $N$, the
effective potential $V$ can be determined by the saddle point approximation where 
$r = i\mu$,
\begin{equation}
V(|\sigma|^2,r) = \frac{r}{g}(|\sigma|^2 - 1) + T\sum_{\wn} \! \int \! \frac{dk}{2\pi} 
\ln ( \D k^2 + |\wn| + r )
\label{eq:EP1}
\end{equation}
along with the constraint equations for $r$ and $\sigma$
\begin{align}
\label{eq:SigmaSP} 
\sigma r &= g h
\\
\label{eq:SSP}
|\sigma|^2 &= 1 -   T g \sum_{\wn}\int \frac{dk}{2\pi} 
\frac{1}{\D k^2 + |\wn| + r} .
\end{align}
We note that an important distinction between Eq.~(\ref{eq:SigmaSP}) and (\ref{eq:SSP}) and the
saddle point equation (\ref{eq:rz}) derived in the last section is that here we integrate over
\emph{all} Matsubara frequencies.  In the absence of an external magnetic field at $T=0$, 
the quantum critical point corresponds to the solution $\sigma=0$, $r=0$, and as before
defines a critical coupling strength $g_c$ as given in Eq.~(\ref{eq:gc}).  Again we will measure
deviations from quantum criticality using the parameter $\delta$ (Eq.~(\ref{eq:delta})).

Using this definition, the solution to Eq.~(\ref{eq:SigmaSP}) in zero conjugate field $h$ is given by 
$r=0$ and thus from Eq.(\ref{eq:SSP}) with $|\sigma|^2 = |\sigma_0|^2$,
\begin{equation}
|\sigma_0|^2 \equiv  1 - \frac{g}{g_c} = -\frac{g}{\sqrt{\D}} \delta 
\label{eq:sigma0delta}
\end{equation}
which is clearly only valid in the ordered phase characterized by $\delta < 0$.
Using Eq.~(\ref{eq:sigma0delta}), Eq.~(\ref{eq:SSP}) can be rewritten as
\begin{equation}
|\sigma|^2 = |\sigma_0|^2 - g\int \frac{dk}{2\pi} \left[T \sum_{\wn} 
\frac{1}{\D k^2 + |\wn| + r} - \int \frac{d\omega}{2\pi} \frac{1}{\D k^2 + |\wn|}
\right] 
\label{eq:finalSSP1}
\end{equation}
and by a method identical to the one used when integrating over all $\wn \ne
0$:
\begin{equation}
|\sigma|^2 = -\frac{g}{\sqrt{\D}} \delta + g\int \frac{dk}{2\pi^2} \left[ 
\psi\left(1 + \frac{\D k^2 + r}{2\pi T}\right) + \ln\left(\frac{2\pi T}{\D k^2}\right)
- \frac{\pi T}{\D k^2 + r} \right] ,
\label{eq:finalSSP2}
\end{equation}
where $\psi$ is the digamma function. This expression can be inverted numerically  to provide $r$ 
as a function of $|\sigma|^2$ and $\delta$.  After this has been accomplished, Eqs.~(\ref{eq:EP1}) 
and (\ref{eq:SSP}) can be combined to give the finite temperature effective potential
\begin{equation}
V(|\sigma|^2,\delta) = T\sum_{\wn} \int \frac{dk}{2\pi} \left[ 
\ln\left(1 + \frac{r(|\sigma|^2,\delta)} {\D k^2 + |\wn|} \right) - 
\frac{r(|\sigma|^2,\delta)}{\D k^2 + |\wn| + r(|\sigma|^2,\delta)} \right] ,
\label{eq:VsigDelta}
\end{equation}
where a constant term independent of $r$ has been discarded.

\subsubsection{Zero temperature effective potential}

At zero temperature, the frequency and momentum integrals in Eq.~(\ref{eq:finalSSP1}) can be written
in an isotropic fashion.  For a finite conjugate field, both $|\sigma|^2$ and $r$ are nonzero, and
dropping the explicit $|\sigma|^2$ and $\delta$ dependence of $r$
\begin{align}
V(|\sigma|^2,\delta) &= \frac{4}{\sqrt{\D}} \int \frac{d^3 p}{(2\pi)^3} \left [
\ln\left(1+\frac{r}{p^2}\right) - \frac{r}{p^2 + r} \right ] \nonumber  \\
&= \frac{r^{3/2}}{3\pi \sqrt{\D}}
\label{eq:VT0}
\end{align}
where the saddle point equation (\ref{eq:SSP}) can now be solved as
\begin{equation}
|\sigma|^2 = |\sigma_0|^2 + \frac{g}{\pi\sqrt{\D}} \sqrt{s} ,
\label{eq:sT0}
\end{equation}
which indicates that a solution exists only for $|\sigma|^2 > |\sigma_0|^2$, requiring the presence
of a non-zero conjugate field $h$.  In the ordered phase, this equation cannot be solved for
$|\sigma|^2 < |\sigma_0|^2$, and hence the effective potential defined below will not be valid in
the weakly ordered regime.

The zero temperature Coleman-Weinberg effective action for a quantum field $\Psi$ 
is defined to be \cite{peskin-schroeder}
\begin{equation}
\Gamma[\Psi_{cl}] = -S_{\mathrm{eff}}[\Psi_{cl}] - \int dx \int d\tau 
\left(h^* \Psi^{\phantom{\ast}}_{cl} + h \Psi^\ast_{cl} \right)
\end{equation}
such that it is the function whose minimum gives exactly $\Psi_{cl} = \langle \Psi
\rangle$.  To lowest order in perturbation theory it is simply the classical potential energy, but
is modified by quantum corrections at higher order.  Using Eqs.~(\ref{eq:SigmaAction}), (\ref{eq:VT0}) 
and (\ref{eq:sT0}), we obtain
\begin{equation}
V_{eff} = \frac{\Gamma(|\sigma|^2,\delta)}{\widetilde{\Omega} (N-1)} =
\frac{\pi^2}{3\sqrt{\D}}\left(\frac{\sqrt{\D}}{g} |\sigma|^2 + \delta \right)^3
\label{eq:CWEffPot}
\end{equation}
where $\widetilde{\Omega}$ denotes the system volume in space-time.  As noted above, this effective potential is
only valid for $|\sigma|^2 > |\sigma_0|^2$, as it has a minima at $|\sigma|^2 = 0$ (for $\delta <
0$) and a point of inflection at $|\sigma|^2 = |\sigma_0|^2$.

\begin{figure}[t]
\centering
\includegraphics[width=1.6in]{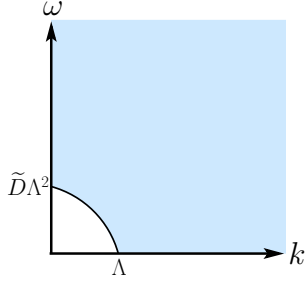}
\caption{\label{fig:cutoff}  The shaded portion shows the region of integration after
the implementation of a symmetric infrared cutoff in frequency and momentum.}
\end{figure}

In order to find a solution for $|\sigma|^2 < |\sigma_0|^2$, and to derive a Ginzburg-Landau
effective potential for the description of any slow degrees of freedom, we should not integrate over
all degrees of freedom, but only over those with a wavelength smaller than some  cutoff
$\Lambda^{-1}$.  This is necessary due to the fact that when integrating over all Matsubara
frequencies we are restricted to $r>0$ and thus can never access $|\sigma|^2 < |\sigma_0|^2$. In
Ref.~\cite{mh} a similar viewpoint was expressed, and the cutoff was taken to be of the
order of the zero temperature superconducting coherence length. We will follow this approach here
with the addition of a step that ensures self-consistency, and the cutoff is implemented
symmetrically in momentum and frequency space, consistent with dynamically critical
exponent $z=2$.  A schematic diagram showing the shaded region of integration can be seen in
Fig.~\ref{fig:cutoff}.
The zero temperature effective potential, with $r < 0$ is now defined as as
\begin{equation}
V_{eff} =  
\frac{4}{\sqrt{\D}} \int_{\sqrt{\D} \Lambda}^\infty \frac{d^3 p}{(2 \pi)^3} 
\left[ \ln \left(1+\frac{r}{p^2}\right) - \frac{r}{p^2} \right] 
+ \frac{r}{g}(|\sigma|^2 -|\sigma_0|^2) .
\label{eq:LGV}
\end{equation}
Differentiation with respect to $r$ gives rise to the modified  saddle point equation 
(for negative $r$)
\begin{equation}
|\sigma|^2 = |\sigma_0|^2 -  \frac{2 g}{\pi^2}\sqrt{\frac{|r|}{\D}}
\int_{\sqrt{\frac{\D}{|s|}}\Lambda}^\infty  dp \frac{1}{p^2 - 1},
\label{eq:SPEcutoff}
\end{equation} 
where $|\sigma_0|^2$ is now modified from Eq.~(\ref{eq:sigma0delta}) as a result of 
the infrared cutoff
\begin{equation}
|\sigma_0|^2 = -\frac{g \delta}{\sqrt{\D}} + \frac{2 g \Lambda}{\pi^2} .
\label{eq:sigma0cutoff}
\end{equation}

\begin{figure}[t]
\centering
\includegraphics*[width=3.2in]{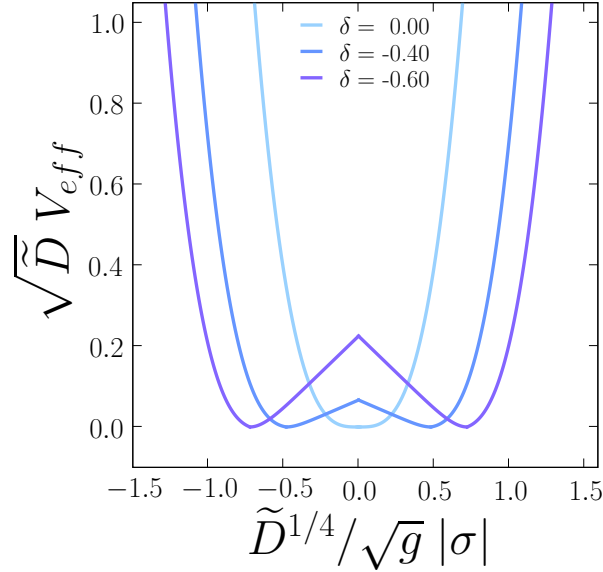}
\caption{\label{fig:Veff}  The effective quantum potential at $T=0$ calculated via the
Coleman-Weinberg procedure for $|\sigma|^2 > |\sigma_0|^2$ and through the self-consistent insertion
of an infrared cutoff for $|\sigma|^2 < |\sigma_0|^2$ with $f(\epsilon) = \pi^2$ for $\delta =
0.0,-0.40,0.60$ (inner to outer curve).}
\end{figure}

In order to gain intuition about the relative size of the cutoff and the effective mass (Lagrange
multiplier) $r$, we calculate the zero temperature superconducting coherence length 
$\xi(0)$ as a function of $r$.  It is determined as usual by the relation 
\begin{equation}
\left.\frac{d V_{eff}}{d |\sigma|^2} \right|_{|\sigma|^2 =  0}  =  - \frac{\D}{g}\frac{1}{\xi^2(0)}.
\end{equation}
By explicitly differentiating the effective potential Eq.~(\ref{eq:LGV}) and using 
Eq.~(\ref{eq:SPEcutoff}), we find
\begin{equation}
\left.\frac{d V_{eff}}{d |\sigma|^2} \right|_{|\sigma|^2 = 0}
=  \frac{1}{g} r(|\sigma|^2 = 0), 
\label{eq:dVLG}
\end{equation}
and combining the last two equations leads to the relation
\begin{equation}
\xi(0)  =  \sqrt{\frac{\D}{|r(|\sigma|^2=0)|}} .
\end{equation}
Using this definition at $T=0$ and $|\sigma|^2 = 0$ we can now determine the coherence length 
$\xi(0)$ self-consistently from Eq.~(\ref{eq:SPEcutoff}) 
\begin{equation}
\xi(0) 
= \frac{\sqrt{\D}}{\pi^2 |\delta|} \left[ \ln \left(\frac{\Lambda \xi(0) + 1}{\Lambda \xi(0) - 1}
\right) - 2 \Lambda \xi(0) \right] ,
\label{eq:xiLambda}
\end{equation}
where Eq.~(\ref{eq:sigma0cutoff}) has been used.  Note that this equation has a solution for all 
choices of $\Lambda $ such that $1 < \xi(0) \Lambda \lesssim 6/5$ with $\xi(0) \to \infty$
logarithmically as $\Lambda \xi(0) \to 1$ and $\xi(0) \to 0$ as $\Lambda \xi(0) \to 6/5$.  
Parameterizing $\Lambda \xi(0) = 1 + \epsilon$ where $\epsilon \ll 1$, and defining
\begin{equation}
f(\epsilon) = \frac{1}{\pi^2} \left[ \ln \left(1 + \frac{2}{\epsilon}\right) - 2(1 +
\epsilon)\right]
\label{eq:fepsilon}
\end{equation}
the zero temperature coherence length is
\begin{equation}
\xi(0) = \frac{\sqrt{\D}}{|\delta|} f(\epsilon).
\label{eq:xi0}
\end{equation}
Due to the logarithmic divergence as $\epsilon \to 0$, one possible choice of $\epsilon
\simeq 1.4 \times 10^{-5}$ gives $f(\epsilon) \simeq 1$ leading to the simple relation 
$\xi(0) = \sqrt{\D} / |\delta|$ or $\Lambda \simeq |\delta|/\sqrt{\D}$.
The preceding arguments now allow us to express the effective potential 
(Eq.~(\ref{eq:LGV})) in terms of $|\sigma(r < 0)|^2 < |\sigma_0|^2$ and $\delta$.
If we choose $\epsilon$ such that $f(\epsilon) = \pi^2$, i.e. $\xi(0) = \sqrt{\D}\pi^2/|\delta|$  
then the saddle point equation (\ref{eq:SPEcutoff}) simplifies to 
\begin{equation}
-\frac{\sqrt{\D}}{g} |\sigma|^2  + \left(1+\frac{2}{\pi^2}\right)|\delta| - \frac{2}{\pi^2}
\sqrt{|r|}\mathrm{arctanh}\left(\sqrt{\frac{|r|}{\delta^2}}\right) = 0
\end{equation}
which when solved for numerically for $r$ can be substituted into Eq.~(\ref{eq:LGV}) to give the
effective potential
\begin{equation}\begin{split}
V_{eff} &= \frac{1}{\sqrt{\D}} \left \{ r\left[\frac{\sqrt{\D}}{g} |\sigma|^2 +
\left(1+\frac{2}{\pi^2}\right)\delta\right] \right.
\\
& \left. \qquad +\; \frac{2}{3\pi^2}\left[-r\delta + \delta^3\ln\left(1 + \frac{r}{\delta^2}\right) 
+ \sqrt{r^3} \ln\left(\frac{\delta + \sqrt{|r|}}{\delta - \sqrt{|r|}}\right) \right]  \right\}.
\label{eq:VeffT0}
\end{split}\end{equation}
This result, valid at $T=0$ and $|\sigma|^2 < |\sigma_0|^2$ can now be combined with 
Eq.~(\ref{eq:CWEffPot}) which is valid for $|\sigma|^2 > |\sigma_0|^2$ to obtain the effective
potential everywhere at $T=0$ and $\delta < 0$ corresponding to the ordered or 
superconducting state, as seen in Fig.~\ref{fig:Veff}.

\subsubsection{Finite temperature Ginzburg-Landau potential} 

\begin{figure}[t]
\centering
\includegraphics*[width=3.2in]{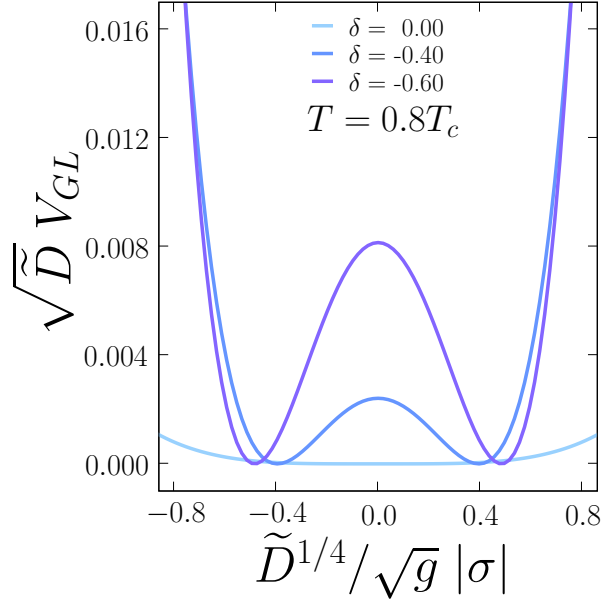}
\caption{\label{fig:VLG}  The Ginzburg-Landau potential of Eq.~(\ref{eq:positVLG}) for
$\delta = 0.0,-0.4,-0.6$ (bottom to top curve) using the values of $T_c$, $\alpha_0$ and $\beta$
found for $f(\epsilon) = \pi^2$ at $T=0.8T_c$.}
\end{figure}

Having computed the form of the infrared momentum cutoff $\Lambda$ self-consistently as a function
of $\delta$ and after investigating the form of the effective potential at zero temperature, we now move
to finite temperatures and consider expanding around some critical temperature $T_c$ for the ordered
phase. The usual form for the potential is posited
\begin{equation}
V_{GL} = V_0 +  \alpha_0 (T-T_c) |\sigma|^2 + \frac{1}{2}\beta |\sigma|^4 + \cdots
\label{eq:positVLG}
\end{equation}
and we will endeavor to evaluate $T_c$, $\alpha_0$ and $\beta$  in terms of the parameters $g$, $\D$
and $\delta$.  The goal of such a procedure is to derive an effective Ginzburg-Landau
theory for the superconducting state near $T_c$ with quantum renormalized coefficients. By
multiplying this potential by the finite temperature Ginzburg-Landau coherence length $\xi(T)$, 
an effective free energy is found from which the barrier height for a thermally 
activated phase slip can be determined directly from the LAMH theory.  The details of the
calculation of the renormalized coupling constants is rather involved and has been included in
Appendix~\ref{app:ginzburgLandau}.  The resulting dimensionless interaction potential is shown in
Fig.~\ref{fig:VLG} for $T = 0.8 T_c$.
We find a relatively steep double well potential which unsurprisingly has both a barrier height and
order parameter expectation value that depends on the distance from the critical point $\delta$.

\subsubsection{Quantum renormalized LAMH theory}
Having derived a effective Ginzburg-Landau potential (with engineering dimensions of energy
divided by length) in Appendix~\ref{app:ginzburgLandau}, we may convert it into a free energy
functional by multiplying by the finite temperature coherence length
\begin{equation}
\xi(T) = \xi(0)\left(1-\frac{T}{T_c}\right)^{-1/2}. 
\end{equation}
Due to the presence of phase slips below $T_c$, phase coherence of the wire is lost and it
necessarily breaks up into $L/\xi(T)$ independent segments, that interact via Josephson coupling
near the finite temperature transition.  After discarding a constant, the functional is given by
\begin{equation}
F_{GL}\{\sigma\} = \overline{\alpha}_0\, \frac{\sqrt{\D}f(\epsilon)}{g \delta}
\left(1-\frac{T}{T_c}\right)^{1/2}|\sigma|^2 
- \overline{\beta} \frac{\sqrt{T_c} \D f(\epsilon)} 
{2 g^2 \delta} \left(1 - \frac{T}{T_c}\right)^{-1/2}|\sigma|^4
\label{eq:FLG}
\end{equation}
with a singular temperature dependence in the quartic term.  We restrict the analysis to
temperatures that place the system in the LAMH region of the phase diagram displayed in
Fig.~\ref{fig:phaseDiagram}, such that $1-T/T_c \ll 1$.  The rescaled distance from the critical
point is negative ($\delta < 0$), and the dimensionless coefficients are
\begin{align}
\overline{\alpha}_0 &= g \alpha_0 \\
\overline{\beta} &= \frac{g^2}{\sqrt{\D T_c}} \beta 
\end{align}
where $\alpha_0$ and $\beta$ are given in Eqs.~(\ref{eq:alpha}) and (\ref{eq:beta}) respectively.

The free energy barrier height for a single phase slip event can now be evaluated using the
coefficients computed directly from the non-linear sigma model version of the full quantum theory, 
\begin{equation}
\Delta F = - \frac{\alpha^2}{2\beta} 
= -\frac{\overline{\alpha}_0^2}{2\overline{\beta}} \frac{f(\epsilon) T_c^{3/2}}{\delta}
\left(1-\frac{T}{T_c}\right)^{3/2} 
\end{equation}
which can be written in terms of a dimensionless scaling function of two unique scaling variables,
the first expressing the classical and the second the quantum nature of the renormalized barrier
height 
\begin{equation}
\Delta F = T_c \Phi_{\Delta F} \left(\frac{T}{T_c},\frac{\delta}{\sqrt{T_c}}\right).
\label{eq:DelFQRLAMH}
\end{equation}
$\Delta F$ can now be directly inserted into the LAMH theory in place of the phase slip barrier height
calculated by Langer and Ambegaokar \cite{la}, with the rest of the arguments leading to the LAMH
resistance remaining unaffected.

\begin{figure}[t]
\centering
\includegraphics*[width=3.2in]{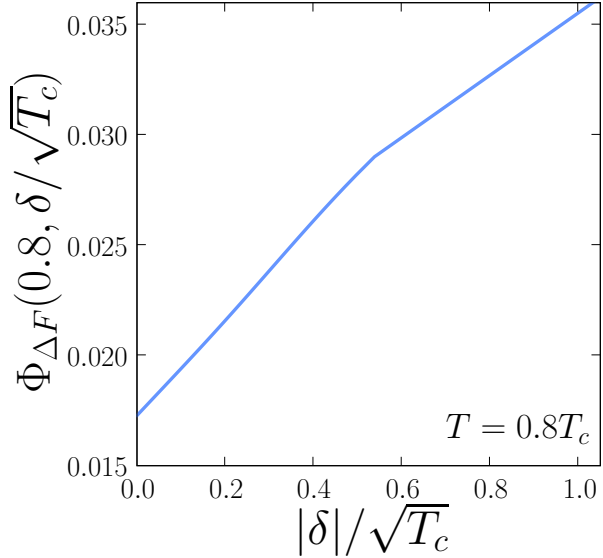}
\caption{\label{fig:deltaF}  The dimensionless free energy barrier height
corresponding to Eq.~(\ref{eq:FLG}) as a function of a single scaling variable $\delta/\sqrt{T_c}$.}
\end{figure}

\begin{figure}[t]
\centering
\includegraphics*[width=3.2in]{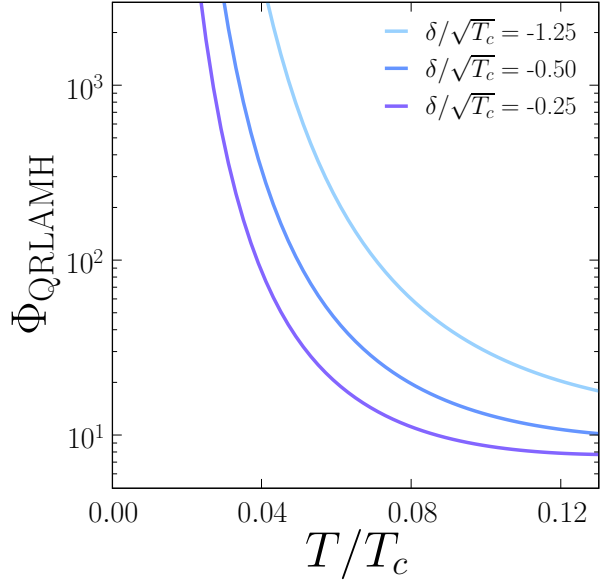}
\caption{\label{fig:PhiQRLAMH}  The quantum renormalized LAMH scaling function for the
conductivity plotted as a function of the reduced temperature for three values of $\delta/\sqrt{T_c}
= -1.25,-0.50,-0.25$ (top curve to bottom).}
\end{figure}

Before doing so, we briefly remark on the seemingly surreptitious form of Eq.~(\ref{eq:DelFQRLAMH}).
The ability to write the barrier height as a cutoff independent scaling function with no explicit
dependence on $\Lambda$ or $f(\epsilon)$ is a consequence of the fact that by fixing the
dimensionless variable $\delta/\sqrt{T_c}$ a unique value of $\epsilon$ and thus $f(\epsilon)$ can
be found from Eq~(\ref{eq:Tc}). This is demonstrated in Fig.~\ref{fig:deltaF} where it appears that
$\Phi_{\Delta F}$ is nearly a linear function of $\delta/\sqrt{T_c}$, with a slight kink for $T =
0.8 T_c$.
Therefore, using the same method as discussed in Section~\ref{subsec:LAMH} we can write the
quantum renormalized LAMH conductivity in terms of the scaling function $\Phi_{\Delta F}$ as
\begin{equation}\begin{split}
\sigma_{\mathrm{QRLAMH}} &= 4\pi\frac{e^2}{h} \xi(0)
\frac{1}{\sqrt{\Phi_{\Delta F}(T/T_c, \delta/\sqrt{T_c})}}
\left(\frac{T}{T_c}\right)^{3/2}\left(1-\frac{T}{T_c}\right)^{-3/2} 
\\
& \qquad \times \; \exp\left[ \Phi_{\Delta F}\left(\frac{T}{T_c},\frac{\delta}{\sqrt{T_c}}\right)
\frac{T_c}{T}\right]
\label{eq:sigmaQRLAMH}
\end{split}\end{equation}
or
\begin{equation}
\sigma_{\mathrm{QRLAMH}} = \frac{e^2}{h} \xi(0) \;
\Phi_{\mathrm{QRLAMH}}\left(\frac{T}{T_c},\frac{\delta}{\sqrt{T_c}}\right)
\label{eq:PhiQRLAMH}
\end{equation}
where $\Phi_{\mathrm{QRLAMH}}$ is shown in Fig.~\ref{fig:PhiQRLAMH} for three values of
$\delta/\sqrt{T_c}$.
The arguments discussed in Section~\ref{subsec:pairbreaking} provide a recipe to convert the
parameter $\delta$ to the physical pairbreaking frequency $\alpha$ for a given experimental
geometry.  The suppression of the critical temperature as a result of an external magnetic field
directed parallel to the wire has already been observed in Ref.~\cite{rogachev-bollinger}.  
Thus, in principle, the relationship between $\delta$ and $T_c$ could be determined experimentally
from a fitting procedure, and the expression for the conductivity in Eq.~(\ref{eq:PhiQRLAMH}) could
be applied to the experimental transport results with one less fitting parameter than the form
currently used in Eq.~(\ref{eq:PhiLAMH}).  

A striking difference between the usual LAMH form of the conductivity and the quantum renormalized
expression in Eq.~(\ref{eq:PhiQRLAMH}) is the absence of any explicit dependence the number of
transverse channels, $N_\perp$ in the latter.  This can be most easily seen by using
Eqs.~(\ref{eq:xi0}) and (\ref{eq:c1}) to write the dimensionless variable $\delta/\sqrt{T_c}$ in
terms of $\D / \xi^2(0) T_c$, the natural energy ratio suitable to our analysis. Due to their
dependence on different microscopic length scales, it is somewhat difficult to quantitatively
compare $\sigma_\mathrm{LAMH}$ with its quantum renormalized value $\sigma_{\mathrm{QRLAMH}}$.
Qualitatively, the inclusion of quantum fluctuations leads to a softening of the free energy barrier
resulting in an enhanced phase slip rate consistent with recent calculations made by Golubev and
Zaikin \cite{golubev}.

\subsection{Large-$N$ expansion}
\label{subsec:largeNExpansion}

\begin{figure}[t]
\centering
\includegraphics*[width=3.2in]{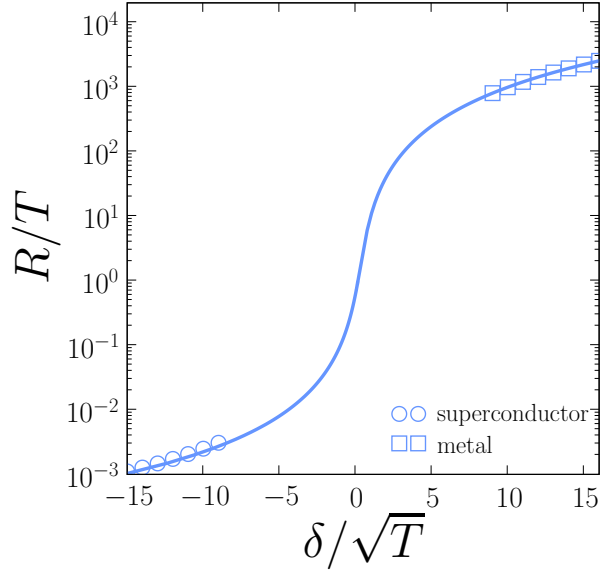}
\caption{\label{fig:R}  The renormalized mass $R$ plotted as a function of the
rescaled distance from criticality $\delta/\sqrt{T}$.  The symbols refer to the analytic results of
Eq.~(\ref{eq:PhiR}) in the metallic and superconducting limits.}
\end{figure}

Moving away from the ordered phase, at low temperatures, at a distance from the quantum critical
point ($\sqrt{T} \ll \delta$), quantum fluctuations are large and there will be a finite number of Matsubara
frequencies which lie below the crossover energy.  Now, the classical model of
Section~\ref{subsec:effClassTheory} is no longer adequate to describe the contribution of 
superconducting fluctuations to transport.  In this regime of the phase diagram however, a  
direct $1/N$ expansion on the full quantum theory can be attempted.  Starting with
Eq.~(\ref{eq:SalphaN}) and decoupling the quartic interaction with a Hubbard-Stratonovich field
$\mu$ we arrive at the effective action
\begin{equation}\begin{split}
\mathcal{S} &= \int d x \int d\tau \left[ \D |\partial_x \Psi_a(x,\tau)|^2
+ i\mu(x,\tau) |\Psi_a(x,\tau)|^2 + \frac{1}{2u} \mu^2(x,\tau) \right.
\\
& \left. \qquad +\; i\frac{\alpha}{u}\mu(x,\tau) \right] 
+ T \sum_{\wn} \int d x\,  |\wn| |\Psi_a(x,\wn)|^2 .  
\label{eq:SalphaNChi}
\end{split}\end{equation}
Integrating out $\Psi_a$ over its now quadratic action in the partition function $\mathcal{Z} =
\mathrm{Tr}\, \exp(-\mathcal{S})$, we as usual recognize an overall factor of $N$ which allows the
functional integral over the Hubbard-Stratonovich field $\mu$ to be performed in the saddle point
approximation where we identify $R = i\mu$.  In the universal limit, the new quadratic effective
action is given by
\begin{equation}
\mathcal{S}_R = T \sum_{\wn} \int \frac{dk}{2\pi} (\D k^2 + |\wn| +  R) |\Psi_a(k,\wn)|^2,
\label{eq:SR}
\end{equation}
where the `mass' $R$ is defined by the saddle point condition 
\begin{equation}
\frac{1}{g} = \int \frac{dk}{2 \pi} T \sum_{\wn} \frac{1}{\D k^2
+ |\wn| + R}.
\label{eq:R}
\end{equation}
The evaluation of $R$ is straightforward and follows our derivation of Eq~(\ref{eq:deltaz})
\begin{equation}
\frac{\delta}{\sqrt{T}} =  \int \frac{dk}{2\pi^2} \left[ \ln \left(
\frac{2 \pi }{ k^2} \right) + \psi \left( 1 + \frac{ k^2 + R/T}{2
\pi } \right) - \frac{\pi }{k^2 + R/T}\right] .
\label{eq:deltaR}
\end{equation}
In general, this expression must be inverted numerically to determine $R/T$ as a function of
$\delta/\sqrt{T}$ as is shown in Fig.~\ref{fig:R}.  However, at the quantum critical point (QC,
$\delta = 0$) and in the metallic (M,$\delta \to \infty$) and superconducting (SC, $\delta \to -\infty$)
we can analyze Eq.~(\ref{eq:deltaR}) along the same lines as was done for Eq.~(\ref{eq:deltaz})
in Section~\ref{subsec:effClassTheory}.  We write $R = T \Phi_R(\delta/\sqrt{T})$ and 
find the following results
\begin{equation}
\Phi_R(x) \simeq \left\{
	\begin{array}{rcl}
	1/4x^{2}& ; & x \to -\infty \\
	0.625 & ; & x \ll 1 \\
	\pi^2 x^2 & ; & x \to \infty
	\end{array} \right. .
\label{eq:PhiR}
\end{equation}
Understanding the behavior of the effective mass will be crucial in the following sections where we
first define and then calculate the thermal and electrical transport coefficients in the zero
frequency limit.

\subsubsection{Thermoelectric transport}
\label{subsec:teTransport}
The electrical ($\sigma$) and thermal ($\kappa$) conductivities and the Peltier coefficient
($\alpha$) are defined in terms of the electrical $\vec{j}_0$ and thermal $\vec{j}_2$ current
densities via the relation
\begin{equation}
\left(\begin{array}{c} \vec{j}_0 \\ \vec{j}_2 \end{array}\right) = 
\left(\begin{array}{cc} \sigma & \alpha \\ \alpha T & \tilde{\kappa} \end{array}\right)
\left(\begin{array}{c} \vec{E} \\ -\vec{\nabla} T \end{array}\right),
\label{eq:JeJq}
\end{equation}
where $\vec{E}$ is an external electric field, $\vec{\nabla} T$ is an imposed temperature gradient and 
$\kappa/T = \tilde{\kappa}/T - \alpha^2/\sigma$.  The current operators can also be defined in terms
of the derivatives of our large-$N$ action $\mathcal{S}_R$
\begin{align}
j_0 &= \frac{\partial\mathcal{S}_R}{\partial A_0} \\ 
j_2 &= \frac{\partial\mathcal{S}_R}{\partial A_2} 
\end{align}
where $A_0$ is the scalar electric potential and $A_2$ is the thermal vector potential, after we
have made it gauge-covariant through the introduction of these two gauge fields
\cite{moreno-coleman} via the replacement 
\begin{equation}
\partial_x \rightarrow \mathcal{D} \equiv \partial_x - ie^* A_0(x,\tau) - iA_2(x,\tau)(i\partial_\tau)
\end{equation}
leading to
\begin{align}
j_0 &= ie^*\D \left[\psi^*\mathcal{D}\psi - \psi(\mathcal{D}\psi)^*\right] \\
j_2 &= \D \left[\partial_\tau\psi(\mathcal{D}\psi)^* + \mathcal{D}\psi\partial_\tau \psi^* \right].
\end{align}
where $e^\ast = 2e$ is the charge of a Cooper pair and we have specialized to $1+1$ dimensions.

The quantum Kubo formula \cite{mahan,fisher-lee} can be used to obtain results for the
thermoelectric conductivities at external complex frequency $i\wn$ (where we ignore the Peltier 
coefficient as its dc part will turn out to be identically zero) \cite{sachdev-troyer}
\begin{align}
\mathcal{G}_p (i\wn)  &= -\frac{1}{\wn T^p} \left. \frac{\partial}{\partial A_p} 
\left \langle \frac{\partial S}{\partial A_p} \right \rangle \right |_{A_0 = A_2 = 0} \nonumber 
\\
& = -\frac{1}{\wn T^p} \left[ \int_0^\beta d\tau 
\langle J_p(\tau)J_p(0)\rangle \mathrm{e}^{i\wn\tau} 
- 2{e^{*}}^{2-p}\D \int dx \left \langle 
\left| \partial^{p/2}_\tau \psi(x,0) \right|^2 \right \rangle \right]
\label{eq:G1}
\end{align}
where the currents are defined by
\begin{equation}
J_p(\tau) = i {e^*}^{1-p/2}\D \!\int dx \left[ \partial^{p/2}_\tau \psi^*(x,\tau)\partial_x \psi(x,\tau)
- (-1)^{p/2} \partial_x \psi^*(x,\tau)\partial^{p/2}_\tau \psi(x,\tau) \right] 
\label{eq:J}
\end{equation}
and $p=0$ corresponds to the electrical conductivity while $p=2$ defines the thermal conductivity,
i.e. $\sigma(i\wn) = \mathcal{G}_0(i\wn)$ and $\kappa(i\wn)/T = \mathcal{G}_2(i\wn)$.

The conductivities are more easily expressed in terms of a one-loop polarization function, 
\begin{equation}
\mathcal{G}_p(i\wn) = -\frac{4\D^2 {e^*}^{2-p}}{\wn T^p} \mathcal{K}_p(i\wn)
\label{eq:G}
\end{equation}
which contains both paramagnetic and diamagnetic contributions
\begin{align}
\mathcal{K}_p(i\wn) &= 
\;
\parbox{43\unitlength}{%
	\begin{fmffile}{KlargeN1}
		\begin{fmfgraph}(40,40)
			\fmfleft{l}
			\fmfright{r}
			\fmfv{decor.shape=circle,decor.size=5,decor.filled=0}{l,r}
			\fmf{plain,right}{l,r}
			\fmf{plain,right}{r,l}
		\end{fmfgraph}
	\end{fmffile}}
\;-\;
\parbox{43\unitlength}{%
	\begin{fmffile}{KlargeN2}
		\begin{fmfgraph}(40,40)
			\fmfleft{l}
			\fmfright{r}
			\fmffreeze
			\fmfforce{(.5w,0.0h)}{b}
			\fmfv{decor.shape=square,decor.size=5,decor.filled=0}{b}
			\fmf{plain,right}{l,r}
			\fmf{plain,right}{r,l}
		\end{fmfgraph}
	\end{fmffile}} \nonumber 
\\
&= T\sum_{\epsilon_n} \nt{k}{2} k^2 \left(\epsilon_n + \frac{\wn}{2}\right)^p 
\left[\frac{1}{(|\wn| + \D k^2 + R)(|\wn + \epsilon_n| + \D k^2 + R)} \right. \nonumber
\\
& \left. \qquad -\; \frac{1}{(|\wn| + \D k^2 + R)^2}\right],
\label{eq:KLargeN}
\end{align}
where a solid line represents the bare propagator $G_0(k,\wn) = (\D k^2 + |\wn| + R)^{-1}$, 
an open circle corresponds to a term linear in the potential $A_p$ and an open square 
to a term quadratic in $A_p$.  Employing the spectral representation of the bare propagator
\begin{align}
\mathcal{A}(k,\omega) &= -2\Im G_0(k,i\wn \to \omega + i\eta) \nonumber 
\\
&= -\frac{2\omega}{\omega^2 + (\D k^2 + R)^2}
\end{align}
where a $|\wn|$ dependence along the imaginary frequency axis becomes $-i\omega$ just above the real
frequency axis.  The polarization function is then given by 
\begin{equation}\begin{split}
\mathcal{K}_p(i\wn) &= T\sum_{\epsilon_n}\nt{k}{2}k^2\nt{\omega_1}{2}\nt{\omega_2}{2} 
	\left(\frac{\omega_1+\omega_2}{2}\right)^p \mathcal{A}(k,\omega_1)\mathcal{A}(k,\omega_2)
\\
& \qquad \qquad \times\; \left[\frac{1}{(i\epsilon_n - \omega_1)[i(\epsilon_n +\wn)-\omega_2]}
	 - \frac{1}{(i\epsilon_n - \omega_1)(i\epsilon_n -\omega_2)}\right],
\end{split}\end{equation}
where we have made the replacement $(\epsilon_n + \wn/2) \rightarrow (\omega_1 + \omega_2)/2$, due
to the temporal non-locality of $j_2$ \cite{moreno-coleman,ambegaokar}.  Performing the Matsubara
summation, and analytically continuing to real frequencies yields
\begin{equation}\begin{split}
\mathcal{K}_p(\omega + i\epsilon) &= \nt{k}{2} \nt{\omega_1}{2}\nt{\omega_2}{2} 
	\left(\frac{\omega_1+\omega_2}{2}\right)^p \mathcal{A}(k,\omega_1)\mathcal{A}(k,\omega_2) k^2 
\\
& \qquad \times \; [n(\omega_1) - n(\omega_2)] 
\left(\frac{1}{\omega_2-\omega_1-\omega-i\eta} - \frac{1}{\omega_2-\omega_1}\right),
\label{eq:RePol}
\end{split}\end{equation}
where $n(\omega) = (\ee{\omega/T} - 1)^{-1}$ is the Bose distribution function, and $\eta$ is a
positive infinitesimal.  After checking that the delta-function contribution to $\Re
\mathcal{G}_p(\omega)$ at zero frequency is proportional to the external frequency, (it vanishes as
$\omega \rightarrow 0$),  we can combine Eqs.~(\ref{eq:G}) and (\ref{eq:RePol}) to give the
remaining regular part
\begin{equation}\begin{split}
\Re \mathcal{G}_p(\omega) &= \frac{4\D^2{e^*}^{2-p}}{T^p} 
\nt{\Omega}{}\frac{[n(\Omega)-n(\Omega+\omega)]}{\omega} \left(\Omega + \frac{\omega}{2}\right)^p 
\\
& \qquad \times \; 
\nt{k}{2}\frac{k^2\Omega(\Omega+\omega)}{[\Omega^2 + (\D k^2 + R)^2][(\Omega+\omega)^2 
+ (\D k^2 + R)^2]}.
\label{eq:ReSig}
\end{split}\end{equation}
The classical limit of Eq.~(\ref{eq:ReSig}) can be found by replacing $n(\omega) \simeq T/\omega$,
but here we directly perform the limit $\omega \rightarrow 0$ and obtain the quantum dc
conductivities
\begin{align}
\Re \mathcal{G}_p &= \frac{{\D^2 e^*}^{2-p}}{T^{p+1}} \nt{\Omega}{}\frac{\Omega^{2+p}}{\sinh^2(\Omega/2T)} 
\nt{k}{2}\frac{k^2}{[\Omega^2 + (\D k^2 + R)^2]^2} \nonumber 
\\
&= \frac{\sqrt{2}{e^*}^{2-p}\sqrt{\D}}{8T^{p+1}} 
\nt{\Omega}{2}\frac{\Omega^{2+p}}{\sinh^2(\Omega/2T)} 
\frac{1}{\sqrt{\Omega^2 + R^2}\left(R + \sqrt{\Omega^2 + R^2}\right)^{3/2}}.
\label{eq:Gdc}
\end{align}
which is the major result of this section.  From this expression it is immediately clear that 
the Peltier coefficient $\alpha$ (corresponding to $p=1$) is identically zero by symmetry, and thus 
$\tilde{\kappa} = \kappa$.  

In addition to the scaling function $\Phi_\sigma$ defined for the electrical conductivity in 
Eq.~(\ref{eq:PhiSigma}) and given by  
\begin{equation}
\sigma = \frac{e^{\ast 2}}{\hbar} \sqrt{\frac{\hbar \D}{\kB T}} \Phi_\sigma
\left( \frac{\delta}{\sqrt{\hbar \kB T}} \right)
\label{eq:largeNsigma}
\end{equation}
the thermal conductivity must obey a similar form
\begin{equation}
\frac{\kappa}{T} = \frac{\kB^2}{\hbar}\sqrt{\frac{\hbar \D}{\kB T}} 
\Phi_{\kappa}\left(\frac{\delta}{\sqrt{\hbar \kB T}}\right) , 
\label{eq:largeNkappa}
\end{equation}
where we have re-inserted the appropriate factors of $\hbar$ and $\kB$ for clarity.

\begin{figure}[t]
\centering
\includegraphics*[width=4.0in]{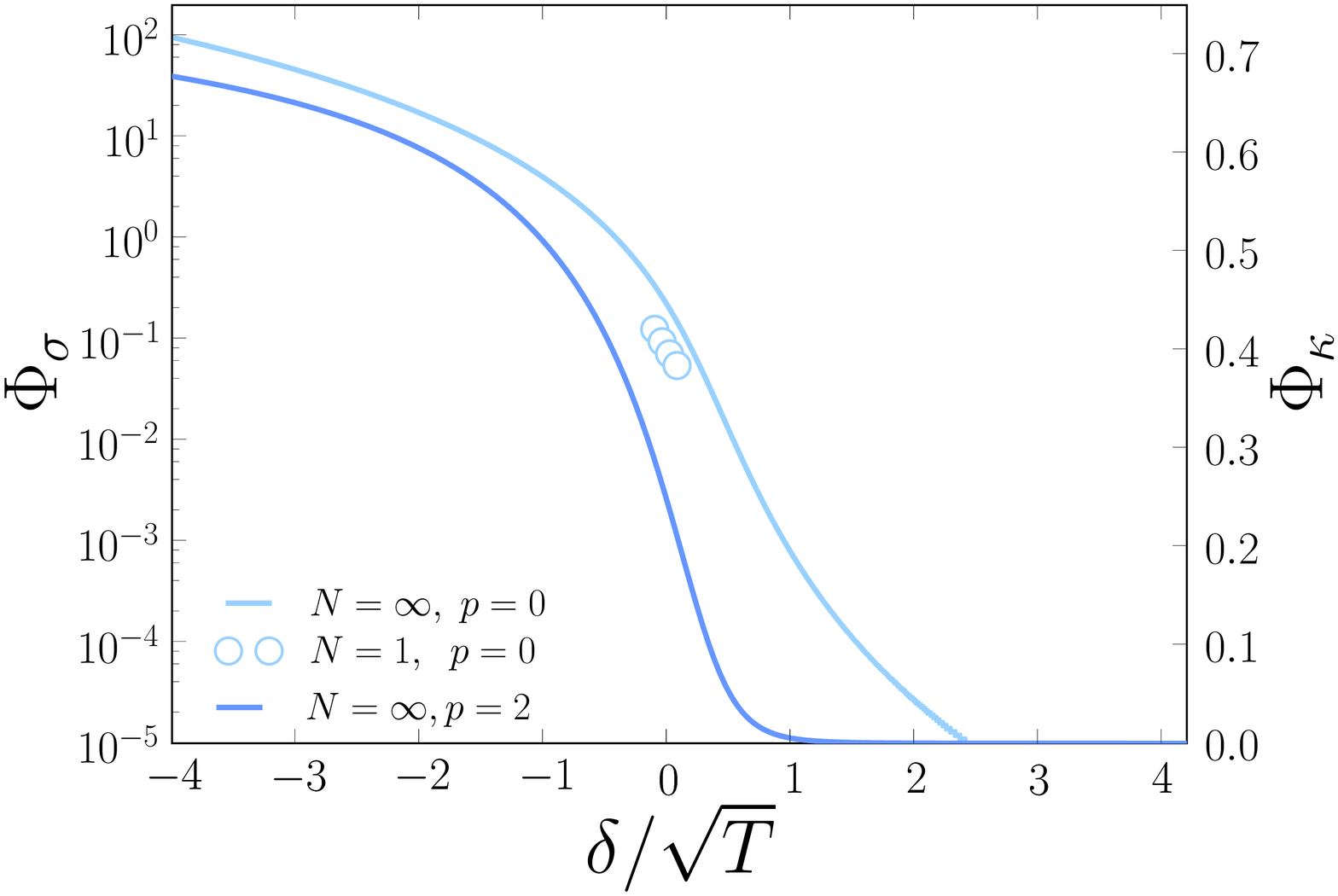}
\caption{\label{fig:PhiSK} The solid lines show the $N=\infty$ universal scaling
functions the electrical ($p=0$, top curve, left axis)  and thermal ($p=2$, bottom curve, right
axis) conductivity calculated by integration of Eq.~(\ref{eq:Gdc}).  The symbols show the effective
classical scaling function for the electrical conductivity calculated in the Langevin formalism in
Section~\ref{subsec:classicalConductivity} and previously shown in Fig.~\ref{fig:PhiSN1} for a one
component complex field.}
\end{figure}

\begin{figure}[t]
\centering
\includegraphics*[width=3.2in]{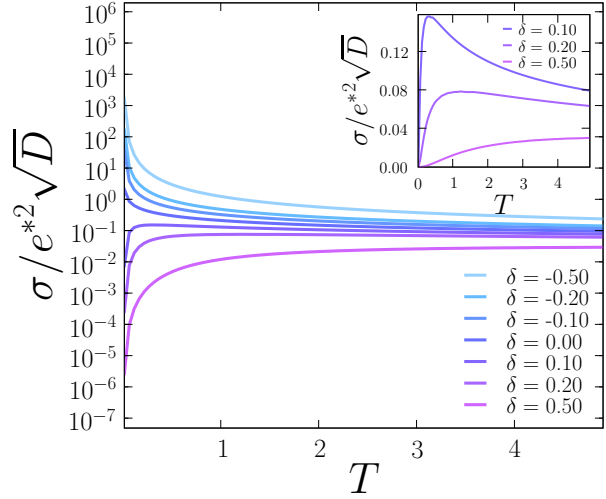}
\caption{\label{fig:sigma}  The temperature dependence of the rescaled dc electrical conductivity
at fixed values of $\delta = -0.50,-0.20,-0.10,0.00,0.10,0.20,0.50$ increasing from top to bottom.
The inset shows behavior non-monotonic in temperature near criticality.}
\end{figure}

\begin{figure}[t]
\centering
\includegraphics*[width=3.2in]{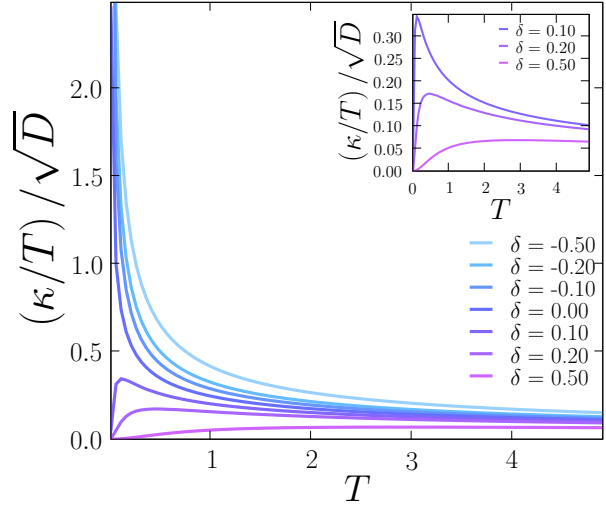}
\caption{\label{fig:kappa}  The temperature dependence of the rescaled dc thermal conductivity 
at fixed values of $\delta = -0.50,-0.20,-0.10,0.00,0.10,0.20,0.50$ increasing from top to bottom.
The inset shows behavior non-monotonic in temperature near criticality.}
\end{figure}

The $\delta/\sqrt{T}$ dependence of $\Phi_\sigma$ and $\Phi_\kappa$ can be found by numerically
inverting Eq.~(\ref{eq:deltaR}), (Fig.~\ref{fig:R}) and the result is shown in Fig.~\ref{fig:PhiSK}.  
Both are smooth functions of $\delta/\sqrt{T}$ throughout the quantum critical regime with the
dimensionless value of the electrical conductivity being two orders of magnitude larger than the
thermal conductivity right above the quantum critical point.  The large-$N$ electrical transport
can also be compared with the previously calculated value found in
Section~\ref{subsec:classicalConductivity} for $N=1$.  There is good  agreement between the two, and 
both results have the same $\delta/\sqrt{T}$ dependence near $\delta=0$ indicating that the
correct physics are manifest even at $N=\infty$.  With the full numerical scaling functions
computed, the temperature dependence of the thermoelectric transport can be determined by fixing
$\delta$ leading to the results displayed in Fig.~\ref{fig:sigma} and \ref{fig:kappa}.
The singular correction to the electrical conductivity clearly shows behavior consistent with a
quantum phase transition between a superconductor (diverging conductivity) and metal (finite or
vanishing conductivity).  Moreover, for a fixed but small positive value of $\delta$, placing the
system in the quantum critical region of Fig.~(\ref{fig:phaseDiagram}), (inset Fig.~\ref{fig:sigma})
as the temperature is lowered, a clear signature of crossover behavior can be seen from the
non-monotonic temperature dependence of the conductivity.  As the system leaves the strongly
fluctuating quantum critical regime for the low temperature metallic regime, AL contributions from
Cooper pairs are suppressed and their positive contribution to the conductivity disappears.  As
mentioned in the introduction, experiments \cite{zgirski-arutyunov,bollinger-bezryadin} have already
seen evidence of such non-monotonic resistance in the metallic regime, strongly supporting
the crossover picture we present here.  The exact form of the temperature-dependent crossovers can
be determined by further investigating the limiting forms of the two scaling functions $\Phi_\sigma$
and $\Phi_\kappa$.

\subsubsection{Finite temperature crossovers}
Analytical forms for the electrical and thermal conductivity can be found in three distinct limits
using the results of Eq.~(\ref{eq:PhiR}).  The first is deep in the superconducting (SC) regime
where $\delta/\sqrt{T} \rightarrow -\infty$ or $R/T \rightarrow 0$. The next is at the quantum
critical point (QC), where $\delta = 0$ or $R/T = 0.624798$ and the final regime is on the metallic
side of the transition (M) where $\delta, R/T \rightarrow \infty$.  For low temperatures,
Eq.~(\ref{eq:Gdc}) can be evaluated in these three limits leading to the approximate analytic
scaling behavior
\begin{align}
\label{eq:dcsigmaScaling}
\Phi_{\sigma}(x) &= \left\{ 
	\begin{array}{rcl}
		x^3 & ; & SC \\
		0.217997\cdots & ; & QC \\
		(12\pi^4)^{-1} x^{-5} & ; & M \\
	\end{array} \right. \\
\label{eq:dckappaScaling}
\Phi_{\kappa}(x) &= \left\{ 
	\begin{array}{rcl}
		\frac{3}{4\sqrt{2\pi}} \zeta\left(\frac{3}{2}\right) & ; & SC \\
		0.24592\cdots & ; & QC \\
		(15\pi^2)^{-1} x^{-5} & ; & M \\
	\end{array} \right. 
\end{align}
which can be used in Eqs.~(\ref{eq:largeNsigma}) and (\ref{eq:largeNkappa}) with $x = \delta /
\sqrt{\hbar\kB T}$.  The leading order temperature dependence of these results is summarized in
Table~\ref{tab:Tdep}.  
\begin{table}
\centering
\renewcommand{\tabcolsep}{10pt}
\renewcommand{\arraystretch}{2}
\begin{tabular}{|c|c|c|c|} \hline 
& \textbf{SC} & \textbf{QC} & \textbf{M} \\ \hline 
$\sigma$ & $1/T^2$ & $1/\sqrt{T}$ & $T^2$ \\ 
$\kappa/T$ & $1/\sqrt{T}$  & $1/\sqrt{T}$ & $T^2$ \\ \hline
\end{tabular}
\caption{A summary of the temperature dependence of the electrical and thermal conductivity in the 
superconducting (SC), quantum critical (QC) and metallic (M) regimes.}
\label{tab:Tdep}
\end{table}
The finite temperature crossover behavior near the $z=2$ SMT is characterized by  the conductivity
increasing like $1/T^{1/z}$ at high temperatures while the system is in the quantum critical regime
of Fig.~\ref{fig:phaseDiagram} and finally decreasing as $T^z$ after the system has fully returned
to metallic behavior.  Although we are about to show that a microscopic theory reproduces the $T^2$
metallic conduction, the $1/\sqrt{T}$ dependence of the conductivity is not present in the simple
Gaussian theory and an accurate determination of the full crossover phase diagram necessitates the
inclusion of interactions between Cooper pairs.

We now comment on a shared regime of validity between our large-$N$ theory and the disordered electron
perturbation theory of Ref.~\cite{lopatin-vinokur}.  A closer investigation of the dc electrical 
conductivity in the metallic regime with careful attention to all prefactor yields
\begin{equation}
\sigma = \frac{{e^*}^2}{\hbar} \frac{\pi\sqrt{\D}T^2}{12 R^{5/2}}.
\label{eq:dcsigmam}
\end{equation}
We note that upon comparing the large-$N$ propagator of Eq.~(\ref{eq:R}) with Eq.~(4) of
Ref.~\cite{lopatin-vinokur} that the mass $R$ used here is exactly double the mass $\alpha$ employed
by Lopatin \emph{et al.}, i.e. $R = 2[\alpha-\alpha_c(T)]$.  Having made this identification, we 
may compare Eq.~(\ref{eq:dcsigmam}) above, with the finite temperature fluctuation correction 
to the normal state conductivity computed via diagrammatic perturbation theory (Eq.~(8) in
Ref.~\cite{lopatin-vinokur}), and find \emph{exact} agreement.  After a rather lengthy calculation
it was confirmed that perfect correspondence is also found for the thermal conductivity in this
limit \cite{wireslett}.  The concurrence between the two theoretical approaches in this limit is a
result of an approximation made in the diagrammatic calculation involving an infinite sum over a
class of ladder diagrams which turns out to be equivalent to the large-$N$ limit taken here.

\subsection{Wiedemann-Franz ratio}
\label{sec:WFratio}
Having computed the temperature dependence of both the thermal and electrical dc conductivity via a
large-$N$ expansion, it is an opportune time to consider what these results may indicate about the
physics of the SMT.  This is mostly easily accomplished through an analysis of the validity of the
Wiedemann-Franz (WF) law, which states that the low temperature limit of the ratio 
\begin{equation}
W \equiv \frac{\kappa}{\sigma T}
\end{equation}
of the thermal and electrical conductivities of metals if finite and given by the universal Lorenz
number \begin{equation}
l_0= \frac{\pi^2}{3} \left(\frac{k_B}{e}\right)^2.
\end{equation}
This remarkable value relates macroscopic transport properties which can be found from an analysis
of current-current correlation functions to fundamental constants of nature.  The prefactor
$\pi^2/3$ is fixed solely by the Fermi statistics and charge of the elementary quasiparticle
excitations of the metal.  Physically, any constant value of $W$ indicates that elastic collisions
dominate as all scattering events are necessarily charge conserving.  Both the temperature
independence of $W$ as well as the actual value of $l_0$ has been experimentally verified to high
precision in a wide range of metals \cite{louis}, and realizes a sensitive macroscopic test of the
quantum statistics of the charge carriers.  

In addition to simple metals, the value of the Wiedemann-Franz ratio is known in some other
important strongly interacting quantum systems. In superconductors, which possess low energy bosonic
quasiparticle excitations, $\sigma$ is infinite for a range of $T>0$, while $\kappa$ is finite in
the presence of impurities \cite{smitha}, and thus $W=0$.  At quantum phase transitions described by
relativistic field theories, such as the superfluid-insulator transition in the Bose Hubbard model,
the low energy excitations are strongly coupled and quasiparticles are not well defined; in such
theories the conservation of the relativistic stress-energy tensor implies that $\kappa$ is
infinite, and so $W=\infty$ \cite{vzs}.  In other words, any quantum critical point which exhibits
Lorentz or Galilean invariance will have an infinite thermal conductivity since the boosted thermal
distribution will never decay \cite{podolsky-sachdev}.  

Catelani and Aleiner \cite{catelani-aleiner,catelani} have recently investigated interaction
corrections to the Lorenz number in disordered metals. They found that neutral bosonic soft modes
resulting from interacting electron-hole pairs contribute to the transport of energy leading to
temperature dependent deviations from $l_0$, with the corrections being larger in lower
dimensions.  Li and Orignac \cite{edmond} computed $W$ in a disordered Luttinger liquid, and found
deviations from $l_0$ leading to a non-zero universal value at the metal-insulator transition for
spinless fermions. Finally, Fazio {\it et al.} have computed the effects of plasmon
scattering on the Lorenz number of thin wires coupled to reservoirs \cite{fazio} where charge-energy
separation leads to a violation of the Wiedemann-Franz law.  

For the SMT considered here, an examination of Table~\ref{tab:Tdep} immediately leads to the
observation that the Wiedemann-Franz ratio is temperature independent in both the quantum critical
and metallic regimes.  Remarkably, all important couplings between bosons and fermions scale to
universal values, and consequently, by studying the analytic form of $\sigma$ and $\kappa/T$ given
in Eqs.~(\ref{eq:largeNsigma}) and (\ref{eq:largeNkappa}) we find the universal constant
\begin{equation}
W_{QC} = (0.28203\cdots)\left(\frac{k_B}{e}\right)^2
\label{eq:WQC}
\end{equation}
in the quantum critical regime whereas in the metallic region of the phase diagram 
\begin{equation}
W_M = \frac{\pi^2}{5}\left(\frac{k_B}{e}\right)^2.
\end{equation}
Both of these corrections are smaller than the Lorenz number and thus it appears that the Cooper
pairs tend to carry more charge than heat.

These results for the Wiedemann-Franz ratio conclude this section, but in the next section the 
theory presented here for $N=\infty$ is extended to the first order in $1/N$. We will exploit
the anomalous scaling dimension of the dynamical critical exponent $z$ to find additional 
universal corrections to Eq.~(\ref{eq:WQC}).

\section{$1/N$ corrections}
\label{sec:Nfluc}
This section is quite heavy on calculational details and can be skipped by the casual reader.  The
main results include the derivation of a critical theory for a finite $N$ component complex field
$\Psi_a$ governing the fluctuations of Cooper pairs near a superconductor-metal transition.  This
theory is used to systematically compute the $1/N$ corrections to critical exponents and the zero
frequency transport coefficients calculated in the previous section at $N=\infty$ when the coupling 
parameter which drives the SMT attains its critical value.  We will find (Eq.~(\ref{eq:W1oN})) that 
although the individual values of the electrical and thermal conductivities are not universal, 
instead depending depend explicitly on an ultra-violet cutoff, their Wiedemann-Franz ratio is a
pure, temperature independent universal number  that characterizes the most singular 
corrections to transport in the strongly fluctuating quantum critical regime
\begin{equation}
W = \left(0.282 + \frac{0.0376}{N}\right)\left(\frac{k_B}{e}\right)^2.
\end{equation}

\subsection{The critical theory}
\label{subsec:criticalTheory}

We begin by reintroducing the strong-coupling effective action of Eq.~(\ref{eq:Sg}) for 
an $N$-component Cooper pair operator $\Psi_a$
\begin{equation}
\mathcal{S}_g = \frac{1}{g} \isk (k^2 + |\wn|)|\Psi_a(k,\wn)|^2
\end{equation}
where distances have been rescaled by a factor of the square root of the effective diffusion
constant $\D$ and ``hard spin'' constraint $|\Psi(x,\tau)|^2 = 1$ must be enforced.  
Imposing the delta-function constraint via a Lagrange multiplier $\mu$ and performing a rescaling
of the field $\Psi_a \to \sqrt{g} \Psi_a$ leads to the partition function
\begin{equation}\begin{split}
&\mathcal{Z} =  \int \mathcal{D} \Psi^{\phantom{\ast}}_a \mathcal{D}\Psi^\ast_a 
\mathcal{D} \mu \exp \Bigg\{
\\
& \times - \; \int dx
\int d\tau \bigg[ \Psi_a^\ast(x,\tau)\left( -\partial_x^2 + |\partial_\tau| + i\mu(x,\tau)\right)
\Psi^{\phantom{\ast}}_a(x,\tau) -  \frac{N}{g}i\mu(x,\tau)\bigg] \Bigg\} ,
\end{split}\end{equation}
where we have again used the notation $|\partial_\tau|$ to infer $|\wn|$ after Fourier transforming.
Integrating out the $\Psi_a$ fields, 
\begin{equation}
\mathcal{Z} = \int \mathcal{D} \mu\ \exp \left \{ -N \left[ \mathrm{Tr} \ln 
\left( -\partial_x^2 + |\partial_\tau| + i\mu(x,\tau)\right) - 
\frac{i}{g} \int dx \int d\tau \mu(x,\tau) \right]\right \}
\end{equation}
and as done previously, for $N$ large, we can approximate the functional integral over $\mu$ by its 
saddle point value defined to be $r = i\mu$ leading to 
\begin{equation}
\frac{1}{g} = \isk \frac{1}{k^2 + |\wn| + r}.
\label{eq:spg}
\end{equation}
This is an auspicious point to make a brief comment regarding the relationship between the notation
introduced here and that of the previous section.  Eq.~(\ref{eq:spg}) is identical to
Eq.~(\ref{eq:R}) with the replacement of $r \to R$.  However, in this section, unless otherwise
stated, we will be considering a critical theory (whether at zero or finite temperatures) with the
coupling $g$ equal to its critical value $g_c$ which will be shifted from its $N=\infty$ value by a
correction of order $1/N$.  As a result, the effective mass $r$ will be also corrected from its
$N=\infty$ saddle point value.  This will be made more explicit soon, but for now we simply indicate
that $r = R + O(1/N)$ with $R$ equal to its $N=\infty$ value defined by Eq.~(\ref{eq:R}) with
$g=g_c$.

Let us now focus on fluctuations around the saddle point by defining $i\mu = r + i\lambda$, and after
expanding to quadratic order in $\lambda$ and noticing that with the help of Eq.~(\ref{eq:spg}) all
linear terms cancel we have
\begin{equation}
\mathcal{Z} = \int \mathcal{D} \lambda\ \exp \left \{ -N \left[ \mathrm{Tr} 
\ln ( -\partial_x^2 + |\partial_\tau| + r) + \frac{1}{2} \sum_{\wn}\int \frac{dk}{2\pi}
\lambda^2 \Pi_T(k,\wn,r) \right] \right\}
\label{eq:Zlambda}
\end{equation}
where
\begin{equation}
\Pi_T(k,\wn,r) = \isq \frac{1}{[(k+q)^2 + |\wn+\epsilon_n| +r](q^2 +
|\epsilon_n| + r)}
\label{eq:PiT}
\end{equation}
can be thought of as the propagator for a $\lambda$ field leading to $1/N$ fluctuations.  Upon
examination of Eq.~(\ref{eq:Zlambda}), it is apparent that we could have simply started from a
partition function for the original field $\Psi_a$ with an additional interaction term such that its
diagrammatic expansion is equivalent to that of Eq.~(\ref{eq:Zlambda}), i.e.
\begin{equation}\begin{split}
&\mathcal{Z} = \int \mathcal{D} \Psi^{\phantom{\ast}}_a \mathcal{D}\Psi^\ast_a 
\mathcal{D} \lambda \exp \left\{ - \int dx \int d\tau \bigg[ \Psi^\ast_a(x,\tau) \left(
-\partial_x^2 + |\partial_\tau| + r\right)\Psi^{\phantom{\ast}}_a(x,\tau) \right.
\\
& \left. \; + \; i \lambda(x,\tau) |\Psi_a(x,\tau)|^2 + 
\frac{N}{2} \int dx' \int d\tau' \lambda(x,\tau)\Pi_T(x-x',\tau-\tau',r) \lambda(x',\tau') \bigg] \right\}
\label{eq:Znfluc}
\end{split}\end{equation}
leading to the effective action in momentum space
\begin{equation}\begin{split}
\mathcal{S}_r &= \isk \left [(k^2 + |\wn| + r)|\Psi_a(k,\wn)|^2 + 
\frac{N}{2} |\lambda(k,\wn)|^2 \Pi_T(k,\wn,r) \right.
\\
& \left. \qquad\qquad +\; \isq 
\Psi_a^\ast(k,\wn)\Psi^{\phantom{\ast}}_a(q,\en)\lambda(k-q,\wn-\en) \right ]
\label{eq:Sr}
\end{split}\end{equation}
where we note that in order to avoid double-counting, the $\lambda$ or fluctuation propagator 
$\Pi_T$ cannot have a self-energy contribution of a single $\Psi_a$ bubble 
(since it has already been included).  Thus, performing a direct $1/N$ expansion 
from $\mathcal{Z}$ for $G(k,\wn) = \langle |\Psi_a(k,\wn)|^2\rangle$ \cite{qpt} we have (to order
$1/N$)
\begin{equation}
G(k,\wn) =
\;
\parbox{53\unitlength}{%
	\begin{fmffile}{G1}
		\begin{fmfgraph}(50,10)
			\fmfleft{l}
			\fmfright{r}
			\fmfpen{thick}
			\fmf{plain}{l,r}
		\end{fmfgraph}
	\end{fmffile}}
+\;\;
\parbox[c][40\unitlength][t]{53\unitlength}{%
	\begin{fmffile}{G2}
		\begin{fmfgraph}(50,40)
			\fmfleft{l}
			\fmfright{r}
			\fmfforce{(.3w,.5h)}{v1}
			\fmfforce{(.7w,.5h)}{v2}
			\fmfdot{v1,v2}
			\fmf{plain}{l,v1,v2,r}
			\fmf{dashes,left=2.0}{v1,v2}
		\end{fmfgraph}
	\end{fmffile}}
+\;
\raisebox{5.3mm}{   
\parbox[c][25mm][b]{53\unitlength}{%
	\begin{fmffile}{G3} 
		\begin{fmfgraph}(50,40)
			\fmfleft{l}
			\fmftop{t}
			\fmftop{u1,u2}
			\fmfright{r}
			\fmffreeze
			\fmfforce{(.5w,1.0h)}{t}
			\fmfforce{(.5w,.50h)}{v}
			\fmfforce{(.3w,1.45h)}{u1}
			\fmfforce{(.7w,1.45h)}{u2}
			\fmfdot{v,t,u1,u2}
			\fmf{plain}{l,v,r}
			\fmf{dashes}{v,t}
			\fmf{dashes,left=1.2}{u1,u2}
			\fmf{plain,right,tension=0.6}{t,t}
		\end{fmfgraph}
	\end{fmffile}}} 
\end{equation}
where a solid line is equal to $(k^2 + |\wn| + r)^{-1}$, a dashed line equal to $\Pi_T/N$ and a
solid dot represents the interaction vertex $i$. There is no tadpole graph as it is already included
in the $1/N$ correction to the effective mass $r$.  The third graph has two loops, but is only of
order $1/N$ as any closed $\Psi_a$ loop gives a factor of $N$. Combining these graphs leads to the
full expression
\begin{align}
G^{-1} (k,\wn) &= k^2 + |\wn| + r + \frac{T}{N} \sum_{\epsilon_n}\! \int \!\frac{dq}{2\pi} 
\frac{1}{\Pi_T(q,\epsilon_n,r)} \frac{1}{[(k+q)^2 + |\wn + \epsilon_n| + r]} \nonumber
\\
& \qquad -\; \frac{1}{N} \frac{1}{\Pi_T(0,0,r)} \isq \isp \frac{1}{\Pi_T(q,\epsilon_n,r)} \nonumber
\\
& \qquad \qquad \times \; \frac{1}{(p^2 + |\nu_n| + r)^2 [(p+q)^2 + |\epsilon_n + \nu_n| + r]} .
\label{eq:Ginv}
\end{align}

\subsection{Quantum critical point}
At $T=0$, the critical point $g_c$ is determined by the condition $G^{-1}
(0,0) = 0$, $r=r_c$. Keeping terms only up to order $1/N$ 
\begin{equation}\begin{split}
r_c &= -\frac{1}{N} \iik \frac{1}{\Pi_0 (k,\omega,0)} \frac{1}{k^2 + |\omega|} 
+ \frac{1}{N} \frac{1}{\Pi_0 (0,0,0)} \iik 
\\
& \qquad \times \; \iiq \frac{1}{(k^2+|\omega|)^2
[(k+q)^2+|\omega+\epsilon|]} \frac{1}{\Pi_0 (q,\epsilon,0)}
\end{split}\end{equation}
where
\begin{align}
\Pi_0 (k,\omega,0) &= \iiq \frac{1}{(q^2+|\epsilon|)[(k+q)^2+|\omega+\epsilon|]} \nonumber 
\\
&= \frac{1}{4\pi |k|} \left[ 2\arcsin \left(\frac{k^2-|\omega|}{k^2+|\omega|}\right) + \pi \right]
\nonumber
\\
& \qquad +\;  \frac{1}{4\pi \sqrt{k^2 + 2|\omega|}} 
\ln \left( \frac{2\sqrt{|\omega|}\sqrt{k^2+2|\omega|} + k^2 + 3|\omega|}
{|2\sqrt{|\omega|}\sqrt{k^2+2|\omega|} - k^2 - 3|\omega||}\right)
\end{align}
with details given in Appendix~\ref{app:Pi}.  Note that $\Pi_0 (0,0,0)$ is infrared divergent, but
this will shortly cancel out of observable quantities. Inserting the expansion for $r_c$ in
Eq.~(\ref{eq:spg}), we obtain
\begin{equation}\begin{split}
\frac{1}{g_c} &= \iik \frac{1}{k^2+|\omega|} + \frac{1}{N} \iik 
\\
& \qquad \times \; \iiq \frac{1}{\Pi_0 (q,\epsilon,0) (k^2+|\omega|)^2 } 
\left[ \frac{1}{q^2+|\epsilon|} - \frac{1}{(k+q)^2+|\omega+\epsilon|} \right] 
\label{eq:Ngc}
\end{split}\end{equation}
which is free of infrared divergences.

\subsection{Quantum critical propagator}

Moving to finite temperatures, but setting $g=g_c$, we write $r = R + \widetilde{R}_1$ with
$\widetilde{R}_1 \sim O(1/N)$. As mentioned previously, $R$ is determined by setting $r=R$ in
Eq.~(\ref{eq:spg}) when $g=g_c$ takes its $N=\infty$ value 
\begin{equation}
\isk \frac{1}{k^2 +|\omega_n| + R} = \iik \frac{1}{k^2+|\omega|}. 
\end{equation}
We have seen this equation before in Eq.~(\ref{eq:R}) with $g=g_c$ and can thus express it as
Eq.~(\ref{eq:deltaR}) with $\delta=0$ giving an equation that can be inverted to 
uniquely determine $R/T$,
\begin{equation}
0 =  \int \frac{dk}{2\pi} \left[ \frac{\pi T}{k^2 + R} - \psi\left(1+\frac{k^2 + R}{2\pi T}\right) +
\ln \left(\frac{k^2}{2\pi T}\right) \right]
\end{equation}
where $\psi(x)$ is the polygamma function.  Solving numerically we find
\begin{equation}
\frac{R}{T} \simeq 0.624798 .
\label{eq:RoT}
\end{equation}

Returning to Eq.~(\ref{eq:spg}) we can write (to order $1/N$)
\begin{align}
\frac{1}{g_c} &= \isk \frac{1}{k^2 + |\omega_n| + R + \widetilde{R}_1} \nonumber \\
&= \isk \frac{1}{k^2 + |\omega_n| + R} - \Pi_T(0,0,R) \widetilde{R}_1
\end{align}
which can be compared with our expression for $1/g_c$ in Eq.~(\ref{eq:Ngc}) order by order to yield
\begin{equation}\begin{split}
\widetilde{R}_1 &= - \frac{1}{N\Pi_T (0,0,R)}  \iik \iiq
\frac{1}{(k^2+|\omega|)^2} \frac{1}{\Pi_0 (q,\epsilon,0)} 
\\
& \qquad \times \; \left[ \frac{1}{q^2+|\epsilon|} - \frac{1}{(k+q)^2 + |\omega + \epsilon|} \right]
\end{split}\end{equation}
and note that $\sqrt{T}\ \Pi_T(0,0,R)$ is a finite universal number given by (see Appendix~\ref{app:Pi})
\begin{equation}
\sqrt{T}\ \Pi_T(0,0,R) = 
\frac{1}{4(2\pi)^{3/2}} \left[ \zeta\left(\frac{3}{2}, \frac{R}{2\pi T}\right) 
+ \zeta\left(\frac{3}{2}, \frac{R}{2\pi T} + 1\right) \right],
\end{equation}
where $\zeta(m,x)$ is the Hurwitz Zeta function.  Inserting everything in Eq.~(\ref{eq:Ginv}) 
\begin{equation}
G^{-1} (k,\omega_n) = k^2 + |\omega_n| + R + R_1 + \Sigma(k,\omega_n)
\label{eq:GinvSig}
\end{equation}
where the self energy $\Sigma(k,\wn)$ is defined to be
\begin{equation}
\Sigma (k,\omega_n) = \frac{T}{N} \sum_{\epsilon_n}\!\int \! 
\frac{dq}{2\pi} \frac{1}{\Pi_T (q,\epsilon_n,R)}
\!\left[\frac{1}{(k+q)^2 + |\omega_n+\epsilon_n| + R} - \frac{1}{q^2 + |\epsilon_n| + R} \right]
\label{eq:Sigma}
\end{equation}
such that $\Sigma(0,0) = 0$, and
\begin{align}
R_1 &= \frac{1}{N \Pi_T (0,0,R)} \left\{ - \iiq \frac{1}{\Pi_0 (q,\epsilon,0)}
\iik \frac{1}{(k^2+|\omega|)^2} \right. \nonumber
\\
& \left. \qquad \qquad \times \; \left[\frac{1}{q^2+|\epsilon|} - \frac{1}{(k+q)^2+|\omega+\epsilon|}
\right]  \right. \nonumber
\\
& \qquad +\;  \isq \frac{1}{\Pi_T (q,\epsilon,R)} \isk \frac{1}{(k^2 + |\omega_n| +  R)^2} \nonumber
\\
& \left. \qquad \qquad \times \; 
\left[\frac{1}{(q^2 + |\epsilon_n| + R)} - \frac{1}{(k+q)^2 + |\omega_n+\epsilon_n| + R)} 
\right] \right\} .
\end{align}
This is equivalent to Eq.~(4.6) in Ref.~\cite{csy}. 
Now, both the inner integral or sum over $(k,\omega)$ and $(k,\omega_n)$ is ultraviolet convergent, 
but the outer integral appears to be divergent. By first introducing an ultraviolet momentum cutoff
and evaluating the integrals numerically using adaptive mesh techniques for fixed $\Lambda$ the UV
limit can be investigated with the result that $R_1/T$ converges to the universal finite value
\begin{equation}
\frac{R_1}{T} \simeq \frac{0.1069}{N}.
\label{eq:R1oT}
\end{equation}
Therefore, as described in Ref.~\cite{pankov} and Section~\ref{subsec:scalingAnalysis},
as a consequence of the scaling relation $z = 2-\eta$ where $\eta$ is the anomalous dimension 
of $\Psi_a$, the uniform static order parameter susceptibility 
\begin{equation}
\chi = \int dx \int d\tau \langle \Psi_a^\ast(x,\tau)\Psi^{\phantom{\ast}}_a(0,0)\rangle
\end{equation}
is determined by the value of $k_B T$ alone.  Using Eqs.~(\ref{eq:RoT}), (\ref{eq:GinvSig}) and
(\ref{eq:R1oT}) we find 
\begin{equation}
\chi^{-1} = k_B T \left(0.6248 + \frac{0.107}{N} \right).
\end{equation}

\subsection{Critical exponents}
\label{subsec:criticalExponents}

With a quantum critical theory firmly established, we may now investigate any possible $1/N$
corrections to the large-$N$ critical behavior characterized by exponents $z=2$ and $\nu=1$.  Such
corrections can be obtained by exploiting the known scaling behavior of the susceptibility in
conjunction with various hyperscaling relations. We begin by computing the anomalous dynamical
scaling dimension $\eta$ which corrects $z$ at order $1/N$.

\subsubsection{The anomalous dimension $\eta$}
It is known that the $\omega=0$ susceptibility should scale with momentum like $G^{-1}(k,0) \sim
k^z$ where the bare dynamical critical exponent $z=2$ will be corrected by the critical exponent
$\eta$ as $z = 2-\eta$.  Therefore, we can write
\begin{equation}
G^{-1}(k,0) \sim k^{2-\eta} \simeq k^2 \left(1+\eta \ln \frac{\Lambda}{k} \right)
\label{eq:Geta1}
\end{equation}
where $\Lambda$ is a large momentum cutoff.  From Eq.~(\ref{eq:Ginv}) at $r=r_c$, $T=0$ and $\omega
= 0$ we have
\begin{equation}
G^{-1} (k,0) = k^2 + \frac{k^2}{N} \int_{-\Lambda / k}^{\Lambda/k} \frac{dq}{2\pi} |q|
\int_0^\infty \frac{d\epsilon}{\pi} \frac{1}{\Pi_0(1,\epsilon,0)} 
\left[ \frac{1}{(1+1/q)^2 + \epsilon} - \frac{1}{1+\epsilon}\right] .
\label{eq:Geta3}
\end{equation}
where we have used Eq.~(\ref{eq:Pi0}) and all variables of integration are dimensionless.  Expanding
the integrand for large $q$ and identifying the logarithmic prefactor leads to
\begin{align}
\eta &= \frac{1}{\pi^2 N} \int_0^{\infty} d\epsilon \frac{3-\epsilon}{\Pi_0(1,\epsilon,0)
(\epsilon+1)^3} \nonumber \\
& \simeq \frac{0.13106}{N} .
\label{eq:eta}
\end{align}
Knowing the value of $\eta$ will be particularly useful because it will fix the cutoff dependence of
the quantum critical conductivity at order $1/N$, since $z=2-\eta$, and we expect $\sigma (T) \sim
T^{-1/z}$.  Thus, if $\sigma (T) = A/\sqrt{T}$ where $A$ is a constant at $N=\infty$, then at order
$1/N$ we should have $\sigma = (A/\sqrt{T})[1 + (\eta/2) \ln (\Lambda/\sqrt{T})]$.

\subsubsection{The correlation length exponent $\nu$}

Calculating the $1/N$ correction to the correlation length exponent $\nu$ is unfortunately not so
simple, but we begin by examining the behavior of the inverse susceptibility at $T=0$  and $k=\omega
= 0$ as one tunes the coupling constant $g$ towards $g_c$  
\begin{equation}
G^{-1}(0,0) \sim (g-g_c)^\gamma 
\label{eq:Ggamma}
\end{equation}
which defines the susceptibility exponent $\gamma$.  At $N=\infty$ we know $\gamma = 2$, and thus
for finite $N$ let us parameterize $\gamma = 2(1-\alpha)$, which can be related to $\nu$ via the
scaling relation $\gamma = (2-\eta)\nu$.  To this end, we define $r_g$ via
\begin{align}
\frac{1}{g_c} - \frac{1}{g} &\equiv \iik \left(\frac{1}{k^2 + |\omega|} - \frac{1}{k^2 + |\omega| +
r_g}\right) \nonumber \\
&=  \frac{\sqrt{r_g}}{\pi},
\label{eq:ggcrg}
\end{align}
where we have exploited the fact that
\begin{equation}
\sqrt{r_g} = \frac{1}{2\pi\Pi_0(0,0,r_g)} .
\end{equation}
Thus, from Eq.~(\ref{eq:ggcrg}) we have $r_g \sim (g-g_c)^2$, and upon comparison with
Eq.~(\ref{eq:Ggamma}) we find
\begin{equation}
G^{-1}(0,0) \sim (g-g_c)^{2(1-\alpha)} 
\sim r_g^{1-\alpha} 
\simeq r_g\left(1 + \alpha \ln \frac{\Lambda^2}{r_g}\right).
\label{eq:Grg}
\end{equation}
So again we can extract a critical exponent by determining the prefactor of a logarithmic divergence
of $G^{-1}$.  At this stage it will useful to quote the following two results (with details given in
Appendix~\ref{app:Pi})
\begin{align}
&\Pi_0(k,\omega,r) = \nonumber
\\ 
&\frac{1}{2\pi |k|} \left[ 
\mathrm{arcsin}\left(\frac{k^2 + |\omega|}{\sqrt{(k^2+|\omega|)^2 + 4 k^2 r}}\right) + 
\mathrm{arcsin}\left(\frac{k^2 - |\omega|}{\sqrt{(k^2+|\omega|)^2 + 4 k^2 r}}\right) \right ]
\nonumber
\\
& \quad +\; \frac{1}{4\pi \sqrt{k^2 + 2|\omega|+4r}} \left[ 
\ln \left( \frac{ 2\sqrt{r+|\omega|}\sqrt{k^2 + 2|\omega| + 4r} + k^2 + 3|\omega| + 4r}
{|2\sqrt{r+|\omega|}\sqrt{k^2 + 2|\omega| + 4r} - k^2 - 3|\omega| - 4r|}\right)  \nonumber \right. 
\\
& \qquad\qquad \left. - \ln \left( \frac{ 2\sqrt{r}\sqrt{k^2 + 2|\omega| + 4r} + k^2 + |\omega| + 4r}
{|2\sqrt{r+|\omega|}\sqrt{k^2 + 2|\omega| + 4r} - k^2 - |\omega| - 4r|}\right)  \right]
\end{align}
and 
\begin{align}
\Pi_0^\prime (k,\omega,r) &\equiv \frac{\partial \Pi_0(k,\omega,r)}{\partial r} \nonumber
\\
&= -2\iiq \frac{1}{(q^2 + |\epsilon|+r)^2 [(k+q)^2+|\omega+\epsilon|+r]}.
\label{eq:dPidr}
\end{align}
Combining Eqs.~(\ref{eq:spg}) with (\ref{eq:Ngc}) and (\ref{eq:ggcrg}) $r$ can be written in terms
of $r_g$ which when used in Eq.~(\ref{eq:Ginv}) leads to the result for the inverse susceptibility
\begin{equation}\begin{split}
G^{-1} (0,0) &= r_g + F(r_g) 
\frac{2 \pi \sqrt{r_g}}{N} \iik \iiq \frac{1}{\Pi_0 (q,\epsilon,0)}  
\frac{1}{(k^2+|\omega|)^2} 
\\
& \qquad \times \; \left[ \frac{1}{q^2+\epsilon} - \frac{1}{(k+q)^2+|\omega+\epsilon|} \right]
\label{eq:GFrg}
\end{split}\end{equation}
where we have been able to replace $r$ with $r_g$ in any term that is already of order $1/N$ and 
\begin{equation}
F(r_g) =  \frac{1}{N} \iik \frac{1}{\Pi_0 (k,\omega,r_g)} \left[
\frac{1}{k^2 + |\omega| + r_g}  + \frac{\Pi_0^\prime (k,\omega,r_g)}{2\Pi_0(0,0,r_g)} \right] .
\label{eq:Frg}
\end{equation}
A useful check is to note that $G^{-1} (0,0)=0$ above for $r_g = 0$.  The next step is to
investigate the small $r_g$ behavior of $F(r_g)$. For this, let us first examine 
$\Pi_0^\prime$ as $r_g \to 0$, we find
\begin{equation}
\Pi_0^{\prime} (k,\omega, r_g) = 
- \frac{1}{\pi (k^2+|\omega|) \sqrt{r_g}} + \frac{1}{|k|^3} \Phi_1 \left(\frac{|\omega|}{k^2}\right)
+ \frac{1}{k^4} \Phi_2 \left(\frac{|\omega|}{k^2}\right) \sqrt{r_g} + \cdots 
\label{eq:dPidrexpan}
\end{equation}
where 
\begin{align}
\Phi_1(x) &= \frac{1}{4\pi(1 + x)} \int_0^1 dy
\frac{y}{(2y-1)^2}\left\{ \frac{y[4y^2 - 6y + 1 + (6y-5)x]}{\sqrt{1-y}(y+x)^{3/2}} \right. \nonumber
\\
& \left. \qquad \qquad \qquad -\; \frac{4y^4 - 10x^3 + 11y^2 - 6y + 1 - (6y^3 - 9y^2 + 6y - 1)x}
{[y(1-y + x)]^{3/2}}\right \}
\label{eq:Phi1}
\\
\Phi_2 (x) &= \frac{2}{\pi} \frac{3 + x}{(1+x)^3}
\label{eq:Phi2}
\end{align}
and we have used the fact that $|\epsilon| \ll \omega$ over the regime important for small $r_g$.
From this expansion we can also determine the small $r_g$ expansion of $\Pi_0$ 
\begin{equation}\begin{split}
& \Pi_0 (k,\omega,r_g) = 
\\
& \qquad \Pi_0(k,\omega,0) - \frac{2 \sqrt{r_g}}{\pi (k^2+|\omega|)} 
+ \frac{1}{|k|^3} \Phi_1\left(\frac{|\omega|}{k^2}\right)r_g 
+ \frac{2}{3 k^4} \Phi_2 \left(\frac{|\omega|}{k^2}\right) r_g^{3/2} + \cdots
\end{split}\end{equation}
and finally that of $F(r_g)$
\begin{equation}\begin{split}
F(r_g) &=  \frac{\pi \sqrt{r_g}}{N} \iik \frac{\Phi_1 (|\omega|/k^2)}{|k|^3 \Pi_0(k,\omega,0)} 
+ \frac{r_g}{N} \iik 
\\
& \; \times \; \left[ \frac{2 \Phi_1 (|\omega|/k^2)}{|k|^3(k^2+|\omega|) \Pi_0^2 (k,\omega,0)} 
+ \frac{\pi \Phi_2 (|\omega|/k^2)}{k^4\Pi_0(k,\omega,0)} - \frac{1}{(k^2+|\omega|)^2 \Pi_0(k,\omega,0)}
\right] .
\label{eq:FrgFinal}
\end{split}\end{equation}
Now, comparing this result with Eq.~(\ref{eq:Grg}) and (\ref{eq:GFrg}) the second term, which is
linear in $r_g$ defines $\alpha$ by
\begin{equation}
F(r_g) = \cdots +  \alpha r_g \ln \left( \frac{\Lambda^2}{r_g} \right) + \cdots
\end{equation}
where $\alpha$ can be related to the correlation length exponent by the scaling relation
\begin{equation}
\gamma = 2 (1-\alpha) = \nu (2-\eta).
\end{equation}
Using Eq.~(\ref{eq:Pi0}) to define
\begin{equation}
\Phi_0\left(\frac{|\omega|}{k^2}\right) = |k| \Pi_0(k,\omega,0)
\end{equation}
and from Eq.~(\ref{eq:FrgFinal}) $\alpha$ is given by
\begin{align}
\alpha &= \frac{1}{2\pi^2 N} \int_0^{\infty} d \omega
 \left[ \frac{2 \Phi_1 (\omega)}{(\omega + 1) \Phi_0^2 (\omega)} + 
 \frac{\pi \Phi_2 (\omega)}{\Phi_0(\omega)} 
- \frac{1}{(1 + \omega)^2 \Phi_0(\omega)} \right] \nonumber 
\\
&\simeq \frac{0.455}{N}.
 \label{eq:alphaCE}
\end{align}
The value of $\nu$ can finally be determined using Eq.~(\ref{eq:eta}) as
\begin{align}
\nu &= 1 - \alpha + \frac{\eta}{2} \nonumber \\
&\simeq 1 - \frac{0.389}{N}.
\label{eq:nu}
\end{align}

The values found in this section for $z=2-\eta$ (Eq.~(\ref{eq:eta})) and $\nu$ (Eq.~(\ref{eq:nu}))
corresponding to a $N$ component complex field are fully consistent with previous calculations
including an $\epsilon$ expansion near 2 dimensions \cite{pankov, sachdev-troyer} 
(Eq.~(\ref{eq:etaeps}) and (\ref{eq:nueps})) and via Monte Carlo simulations where
$z=1.97(3)$ and $\nu=0.689(6)$ \cite{werner-troyer}.

\subsection{Quantum transport at finite N}
\label{subsec:quantumTransport}
We now endeavor to compute the dc values of the thermal and electrical conductivity in
a $1/N$ expansion in the quantum critical regime.  Transport is again calculated via the Kubo
formula, and the initial steps are identical to those presented in Section~\ref{subsec:teTransport} for
the derivation of Eq.~(\ref{eq:G}).  However, unlike the case where the number of components of 
our order parameter field was infinite, we now have the modified propagator of
Eq.~(\ref{eq:GinvSig}) and the single polarization bubble diagrams will be corrected by
additional loops giving rise to corrections of order $1/N$.

It will turn out that although the individual values of the thermal ($\kappa$) and electrical 
($\sigma$) conductivities are not by themselves universal to order $1/N$, their ratio is a universal
number, solely as a result of the appearance of an anomalous dimension that alters the critical
dynamic scaling.

\subsubsection{Diagrammatic expansion}

We ask the reader to recall Eqs.~(\ref{eq:JeJq}) to (\ref{eq:G}) and begin by writing down the
expression for the transport coefficients ($p=0$ for electrical conductivity and $p=2$
for thermal conductivity) obtained from the Kubo formula in terms of a polarization function at	
external imaginary frequency $i\wn$
\begin{equation}
\mathcal{G}_p(i\wn) = -\frac{4 {e^*}^{2-p}}{\wn T^p} \mathcal{K}_p(i\wn).
\label{eq:GKpara}
\end{equation}
The current-current correlation function is evaluated with respect to $\mathcal{S}_r$,
Eq.~(\ref{eq:Sr}), and the result contains both a dia and paramagnetic part
\begin{equation}\begin{split}
\mathcal{K}_p(i\wn) &= -\frac{T}{2}\sum_{\en}\nt{k}{2} \  (\en + \wn/2)^p  \left[
\langle |\psi(k,\en)|^2\rangle \right.
\\
& \left. \qquad -\; k^2 \langle \Psi_a^\ast(k,\en) 
\Psi^{\phantom{\ast}}_a(k,\en+\wn) \Psi^{\ast}_a(k,\en+\wn)\Psi^{\phantom{\ast}}_a(k,\en)\rangle 
\right].
\label{eq:Kp}
\end{split}\end{equation}
As we are only interested in the real dc thermal and electric transport coefficients, an examination
of Eqs.~(\ref{eq:GKpara}) and (\ref{eq:Kp}) indicates that after analytic continuation to real
frequencies we will need the imaginary part of $\mathcal{K}_p$, and can thus focus only 
on the paramagnetic contribution to Eq.~(\ref{eq:Kp}) corresponding to the four-point correlation
function.  This term arises as a result of quadratic insertions of the scalar or thermal potentials
$A_j$.  The resulting paramagnetic polarization function has the diagrammatic expansion to order
$1/N$ given by
\begin{equation}
\mathcal{K}^{\mathrm{para}}_p(i\wn) = 
\;
\parbox{43\unitlength}{%
	\begin{fmffile}{K5}
		\begin{fmfgraph}(40,40)
			\fmfleft{l}
			\fmfright{r}
			\fmfv{decor.shape=circle,decor.size=5,decor.filled=0}{l,r}
			\fmf{dbl_plain,right}{l,r}
			\fmf{dbl_plain,right}{r,l}
		\end{fmfgraph}
	\end{fmffile}}
\;+\;\;
\parbox{43\unitlength}{%
	\begin{fmffile}{K4}
		\begin{fmfgraph}(40,40)
			\fmfleft{l}
			\fmfright{r}
			\fmfv{decor.shape=circle,decor.size=5,decor.filled=0}{l,r}
			\fmffreeze
			\fmftop{t1,t2}
			\fmfforce{(.5w,1.0h)}{t1}
			\fmfforce{(.5w,0.0h)}{t2}
			\fmfdot{t1,t2}
			\fmf{plain,right}{l,r}
			\fmf{plain,right}{r,l}
			\fmf{dashes}{t1,t2}
		\end{fmfgraph}
	\end{fmffile}}
\end{equation}
where 
\begin{align}
\begin{fmffile}{K8}
\begin{fmfgraph}(30,5)
	\fmfleft{l}
	\fmfright{r}
	\fmf{plain}{l,r}
\end{fmfgraph}
\end{fmffile}
&= G_0(k,\en) = \frac{1}{k^2 + |\en| + R} \\
\begin{fmffile}{K9}
\begin{fmfgraph}(30,5)
	\fmfleft{l}
	\fmfright{r}
	\fmf{dbl_plain}{l,r}
\end{fmfgraph}
\end{fmffile}
&= \frac{1}{k^2 + |\en| + R + R_1 + \Sigma(k,\en)}, 
\end{align}
and an open circle indicates a factor of $k(\en+\wn/2)^{p/2}$ where $p=0$ for electrical transport
and $p=2$ for thermal transport and a closed dot represents the interaction $i$.  The dashed line is
the fluctuation propagator $\Pi_T/N$ and $R_1$ is the finite shift in the critical point to order
$1/N$ given by Eq.~(\ref{eq:R1oT}). The self-energy is defined in Eq.~(\ref{eq:Sigma}) such that
$\Sigma(0,0) = 0$.  To identify the role of various $1/N$ corrections to transport it will be useful
to present the full integral form
\begin{align}
& \mathcal{K}^{\mathrm{para}}_{p}(i\wn) =  \iske k^2 (\en + \wn/2)^p \G{k}{\en}{}\G{k}{\en+\wn}{} 
\nonumber \\ 
& \;-\; 2 R_1 \iske k^2 (\en+\wn/2)^p \G{k}{\en}{2}\G{k}{\en+\wn}{} \nonumber \\
& \;-\; 2 \iske k^2 (\en+\wn/2)^p \G{k}{\en}{2}\G{k}{\en+\wn}{}\Sigma(k,\en) \nonumber \\
& \;-\; \frac{2}{N} \iskq \frac{k(k+q) (\en+\wn/2)^{p/2}(\en+\On+\wn/2)^{p/2}}{\Pi_T(q,\On,R)} 
\nonumber \\ 
& \; \times \; \G{k}{\en}{}\G{k}{\en+\wn}{} \G{k+q}{\en+\On}{} \G{k+q}{\en+\On+\wn}{}.
\label{eq:Kpara}
\end{align}
The first term is just the paramagnetic contribution in the large-$N$ polarization function
previously defined in Eq.~(\ref{eq:KLargeN}).  The second term arises from the $1/N$ 
correction to the mass $R$, and the final two terms from the self-energy and vertex
corrections respectively.

\subsubsection{Frequency summations}
The Matsubara summations in the various terms of Eq.~(\ref{eq:Kpara}) can be performed by solving
integrals in the complex plane with repeated use of the basic identity \cite{agd}
\begin{equation}
T\sum_{\en} \mathcal{F}(i\en) = 
\frac{1}{2} \int \frac{d\epsilon}{2\pi i} \coth\left(\frac{\epsilon}{2 T}\right) 
\left[ F(\epsilon + i\eta) - F(\epsilon - i\eta)\right]
\end{equation}
where we note that if $F(i\en) = \mathcal{F}(|\en|)$ then after analytic continuation $F(\epsilon
\pm i\eta) = \mathcal{F}(\mp i\epsilon)$.  The full details on the derivation of various summation
formulae used in this section are given in an Appendix~\ref{app:sum}.  The general approach will be 
as follows: use the relevant summation formula to obtain an expression for each term in 
Eq.~(\ref{eq:Kpara}) analytically continued to real frequencies.  Since we are only interested 
in dc transport, an examination of Eq.~(\ref{eq:GKpara}) tells us that we will require the 
imaginary part of the term that is linear in the external frequency, $\omega$.  Thus by Taylor 
expanding our analytically continued result, we can extract the relevant transport coefficients.  
We will examine each term separately.

For the first term in Eq.~(\ref{eq:Kpara}) we could just as easily perform the Matsubara sum using
the spectral representation of the bare Green function, which was done in
Section~\ref{subsec:teTransport} and led to Eq.~(\ref{eq:Gdc}).  This allows for a test and 
benchmark of the contour integration approach.  We need to evaluate:
\begin{align}
I_{2,p}(i\wn) &= T\sum_{\en} \frac{(\en+\wn/2)^p}{(k^2+R + |\en|)(k^2+R+|\en+\wn|)} \nonumber 
\\
&\equiv T\sum_{\en} \mathcal{F}_{2,p}(i\en,i(\en+\wn))
\label{eq:I2sum}
\end{align}
where we have suppressed the momentum dependence of $\mathcal{F}_{2,p}$ for compactness.
Using Eq.~(\ref{eq:I2}) we find
\begin{equation}
\lim_{\omega\to0} \frac{\Im I_{2,p}(\omega + i\eta)}{\omega} =  
\frac{1}{2T} \ie \frac{\epsilon^{2+p}}{\sinh^2(\epsilon/2T)} \frac{1}{[(k^2+R)^2+\epsilon^2]^2},
\end{equation}
which does indeed agree with our previous result, Eq.~(\ref{eq:Gdc}).
Substituting into Eq.~(\ref{eq:GKpara}) and defining 
\begin{align}
\mathcal{G}_p & \equiv \lim_{\omega\to0} \mathcal{G}_p(\omega+i\eta) 
\\
& = \mathcal{G}_p^{N=\infty} + \mathcal{G}_p^{R_1} + \mathcal{G}_p^{\Sigma} + \mathcal{G}_p^{\Gamma}
\end{align}
where we have broken the total dc transport into a sum of four contributions coming from the four
types of terms in Eq.~(\ref{eq:Kpara}). Because we have ignored the diamagnetic part of the
polarization function, $\mathcal{K}_p^{\mathrm{para}}$ is purely imaginary and thus after analytic
continuation $\mathcal{G}_p$ is a real number.  The $N=\infty$ contribution is
\begin{align}
\mathcal{G}_p^{N=\infty} &= \frac{4{e^*}^{2-p}}{2T^{p+1}} \ie
\frac{\epsilon^{2+p}}{\sinh^2(\epsilon/2T)} \nonumber 
\int \frac{dk}{2\pi} \frac{k^2}{[(k^2+R)^2+\epsilon^2]^2} \nonumber 
\\
&= \frac{1}{\sqrt{T}} \left\{
\begin{array}{lcr}
0.217997 \cdots \ {e^*}^2 & ; & p = 0 \\
0.24592 \cdots & ; & p = 2 \\
\end{array}\right. .
\label{eq:GNinfty}
\end{align}

Due to the finite shift in the critical point, coming from $R_1 \sim O(1/N)$, we need to 
evaluate a correction of the form
\begin{align}
\widetilde{I}_{2,p}(i\wn) &= T\sum_{\en} 
\frac{(\en+\wn/2)^p}{(k^2+R + |\en|)^2(k^2+R+|\en+\wn|)} \nonumber 
\\
&\equiv T\sum_{\en} \widetilde{\mathcal{F}}_{2,p}(i\en,i(\en+\wn))
\end{align}
however, upon examination of Eq.~(\ref{eq:I2sum}) it is clear that in the dc limit, this can
be evaluated by taking a derivative of Eq.~(\ref{eq:GNinfty}) with respect to $R$.  
\begin{align}
\mathcal{G}_p^{R_1} &= -\frac{1}{2}\frac{\partial}{\partial R}\left[ (-2 R_1)
\mathcal{G}_p^{N=\infty}\right] \nonumber 
\\
&= -\frac{1}{\sqrt{T} N} \left\{
\begin{array}{lcr}
0.062251 \cdots \ {e^*}^2 & ; & p = 0 \\
0.026867 \cdots & ; & p = 2 \\
\end{array}\right. 
\label{eq:GR1}
\end{align}
where we have used the previously calculated values of $R/T = 0.6248$ and $R_1 / T = 0.1069 / N$.

Examining the third term in Eq.~(\ref{eq:Kpara}) we now have to perform a dual Matsubara sum over a
function with four separate frequency arguments
\begin{align}
I_{4,p}(i\wn) &= T\sum_{\en} \frac{(\en+\wn/2)^p}{(k^2+R + |\en|)^2(k^2+R+|\en+\wn|)}
\Sigma(k,\en) \nonumber \\
\\
&= T^2\sum_{\en,\On} \mathcal{F}_{4,p}(i\en,i\On,i(\en+\On),i(\en+\wn)).
\end{align}
Using Eq.~(\ref{eq:I4}) we can write:
\begin{align}
& \lim_{\omega\to0}\frac{\Im I_{4,p}(\omega+i\eta)}{\omega} = \nonumber
\\
& \qquad  \frac{1}{T} \iO\ie
\frac{i^p \epsilon^{2+p} (k^2+R) \csch{\epsilon}{2T}{2} \Re[\Pi_T(q,\Omega,R)]^{-1}}
{[(k^2+R^2)^2+\epsilon^2]^3} \nonumber \\
& \qquad \qquad \qquad \times 
\left\{ \frac{(\epsilon+\Omega)\coth\left(\frac{\epsilon+\Omega}{2T}\right)}
{[(k+q)^2+R]^2+(\epsilon+\Omega)^2} - \frac{\Omega\coth\left(\frac{\Omega}{2T}\right)}
{(q^2+R)^2+\Omega^2}\right\}
\end{align}
which leads to the self-energy corrections to the dc conductivities
\begin{align}
\mathcal{G}_p^{\Sigma} &= -\frac{8{e^*}^{2-p}i^p}{NT^{p+1}} \iiqO \iike
\frac{\epsilon^{2+p} k^2 (k^2+R) \csch{\epsilon}{2T}{2}} {[(k^2+R^2)^2+\epsilon^2]^3} \nonumber 
\\
& \; \times \; \Re[\Pi_T(q,\Omega,R)]^{-1}
\left\{ \frac{(\epsilon+\Omega)\coth\left(\frac{\epsilon+\Omega}{2T}\right)}
{[(k+q)^2+R]^2+(\epsilon+\Omega)^2} - \frac{\Omega\coth\left(\frac{\Omega}{2T}\right)}
{(q^2+R)^2+\Omega^2}\right\}.
\label{eq:GSigma}
\end{align}

The final term in Eq.~(\ref{eq:Kpara}) has five separate frequency arguments
\begin{align}
I_{5,p}(i\wn) &= T^2\sum_{\en,\On} \frac{(\en+\wn/2)^{p/2}(\en+\On+\wn/2)^{p/2}}
{\Pi_T(q,\On,R) (k^2+R + |\en|) (k^2+R+|\en+\wn|)}
\nonumber \\
& \qquad \times \; \frac{1}{[(k+q)^2+R+|\en+\On|][(k+q)^2+R+|\en+\On+\wn|]}
\nonumber \\
&= T^2\sum_{\en,\On} \mathcal{F}_{5,p}(i\en,i(\en+\wn),i(\en+\On),i(\en+\On+\wn),i\On).
\end{align}
Using Eq.~(\ref{eq:I5}) we can write:
\begin{align}
& \lim_{\omega\to0}\frac{\Im I_{5,p}(\omega+i\eta)}{\omega} = \nonumber
\\
& \quad \frac{1}{2T} \iO\ie \frac{i^p [\epsilon(\epsilon+\Omega)]^{1+p/2} \csch{\epsilon}{2T}{2} 
\csch{\epsilon+\Omega}{2T}{2} \Re[\Pi_T(q,\Omega,R)]^{-1}}
{[(k^2+R^2)^2+\epsilon^2]^2\{[(k+q)^2+R]^2+(\epsilon+\Omega)^2\}^2} \nonumber
\\
& \qquad \times\; \left\{ (k^2+R) (\epsilon+\Omega)\sinh\left(\frac{\epsilon}{T}\right)
+ [(k+q)^2+R] \epsilon\sinh\left(\frac{\epsilon+\Omega}{T}\right)\right\}
\end{align}
giving the vertex contribution to the $\omega=0$ transport coefficients
\begin{align}
\label{eq:GVertex}
& \mathcal{G}_p^{\Gamma} = -\frac{4{e^*}^{2-p}i^p}{NT^{p+1}} \iiqO \iike
\frac{k(k+q) [\epsilon(\epsilon+\Omega)]^{1+p/2}}
{[(k^2+R^2)^2+\epsilon^2]^2} \nonumber
\\
& \qquad \times \; \frac{\csch{\epsilon}{2T}{2} \csch{\epsilon+\Omega}{2T}{2}
\Re[\Pi_T(q,\Omega,R)]^{-1}}{\{[(k+q)^2+R]^2+(\epsilon+\Omega)^2\}^2} \nonumber
\\
& \qquad \qquad \times \; \left\{ (k^2+R) (\epsilon+\Omega)\sinh\left(\frac{\epsilon}{T}\right)
+ [(k+q)^2+R] \epsilon\sinh\left(\frac{\epsilon+\Omega}{T}\right)\right\}.
\end{align}

\subsubsection{Numerical evaluation}
The $1/N$ corrections to thermoelectric transport coming from the self-energy and vertex corrections
are written in Eqs.~(\ref{eq:GSigma}) and (\ref{eq:GVertex}) as two \emph{four} dimensional integrals
that cannot be evaluated analytically. Before we attempt to compute them numerically, we first
present a simple argument concerning their expected ultra-violet behavior. From scaling we
understand
\begin{equation}
\mathcal{G}_p \sim \frac{1}{T^{1/z}}
\end{equation}
where in Section~\ref{subsec:criticalExponents} we found that $z = 2-\eta$ with $\eta \sim O(1/N)$.  Thus
we can write
\begin{align}
\mathcal{G}_p &\sim \frac{1}{T^{1/(2-\eta)}} \nonumber
\\
&= \mathcal{G}_p^{N=\infty}\left(1 + \frac{C_p}{N} + \frac{\eta}{2}\ln \frac{\Lambda}{\sqrt{T}}
\right)
\label{eq:Gdiv}
\end{align}
where $C_p$ are universal constants and $\Lambda$ is a non-universal ultra violet cutoff.
Immediately we see that to order $1/N$, the ratio of the thermal to electrical conductivity divided
by temperature --- the Wiedemann-Franz ratio --- will be independent of any cutoff as $\Lambda \to
\infty$:
\begin{align}
W &\equiv \frac{\mathcal{G}_2}{\mathcal{G}_0} \nonumber 
\\
&= \frac{\mathcal{G}_2^{N=\infty}}{\mathcal{G}_0^{N=\infty}}
\left[1 + \frac{C_2-C_0}{N} + O\left(\frac{\sqrt{T}}{\Lambda}\right) \right] .
\label{eq:West}
\end{align}
This is an important equation that guarantees the universality of our final result, and will allow us
to test the accuracy of our numerical integration procedure.

\begin{figure}[t]
\centering
\includegraphics*[width=3.2in]{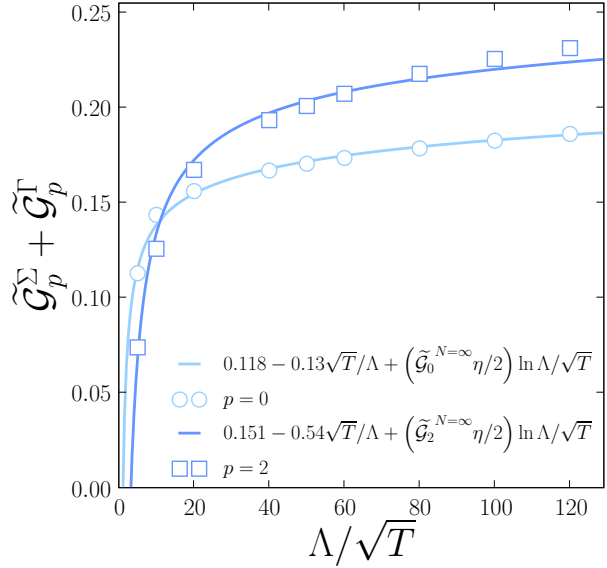}
\caption{\label{fig:GN}  The $1/N$ corrections to the rescaled thermoelectric transport coefficients
coming from self-energy and vertex corrections ($p=0$, bottom curve for $\sigma$ and $p=2$, top
curve for $\kappa/T$) plotted as a function of a dimensionless external ultra violet momentum cutoff
$\Lambda/\sqrt{T}$.  The solid lines are fits to the expected divergent behavior, from which the
non-divergent corrections as $\Lambda \to \infty$ can be extracted.}
\end{figure}

We begin by combining the expressions for the self-energy and vertex corrections such that 
$\Re[\Pi_T(q,\Omega,R)]^{-1}$ (the most costly function to compute, as described in
Appendix~\ref{app:Pi}) is in the outermost integral.  
\begin{equation}\begin{split}
\mathcal{G}_p^{\Sigma}+\mathcal{G}_p^{\Gamma} &= -\frac{4{e^*}^{2-p} i^p}{NT^{p+1}} \iiq 
\Re\left[\frac{1}{\Pi_T(q,\Omega,R)}\right]
\\
& \qquad \times \; 
\iik \left[ Y_p^{\Sigma}(k,\epsilon,q,\Omega) + Y_p^{\Gamma}(k,\epsilon,q,\Omega)\right]
\label{eq:GSEV}
\end{split}\end{equation}
where
\begin{align}
& Y_p^{\Sigma}(k,\epsilon,q,\Omega) = 
\frac{2 k^2 (k^2+R) \epsilon^{2+p}\csch{\epsilon}{2T}{2}}{[(k^2+R)^2+\epsilon^2]^3} \nonumber
\\
& \qquad \times \; \left\{ \frac{(\epsilon+\Omega)\coth\left(\frac{\epsilon+\Omega}{2T}\right)}
{[(k+q)^2+R]^2+(\epsilon+\Omega)^2} - \frac{\Omega\coth\left(\frac{\Omega}{2T}\right)}
{(q^2+R)^2+\Omega^2}\right\} 
\\
& Y_p^{\Gamma}(k,\epsilon,q,\Omega) = 
\frac{k(k+q)[\epsilon(\epsilon+\Omega)]^{1+p/2}\csch{\epsilon+\Omega}{2T}{2}\csch{\epsilon}{2T}{2}}
{[(k^2+R)^2 + \epsilon^2]^2 \{[(k+q)^2+R]^2+(\epsilon+\Omega)^2\}^2} \nonumber 
\\
& \qquad \times \; \left\{ (k^2+R) (\epsilon+\Omega)\sinh\left(\frac{\epsilon}{T}\right) + 
[(k+q)^2+R] \epsilon\sinh\left(\frac{\epsilon+\Omega}{T}\right)\right\}.
\end{align}
Performing the outermost integral numerically using an adaptive routine we arrive at the final results
shown in Fig.~\ref{fig:GN} where a tilde indicates that a quantity has been multiplied by a factor
of $N \sqrt{T}/{e^*}^{2-p}$.
After fitting to the expected divergent form in Eq.~(\ref{eq:Gdiv}), we find (as $\Lambda \to \infty$)
\begin{align}
\mathcal{G}_0^{\Sigma}+\mathcal{G}_0^{\Gamma} = \frac{{e^*}^2}{\sqrt{T}} \frac{0.118}{N}
+ \mathcal{G}_0^{N=\infty} \frac{\eta}{2} \ln \frac{\Lambda}{\sqrt{T}} \\
\mathcal{G}_2^{\Sigma}+\mathcal{G}_2^{\Gamma} = \frac{1}{\sqrt{T}} \frac{0.151}{N}
+ \mathcal{G}_2^{N=\infty} \frac{\eta}{2} \ln \frac{\Lambda}{\sqrt{T}}, 
\end{align}
and combining with the previous results of Eq.~(\ref{eq:GNinfty}) and (\ref{eq:GR1}) we have the
dc thermoelectric transport coefficients to order $1/N$ 
\begin{align}
\label{eq:sigN}
\sigma = \frac{{e^*}^2}{\sqrt{T}} \left( 0.218 + \frac{0.0561}{N} + 
\frac{0.0142}{N} \ln \frac{\Lambda}{\sqrt{T}} \right ) \\
\label{eq:kapN}
\frac{\kappa}{T} = \frac{1}{\sqrt{T}} \left( 0.246 + \frac{0.124}{N}
+ \frac{0.0161}{N} \ln \frac{\Lambda}{\sqrt{T}} \right ),
\end{align}
which both explicitly depend on $\Lambda$ as expected.

\subsection{Wiedemann-Franz ratio in the quantum critical regime}
\begin{figure}[t]
\centering
\includegraphics*[width=3.2in]{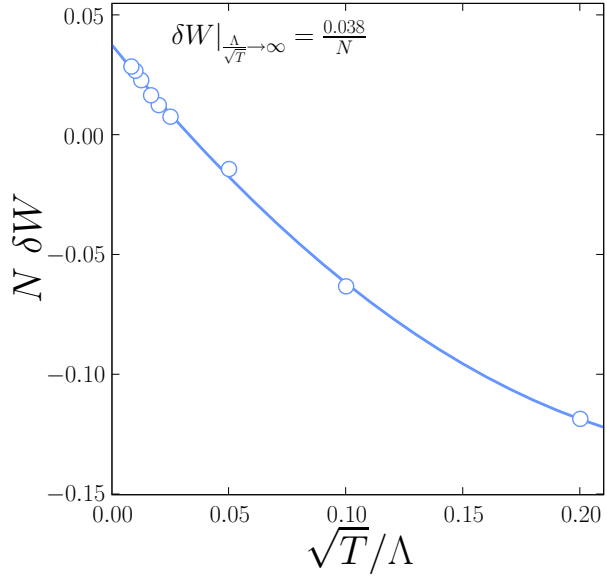}
\caption{\label{fig:WFc} The $1/N$ corrections to the Wiedemann-Franz ratio plotted as
a function of the inverse external rescaled ultra violet momentum cutoff in the quantum critical
regime.  The solid line is a fit to a second order polynomial and the individual divergences of the
electrical and thermal conductivity are exactly canceled as $\Lambda/\sqrt{T} \to \infty$ giving the
universal correction $\delta W = 0.0376 / N$.}
\end{figure}

We have evaluated the full fluctuation corrections to thermoelectric transport up to order $1/N$
(Eqs.~(\ref{eq:sigN}) and (\ref{eq:kapN})) coming from the direct contributions of Cooper pairs due
to the proximate superconducting state.  Initially dismayed by their cutoff dependence, we now
recall the  previous argument that led to Eq.~(\ref{eq:West}). We found from scaling that the
required $T^{-1/z}$ temperature dependence of $\kappa/T$ and $\sigma$ implied that when dividing
them to form the Wiedemann-Franz ratio, all divergent $\Lambda$-dependence must exactly cancel.
This exact cancellation is seen in Fig.~(\ref{fig:WFc}) where we plot the total correction to the
WF ratio, $\delta W$ as a function of the inverse of the rescaled dimensionless cutoff. As $\Lambda
\to \infty$,  $\delta W$ approaches a constant.
Extracting the infinite cutoff result via a polynomial fit we find (after inserting the proper
power of the Boltzmann constant)
\begin{equation}
W = \frac{\kappa}{\sigma T} = \left(0.282 + \frac{0.0376}{N}\right)\left(\frac{k_B}{e}\right)^2.
\label{eq:W1oN}
\end{equation}
Therefore, the WF law is indeed obeyed, (i.e. is temperature independent) indicating the presence 
of only fully elastic scattering and is independent of any microscopic constants.  The term 
proportional to $1/N$ is quite small, and for the physical case, $N=1$ it corresponds to 
a correction on the order of ten percent.

\section{Conclusions}

This paper has been concerned with a topic that could be mistakingly confused with one of limited
scope, the pairbreaking quantum phase transition between a superconductor and a metal in an
ultra-narrow wire as modeled by a continuum quantum field theory.  Instead, we have discovered a
remarkably rich phase diagram full of interesting phases and crossovers.

Experimental motivations for a theoretical analysis of this transition exist in the form of
transport experiments on metallic nanowires, formed through molecular templating by sputtering
material on top of a long rigid ``bridge'' or ``backbone'' molecule lying over a trench
\cite{bezryadin-lau}. In this way, wires with diameters of less than $10$~nm can be fabricated; a
giant step towards reaching the limit where the length scale characterizing quantum fluctuations
approaches the finite radius of the wire. In an applied current and at fixed temperatures, below the
bulk superconducting transition temperature for the wires composite material, a given wire can
display either metallic or superconducting behavior depending on its radius, with the general trend
that thinner wires are less superconducting.  In addition, for a particular wire which does exhibit
electrical transport without resistance, superconductivity can be destroyed by turning on a suitably
strong magnetic field oriented along its parallel axis.  In both of these cases, it is some
non-thermal parameter which suppresses the Cooper pairing instability and tunes between the
superconducting and normal metallic state at zero temperature, providing an excellent manifestation
of a quantum phase transition.

The description of the transition that we have adopted in this paper is a critical theory of
strongly repulsive, fluctuating Cooper pairs, written in terms of a complex order parameter,
overdamped by its coupling to a bath of unpaired fermions resulting from the presence of some type
of pairbreaking interaction.  The existence of the bath, imagined as a large number of unpaired
electrons residing in the transverse conduction channels of the wire, leads to a long range
interaction in imaginary time providing Ohmic dissipation in the form of a non-analytic $|\wn|$ term
in the effective action.  The presence of such an anisotropic relationship between space and time
coupled with a continuous symmetry order parameter fixes the dynamical critical exponent at $z=2$, and
the resulting upper critical dimension is $d_{\mathrm{UCD}}=2$.

The thinness of the experimentally investigated wires provided us with a useful theoretical
framework in the form of the quasi-one dimensional limit, where the radius of the wire $R$ is on the
order of, or much smaller than the superconducting coherence length $\xi$ at low temperatures.  The
length scale $\xi$ measures the average separation between the electrons in a Cooper pair, and for
$R < \xi$ the composite object begins to \emph{feel} the boundary.  If the wire is sufficiently
long, the paired states can be described in terms of a quantum field theory in one space and one
imaginary time dimension.  In $1+1$ dimensions, we found ourselves below $d_{\mathrm{UCD}}$, with
the repulsive self-interactions between Cooper pairs being strongly relevant.   As a result, any
perturbative or mean field approaches are unable to provide a complete and accurate picture of the
physical phenomena.

To deal with strong interactions, we have employed a variety of field theoretic and numerical
techniques suitable in different regions of the pairbreaking phase diagram in conjunction with a
careful scaling analysis. In the quantum critical regime, at temperatures limited to those where
$\hbar \omega \ll \kB T$, dynamical observables were computed from the real time Langevin dynamics
of an effective classical theory renormalized by quantum fluctuations in the limit where the number
of complex order parameter components ($N=1$ in the physical case) was equal to infinity.  Near the
finite temperature classical phase boundary, the effective theory was used to derive a modified
version of the thermally activated phase slip theory of Langer, Ambegaokar, McCumber and Halperin
\cite{la,mh} which does not explicitly depend on the number of transverse channels of the wire.  At
low temperatures, the full quantum theory was analyzed in a systematic expansion to first order in
$1/N$, with extensive detail provided on the calculation of critical exponents and various physical
quantities near $z=2$.

The combination of all these approaches allowed us to determine the full form of the zero frequency
(dc) electrical and thermal conductivities as a function of temperature and the pairbreaking
parameter which drives the transition.  Our first experimentally testable result is a complete
crossover phase diagram for the quantum superconductor-metal transition (SMT).  We predict that upon
fixing the source of pairbreaking (either magnetic field or wire radius) at a value near
criticality, as function of decreasing temperature, the electrical conductivity should first
increase as $1/\sqrt{T}$ due to pairing fluctuations near the quantum critical point, then change to
decreasing as $T^2$ once the low temperature metallic phase has been reached.  The $T^{-1/z}$
dependence of the dc electrical conductivity appears as a natural consequence of scaling near
criticality and was not found in previous non-interacting microscopic perturbative calculations
\cite{shah-lopatin}.  There may already be qualitative experimental evidence for transport that is
non-monotonic in temperature \cite{zgirski-arutyunov,bollinger-bezryadin} and a further analysis of
the reported resistance data along these lines seems apropos.

The second prediction is that in the quantum critical regime at finite temperatures, defined by a
pairbreaking strength that is close to the one that would destroy superconducting order at zero
temperature, the ratio of dc thermal ($\kappa$) to electrical ($\sigma$) conductivity divided by
temperature (the Wiedemann-Franz ratio) should be a temperature independent constant measuring
deviations from the Lorenz number for a normal metal $l_0 = \pi^2/3 (\kB/e)^2$.  We have computed
the exact value of this correction in a $1/N$ expansion and found
\begin{equation} 
W \equiv \frac{\kappa}{\sigma T} = \left(0.282 + \frac{0.0376}{N}\right)\left(\frac{k_B}{e}\right)^2.  
\end{equation} 
Although electrical transport measurements already exist on a variety of metallic nanowires, their
thermal contact with both the substrate as well as the two dimensional leads remains a challenging
problem in the suspended molecular geometry, and would inhibit experimental access to this
prediction.  However, various intriguing and promising directions are currently being explored
including the use of a scanning tunneling microscope \cite{hoffman-pc}.

Further avenues for theoretical progress still remain in superconducting systems in confined
geometries, including a full understanding of the superfluid density in the low temperature ordered
phase which has not been attempted here. Such a description would require proper inclusion of the
pairing interaction as well as Coulomb repulsion in the presence of disorder, leading to a plasmon
mode describing the strongly fluctuating phase of the superconducting order parameter.  In addition,
the presence of a recently identified infinite randomness fixed point at the disordered
superconductor-metal transition \cite{diswireslett,hoyos} opens up a plethora of questions
surrounding the nature of hydrodynamic transport in the presence of unconventional activated dynamic
scaling and a detailed investigation by the authors is currently underway.


\section{Acknowledgments}
It is a pleasure to thank E.~Demler, B.~Halperin, G.~Rafael, V.~Galitski, G.~Catelani, N.~Shah and
B. Spivak, for discussions related to multiple aspects of this work.  This research was supported
by NSF Grant DMR-0537077. A.~D. would like to thank NSERC of Canada for financial support through
Grant PGS D2-316308-2005.  B.~R. acknowledges support through the Heisenberg program of DFG.  All
computer simulations were carried out using resources provided by the Harvard Center for Nanoscale
Systems, part of the National Nanotechnology Infrastructure Network.

\appendix

\section{Computation of $I(\Omega)$}
\label{app:phas}
This appendix provides details on the evaluation of the scaling dimension of $\rho$ characterizing
the strength of particle-hole symmetry breaking in the perturbing action $S_\rho$. The combination
of the two-loop diagrams in Eq.~(\ref{eq:phasGraphs}) evaluated at zero external frequency and
momentum lead to
\begin{equation}\begin{split}
I(\Omega) &= -i 2 \rho u^2 \int \frac{d \omega_1}{2\pi} \int \frac{d \omega_2}{2\pi}
\int \frac{d^d k}{(2\pi)^d} \int \frac{d^d q}{(2\pi)^d} \\
& \qquad \times\; \frac{\omega_1}
{(k^2 + |\omega_1|)^2(q^2+|\omega_2|)[(k+q)^2+|\omega_1+\omega_2+\Omega|]}.
\label{eq:IOmega1}
\end{split}\end{equation}
The two momentum integrals can be evaluated by employing Feynman parameters
\begin{equation}\begin{split}
I(\Omega) &= -i 2 \rho u^2 \frac{\Gamma(4-d)}{(4\pi)^d} \int_0^1 d x \int_0^1 dy 
\int \frac{d \omega_1}{2\pi} \int \frac{d \omega_2}{2\pi} \frac{(1-y)y^{1-d/2}}{[1-y(x^2-x+1)]^{d/2}} 
\\
& \qquad \qquad \qquad \times \; \frac{\omega_1}{[(1-y)|\omega_1| + y(1-x)|\omega_2| + 
xy |\omega_1 + \omega_2 + \Omega|]^{4-d}}.
\label{eq:IOmega2}
\end{split}\end{equation}
where $\Gamma(x)$ is the gamma function.  The double frequency integral, can be done by determining 
the sign of the various absolute values in the relevant seven regions of the $\omega_1 - \omega_2$ 
plane delineated by changes in signs $\omega_1$, $\omega_2$ and $\omega_1+\omega_2$  leading to the
following useful but complicated looking result 
\begin{align}
I_\omega(A,B,C,\sigma) &=\int \frac{d \omega_1}{2\pi} \int \frac{d \omega_2}{2\pi}
\frac{\omega_1}{(A|\omega_1| + B|\omega_2| + C |\omega_1 + \omega_2 + \Omega|)^{\sigma}} \nonumber
\\
&= - \frac{A B C \Gamma(\sigma - 3)\Omega^{3-\sigma}}{\pi^2 \Gamma(\sigma)}
\left \{ A^{3-\sigma} \left[ \frac{2(2A^2-B^2-C^2)}{(A^2-B^2)^2(A^2-C^2)^2} \right. \right.  \nonumber
\\
& \left. \qquad \qquad - \frac{3-\sigma}{A^2(A^2-C^2)(A^2-B^2)} \right] \nonumber
- \frac{2B^{3-\sigma}}{(B^2-C^2)(A^2-B^2)^2} \nonumber
\\
& \left. \qquad +\; \frac{2C^{3-\sigma}}{(B^2-C^2)(A^2-C^2)^2} \right\}.
\label{eq:Iomega}
\end{align}
In $d=2-\epsilon$ and  $\sigma = 2 + \epsilon$, $I_\omega$ has a pole at $\epsilon = 0$ with a
residue that can be read off from
\begin{equation}
I_\omega(A,B,C,2+\epsilon) = - \frac{\Omega}{\pi^2}\left[\frac{B
C(2A+B+C)}{(A+B)^2(A+C)^2(B+C)}\right] \frac{1}{\epsilon} + O(1).
\label{eq:Iomegapole}
\end{equation}
Using Eq.~(\ref{eq:Iomegapole}) in Eq.~(\ref{eq:IOmega2}) we find
\begin{align}
I(\Omega) &= \frac{i\Omega u^2 \rho}{8\pi^4 \epsilon} \int_0^1 d x \int_0^1 dy \frac{xy(1-x)(1-y)(2-y)}
{[1-y(x^2-x+1)][1-y+xy^2(1-x)]^2} \nonumber \\
&= \frac{i\Omega u^2 \rho}{8\pi^4 \epsilon} \left(\frac{\pi^2}{4}-2\right).
\end{align}

\section{Connection to microscopic BCS theory}
\label{app:microBCS}

In order to motivate the experimental relevance of the effective action $\mathcal{S}_{\alpha}$, the
microscopic values of the renormalized pairbreaking frequency $\alpha$, the bare diffusion constant
$\D$, dissipation strength $\gamma$ and quartic coupling $u$ can be determined in both the clean and
dirty limits.  We begin with the connection of the pairbreaking frequency to various experimentally
relevant geometries, then move on to the relationship between the theory presented here and a time
dependent Ginzburg-Landau (TDGL) theory for a conventional superconductor.

\subsection{Pair-breaking in quasi-one dimensional wires}
\label{subsec:pairbreaking}
As mentioned in the introduction, there are various origins of pairbreaking perturbations relevant
to experiments on ultra-narrow wires.  The most theoretically appealing consists of
magnetic impurities localized on the surface of a metallic wire leading to an inhomogeneous BCS
coupling (Fig.~\ref{fig:wCS}). In this case, the microscopic value of $\alpha$ is not known
exactly, but it can be related to the inverse of the spin-flip scatting time.  However, there are a
number of well-defined experimental geometries where the actual value of $\alpha$ can be computed in
terms of the physical properties of the system.  One case which is of interest here,
is a narrow metallic wire in a parallel magnetic field.

In the dirty limit, Shah and Lopatin \cite{shah-lopatin} have computed the precise form of $\alpha$
through the use of the Usadel equation formalism. They find that for a narrow diffusive wire with
radius $R$ smaller than both the superconducting coherence length and the magnetic penetration depth 
placed in a parallel magnetic field $H$,
\begin{equation}
\alpha_{\mathrm{wire}} = \frac{\D}{2}\left(\frac{e H R}{c}\right)^2
\label{eq:alphaWire}
\end{equation}
where $\D$ is the diffusion constant and $c$ the speed of light.  

\subsection{Microscopic parameters in the clean and dirty limits}
The microscopic values of $\D$ ,$\gamma$ and $u$  can be found through an analysis of the 
time dependent Ginzburg-Landau theory studied by Tucker and Halperin \cite{tucker}.  There, 
the three dimensional equation of motion for the Cooper pair operator $\Psi(\vec{x},t)$ in real time
is given by 
\begin{equation}
\hbar \gamma \frac{\partial}{\partial t} \Psi(\vec{x},t) = - \left[ a + b|\Psi(\vec{x},t)|^2 +
\delta (-i\nabla)^2 \right] \Psi(\vec{x},t).
\label{eq:thEOM}
\end{equation}
Rescaling, to ensure that the coefficient of the time derivative term is unity and performing an
integral over the cross-sectional area of the wire to move to the quasi-one dimensional case of
interest ($\Psi(x,y,z,t) \sim \Psi(x,t)$), we read off the value of the coupling constants to be 
\begin{align}
\D &= \frac{\delta}{\hbar \gamma} \\
u &= \frac{b}{A \hbar^2 \gamma^2}
\end{align}
where $A$ is the cross-sectional area of the wire.  Appendix A of Ref.~\cite{tucker} gives
the microscopic values of $\delta$, $b$ and $\gamma$ as
\begin{align}
\delta &= \frac{\hbar^2}{2m}, \\
b &= \frac{\hbar^2}{2m\xi^2(0)} \frac{2}{n \chi(0.882\xi_0 / l)}, \\
\gamma &= \frac{\pi \hbar^2}{16 m \xi^2(0) k_B T_{c0}}.
\end{align}
where $\xi(T)$ is the Ginzburg-Landau coherence length, $\xi_0$ the BCS coherence length, $\ell$ the
mean free path and $\chi(\rho)$ the Gor'kov function defined by
\begin{equation}
\chi(\rho) = \sum_{n=0}^{\infty} \frac{1}{(2n+1)^2(2n + 1 + \rho)}\left[\sum_{n=0}^{\infty}
\frac{1}{(2n+1)^3}\right]^{-1}.
\end{equation}
The critical temperature ($T_c$) and density of conduction electrons in the normal state ($n$) are known
to be 
\begin{align}
\label{eq:Tc0}
\frac{1}{k_B T_{c0}} &= \frac{\xi_0}{0.18 \hbar v_\mathrm{F}};  \\
\label{eq:N}
n &= \frac{k_\mathrm{F}^3}{3\pi^2}
\end{align}
respectively,  while, the zero temperature coherence length and relevant Gor'kov function depend on
whether we are in the clean or dirty limit
\begin{align}
\xi(0) &= \left\{\begin{array}{rcl}
0.74 \xi_0 & ; & \xi_0 \ll \ell \\
0.85 \sqrt{\xi_0 \ell} & ; & \xi_0 \gg \ell
\end{array} \right., \\
\chi(0.882\xi_0/l) &= \left\{\begin{array}{rcl}
 1 & ; & \xi_0 \ll \ell \\
1.33 \ell/\xi_0 & ; & \xi_0 \gg \ell
\end{array} \right.\, . 
\end{align}
We have now gathered all the required information to compute the actual microscopic values of our model 
parameters in the dirty $(\xi_0 \gg \ell)$ and clean $(\xi_0 \ll \ell)$ limits differentiated by the 
subscripts $d$ for \emph{dirty} and $c$ for \emph{clean}.

Using the above relations, we find that for the dirty limit
\begin{align}
\label{eq:DDirty}
\D_d &= D = \frac{1}{3} v_\mathrm{F} \ell \\
\label{eq:gammaDirty}
\gamma_d &\simeq \frac{1.5}{k_\mathrm{F} \ell} \\
\label{eq:uDirty}
u_d &\simeq 2.9\frac{v_F}{\hbar N_{\perp}}
\end{align}
where the number of transverse conduction channels is assumed to be large, and is given by 
\begin{equation}
N_\perp = \frac{2k_\mathrm{F}^2 A}{3\pi}. 
\end{equation}

Similarly, in the clean limit
\begin{align}
\label{eq:DClean}
\D_c &= \frac{1}{4} v_F \xi_0  \\
\label{eq:gammaClean}
\gamma_c &\simeq \frac{2.0}{k_\mathrm{F} \xi_0} \\
\label{eq:uClean}
u_c &= u_d \simeq 2.9\frac{v_F}{\hbar N_{\perp}}
\end{align}
where we note that the bare value of the quartic coupling is \emph{identical} in both limits.  The
value of these parameters clearly depends on the particular normalization scheme chosen for the
order parameter, but the final results for all physically measurable quantities, such as the
conductivity, will obviously be normalization independent.

\section{Ginzburg-Landau theory near $T_c$}
\label{app:ginzburgLandau}

The insertion of a self-consistent infrared cutoff allowed for the derivation of an effective
potential at zero temperature within the ordered phase (Eq.~(\ref{eq:VeffT0})).  At finite
temperatures, near $T_c$ Eq.~(\ref{eq:VsigDelta}) can be expanded in powers of $|\sigma|^2$ in order
to derive an effective Ginzburg-Landau theory for the superconducting phase with coefficients
renormalized by quantum fluctuations.  This is accomplished by first computing the critical
temperature where $\langle \sigma \rangle$ becomes non-zero.  Suitable derivatives are then taken to
determine the values of the quadratic ($\alpha_0$) an quartic ($\beta$) coefficients in the expansion
\begin{equation}
V_{GL} = V_0 +  \alpha_0 (T-T_c) |\sigma|^2 + \frac{1}{2}\beta |\sigma|^4 + \cdots .
\label{eq:positVGLApp}
\end{equation}


\subsection{Evaluation of the critical temperature $T_c$}
\begin{figure}[t]
\centering
\includegraphics*[width=3.2in]{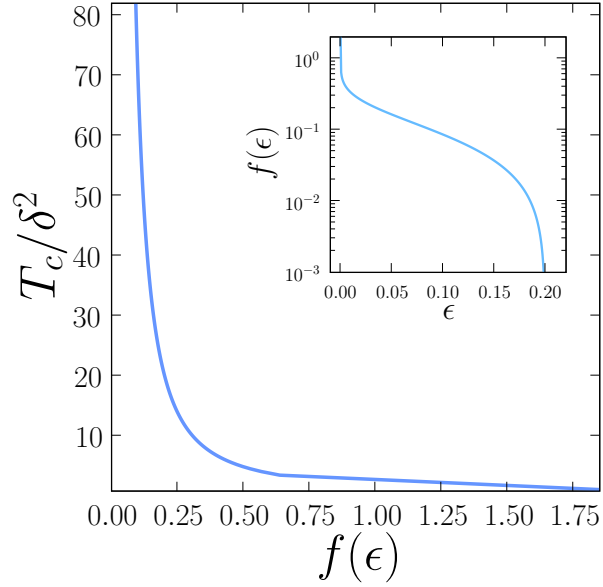}
\caption{\label{fig:Tc}  The rescaled critical temperature found from the solution of
Eq.~(\ref{eq:Tc}) as a function of $f(\epsilon)$ given in Eq.~(\ref{eq:fepsilon}) which is shown in
the inset.} \end{figure}

We begin by considering the saddle point equation (Eq.~(\ref{eq:SSP})) in the presence of the
symmetric cutoff.  A similar procedure that led to Eq.~(\ref{eq:finalSSP2}) can be used here, giving 
\begin{equation}\begin{split}
 |\sigma|^2 &= \frac{g|\delta|}{\sqrt{\D}} + \frac{g\Lambda}{\pi^2} \left[ 2 - \ln\left(\frac{\D
\Lambda^2}{2 \pi T} \right) + \psi \left( \frac{\D \Lambda^2 + r}{2 \pi T}\right) \right] 
\\
& \qquad - \;  \frac{g}{\pi^2} \int_\Lambda^\infty dk \left[ \frac{\pi T}{\D k^2 + r} - \psi\left(1 +
\frac{\D k^2 + r}{2 \pi T}\right) + \ln \left(\frac{\D k^2}{2 \pi T}\right) \right].
\label{eq:cutoffSPE}
\end{split}\end{equation}
Returning to Eq.~(\ref{eq:dVLG}) and noting that at $T=T_c$, $r(|\sigma|^2=0) = 0$, we can derive an
equation for $T_c$ from Eq.~(\ref{eq:cutoffSPE})
\begin{equation}\begin{split}
0 & = \frac{|\delta|}{\sqrt{2 \pi T_c}} + \frac{(1+\epsilon)}{\pi^2 f(\epsilon)} 
\frac{|\delta|}{\sqrt{2 \pi T_c}} \left\{ 2 - \ln\left[\frac{(1+\epsilon)^2 \delta^2}
{2 \pi T_c f^2(\epsilon)} \right] + \psi \left[ \frac{(1+\epsilon)^2 \delta^2}{2 \pi T_c
f^2(\epsilon)} \right] \right\} 
\\
& \qquad -\; \frac{1}{\pi^2} \int_{\frac{(1+\epsilon)|\delta|}
{\sqrt{2\pi T_c}f(\epsilon)}}^\infty dk \left[ \frac{1}{2 k^2} - \psi(1+k^2) + \ln k^2 \right],
\label{eq:Tc}
\end{split}\end{equation}
where $k$ is a dimensionless momentum, and the cutoff can be fixed by setting $\epsilon$ and using
Eq.~(\ref{eq:fepsilon}).  Solving numerically using a secant method gives the result seen in
Fig.~\ref{fig:Tc}, where $f(\epsilon)$ is plotted as an inset.
The critical temperature is found to be proportional to the square of the distance from criticality, 
$T_c \propto \delta^2$ and more specifically
\begin{equation}
T_c = c_1(\epsilon) \delta^2
\label{eq:c1}
\end{equation}
where $c_1 \simeq 1.90$ for $f(\epsilon) = 1$.  One could either fix $\epsilon$ at this value,
or choose a value of $\epsilon$ using a plot like Fig.~\ref{fig:Tc} that reproduced the 
relationship between $T_c$ and $\delta$ measured in an experiment.

\subsection{Evaluation of the quadratic coefficient $\alpha_0$}
In order to evaluate $\alpha_0$ in Eq.~(\ref{eq:positVGLApp}) we again appeal to Eq.~(\ref{eq:dVLG}) 
and note that
\begin{equation}
\alpha_0 (T - T_c) = \frac{1}{g} r(|\sigma|^2 = 0).
\label{eq:alpha}
\end{equation}
Near $T_c$, $r \ll 1$ and a double expansion of Eq.~(\ref{eq:cutoffSPE}) can be performed 
in $r$ and the reduced temperature $t = (T - T_c)/T_c$ leading to (after some considerable
algebra)
\begin{equation}\begin{split}
0 &= 
\frac{r}{2 \pi T_c} \left\{ \frac{(1+\epsilon)}{\pi^2 f(\epsilon)} \frac{|\delta|}{\sqrt{2 \pi T_c}} 
\psi^{(1)}\left[\frac{(1+\epsilon)^2 \delta^2}{2 \pi T_c f^2(\epsilon) }\right] + c_2(\epsilon) 
\right\}  
\\
& \qquad +\;  
t\left\{ \frac{|\delta|}{2\sqrt{2 \pi T_c}} + \frac{1}{\pi^2}\left[\frac{(1+\epsilon)|\delta|}
{\sqrt{2 \pi T_c} f(\epsilon)}\right]^3  \psi^{(1)}\left[\frac{(1+\epsilon)^2 \delta^2}
{2 \pi T_c f^2(\epsilon) }\right] - \frac{\sqrt{2 \pi T_c} f(\epsilon)}{4\pi^2 |\delta|} \right\}
\label{eq:st}
\end{split}\end{equation}
where only linear terms in $r$ and $t$ have been retained, $\psi^{(1)}(x)$ is the first polygamma
function and 
\begin{equation}
c_2(\epsilon) = \frac{1}{\pi^2} \int \limits_{\frac{(1+\epsilon)|\delta|}
{\sqrt{2\pi T_c}f(\epsilon)}}^\infty dk \left[ \frac{1}{2 k^4} + \psi^{(1)}(1+k^2) \right] .
\label{eq:c2}
\end{equation}
Comparing Eq.~(\ref{eq:st}) with Eq.~(\ref{eq:alpha}) we find
\begin{equation}\begin{split}
\alpha_0 &= \frac{2\pi}{g} \left\{ \frac{|\delta|}{2\sqrt{2 \pi T_c}} 
+ \frac{1}{\pi^2}\left[\frac{(1+\epsilon)|\delta|}
{\sqrt{2 \pi T_c} f(\epsilon)}\right]^3  \psi^{(1)}\left[\frac{(1+\epsilon)^2 \delta^2}
{2 \pi T_c f^2(\epsilon) }\right] - \frac{\sqrt{2 \pi T_c} f(\epsilon)}{4\pi^2 |\delta|}\right\}
\\
& \qquad \times \; \left\{ \frac{(1+\epsilon)}{\pi^2 f(\epsilon)} \frac{|\delta|}{\sqrt{2 \pi T_c}} 
\psi^{(1)}\left[\frac{(1+\epsilon)^2 \delta^2}{2 \pi T_c f^2(\epsilon) }\right] +
c_2(\epsilon)\right\}^{-1}
\label{eq:alpha0}
\end{split}\end{equation}
and upon choosing $f(\epsilon) = 1$,  $c_2 \simeq 0.856$ and $\alpha_0 \simeq 0.509385/g$.

\subsection{Evaluation of the quartic coefficient $\beta$}
In order to determine the value of the quartic coefficient, we examine Eqs.~(\ref{eq:dVLG}) and
(\ref{eq:positVLG}) at $T=T_c$ leading to
\begin{equation}
\beta \equiv \frac{d^2 V_{LG}}{d(|\sigma|^2)^2} = \frac{1}{g} 
\left.\frac{d r}{d|\sigma|^2} \right|_{T=T_c}.
\label{eq:d2VLG}
\end{equation}
Taking a derivative with respect to $|\sigma|^2$ of Eq.~(\ref{eq:cutoffSPE}) we find
\begin{equation}\begin{split}
1 &= \frac{g}{2\pi T_c}\left\{ \frac{\Lambda}{\pi^2} \psi^{(1)}\left(\frac{\D\Lambda^2}{2 \pi
T_c}\right) \right.
\\
& \left. \qquad +\;  
\sqrt{\frac{2\pi T_c}{\D}} \frac{1}{\pi^2} \int_{\sqrt{\frac{\D}{2\pi T_c}} \Lambda}^\infty
dk \left[ \frac{1}{2 k^4} + \psi^{(1)}(1+k^2) \right]  \right\}
\left.\frac{d r}{d|\sigma|^2}\right|_{T=T_c}
\end{split}\end{equation}
and using the relation between $\Lambda$ and $\xi(0)$, Eqs.~(\ref{eq:xiLambda}) and
(\ref{eq:fepsilon}) as well as Eq.~(\ref{eq:d2VLG}), the quartic coefficient is given by
\begin{equation}
\beta = \frac{\sqrt{\D}}{g^2} |\delta| \; 
\left\{ \frac{(1+\epsilon)}{\pi^2 f(\epsilon)} \psi^{(1)}\left[\frac{(1+\epsilon)^2 \delta^2}
{2 \pi T_c f^2(\epsilon) }\right] + \frac{\sqrt{2 \pi T_c} c_2(\epsilon)}{|\delta|} \right\}^{-1}.
\label{eq:beta}
\end{equation}
where Eq.~(\ref{eq:c2}) has been used.  For $f(\epsilon) = 1$, $\beta \simeq 0.495 \sqrt{\D}
|\delta| / g^2$.

The results of Eqs.~(\ref{eq:Tc}), (\ref{eq:alpha0}) and (\ref{eq:beta}) constitute all the required
ingredients needed to construct the Ginzburg-Landau potential of Eq.~(\ref{eq:positVGLApp}) with the
final result shown in Fig.~\ref{fig:VLG}.  This appendix has presented numerical results for a
fixed value of $\epsilon$ that was determined by setting $f(\epsilon)=1$.  However, the
$f(\epsilon)$ dependence of all coefficients is relatively weak for $f(\epsilon) > 1$ which is
well satisfied by a physical value of the cutoff.  

\section{The Fluctuation Propagator}
\label{app:Pi}

In this appendix, we will provide details on various results related to the evaluation of the fluctuation 
propagator at both zero and finite temperatures.

\subsection{Zero temperature}

At zero temperature, and coupling $g$ the fluctuation propagator is given by
\begin{equation}
\Pi_0(k,\omega,r) = \iiq \frac{1}{(q^2+|\epsilon|+r)[(k+q)^2+|\omega+\epsilon|+r]}.
\end{equation}
The momentum and frequency integrals can be done by employing Feynman parameters to yield
\begin{align}
&\Pi_0(k,\omega,r) = \nonumber
\\
& \quad \frac{1}{2\pi} \int_0^1 dx \frac{1}{2x-1} 
\left( \frac{x}{\sqrt{(k^2-|\omega|)x - k^2 x^2 + |\omega|}} + 
\frac{x-1}{\sqrt{(k^2 + |\omega|)x - k^2 x^2}} \right) \nonumber 
\\
&= \frac{1}{2\pi |k|}\! \left[ 
\mathrm{asin}\left(\frac{k^2 + |\omega|}{\sqrt{(k^2+|\omega|)^2 + 4 k^2 r}}\right) + 
\mathrm{asin}\left(\frac{k^2 - |\omega|}{\sqrt{(k^2+|\omega|)^2 + 4 k^2 r}}\right) \right ] \nonumber 
\\
&\quad + \frac{1}{2\pi \sqrt{k^2 + 2|\omega|+4r}} \left \{
\Re \left[ \mathrm{atanh}\left(\frac{k^2 + 3|\omega| + 4r}
{2\sqrt{k^2 + 2|\omega|+4r}\sqrt{r+|\omega|}} \right) \right] \right. \nonumber 
\\
&\left. \qquad\qquad -\; \Re \left[ \mathrm{atanh}\left(
\frac{k^2 + |\omega| + 4r}{2\sqrt{k^2 + 2|\omega|+4r}\sqrt{r}} \right) \right]
\right \}.
\end{align}
Using the relation
\begin{equation}
\Re[\mathrm{atanh} (z)] = \frac{1}{2}\ln \left| \frac{1+z}{1-z} \right|
\end{equation}
we can write
\begin{align}
&\Pi_0(k,\omega,r) = \nonumber
\\
& \quad \frac{1}{2\pi |k|} \left[ 
\mathrm{asin}\left(\frac{k^2 + |\omega|}{\sqrt{(k^2+|\omega|)^2 + 4 k^2 r}}\right) + 
\mathrm{asin}\left(\frac{k^2 - |\omega|}{\sqrt{(k^2+|\omega|)^2 + 4 k^2 r}}\right) \right ]
\nonumber
\\
& \; \quad +\;  \frac{1}{4\pi \sqrt{k^2 + 2|\omega|+4r}} \! \left[ 
\ln \left( \frac{ 2\sqrt{r+|\omega|}\sqrt{k^2 + 2|\omega| + 4r} + k^2 + 3|\omega| + 4r}
{|2\sqrt{r+|\omega|}\sqrt{k^2 + 2|\omega| + 4r} - k^2 - 3|\omega| - 4r|}\right)  \right. \nonumber
\\
& \qquad \left. - \ln \left( \frac{ 2\sqrt{r}\sqrt{k^2 + 2|\omega| + 4r} + k^2 + |\omega| + 4r}
{|2\sqrt{r+|\omega|}\sqrt{k^2 + 2|\omega| + 4r} - k^2 - |\omega| - 4r|}\right)  \right]
\label{eq:Pi0r}
\end{align}
and at the critical coupling where $r=0$ this simplifies to
\begin{equation}\begin{split}
\Pi_0 (k,\omega,0) &= \frac{1}{4\pi |k|} \left[ 
2\mathrm{asin}\left(\frac{k^2-|\omega|}{k^2+|\omega|}\right) + \pi \right] 
\\
& \qquad +\; \frac{1}{4\pi \sqrt{k^2 + 2|\omega|}} \ln \left( 
\frac{2\sqrt{|\omega|}\sqrt{k^2+2|\omega|} + k^2 + 3|\omega|}
{|2\sqrt{|\omega|}\sqrt{k^2+2|\omega|} - k^2 - 3|\omega||}\right).
\label{eq:Pi0}
\end{split}\end{equation}

\subsection{Finite temperature}

A key step required for the evaluation of the shift in the critical point coming from $1/N$
corrections (Eq.~(\ref{eq:R1oT})) and the thermoelectric transport coefficients (Eq.~(\ref{eq:GSEV}))
is the fast and accurate computation of $\Pi_T(k,\omega_n,R)$ as well as the real and imaginary 
parts of its analytically continued value just above the real axis where 
$i \omega_n \to \omega + i\eta$. 

\subsubsection{Numerical evaluation}
Starting from Eq.~(\ref{eq:PiT}) and performing the momentum integral we have
\begin{align}
\Pi_T(k,\omega_n,R) &= \isq \frac{1}{[(k+q)^2 + |\omega_n+\epsilon_n| +R](q^2 +
|\epsilon_n| + r)} \nonumber 
\\
&= T \sum_{\epsilon_n} \frac{ \sqrt{|\epsilon_n| + R} +
\sqrt{|\epsilon_n + \omega_n| + R}}{2 \sqrt{(|\epsilon_n| + R)(|\epsilon_n + \omega_n| + R)} }
\nonumber
\\
& \qquad \qquad \times \; 
\frac{1}{k^2 + (\sqrt{|\epsilon_n| + R} + \sqrt{|\epsilon_n + \omega_n| + R})^2} .
\label{eq:PiTR}
\end{align}
Let us first find the value of the finite temperature fluctuation propagator at $k = \wn =
0$.  Starting form Eq.~(\ref{eq:PiTR}) above we can derive a simple result
\begin{align}
\Pi_T(0,0,R) &= \frac{1}{8\sqrt{2}\pi^{3/2} \sqrt{T}} \sum_{n =
-\infty}^{\infty}\frac{1}{(|n| + R/2\pi T)^{3/2}} \nonumber \\
&= \frac{1}{8\sqrt{2}\pi^{3/2} \sqrt{T}} \left[ \zeta\left(\frac{3}{2}, \frac{R}{2\pi
T}\right) + \zeta\left(\frac{3}{2}, \frac{R}{2\pi T} + 1\right) \right],
\label{eq:PiT0}
\end{align}
where $\zeta(m,x)$ is the Hurwitz zeta function.  In order to evaluate the sum in
Eq.~(\ref{eq:PiTR}) at finite frequencies and wavevectors, we explicitly sum the terms up to some
large value of $|\epsilon_n| < 2 \pi L$ where $2 \pi L \gg |\omega_n|$. For the remaining terms, we
perform a series expansion of the summand in powers of $1/|\epsilon_n|$ and then use the asymptotic
series
\begin{align}
\sum_{n=L}^{\infty} \frac{1}{n^s} &= \frac{L^{-s+1}}{\Gamma(s)} \int_0^{\infty} \frac{y^{s-2} e^{-y}
dy}{1 - e^{-s/L}} \nonumber 
\\
&= L^{-s+1} \Biggl[ \frac{1}{s-1} + \frac{1}{2L} + \frac{s}{12
L^2} - \frac{\Gamma(s+3)}{720 \Gamma(s) L^4} + \frac{\Gamma(s+5)}{30240
\Gamma(s) L^6} \nonumber
\\
& \qquad -\; \frac{\Gamma(s+7)}{1209600 \Gamma(s) L^8} +
\frac{\Gamma(s+9)}{47900160 \Gamma(s) L^{10}} + \ldots \Biggr].
\end{align}
As discussed in Ref.~\cite{csy}, we must use the value of $R$ given in Eq.~(\ref{eq:RoT}) 
for the resulting $\Pi_T (k, \omega_n,R)$ to be well behaved at large $k$ and $\omega_n$.

\subsubsection{$\Re [\Pi_T(q,\Omega,R)]^{-1}$}

We now provide details on the use of the summation formulas described in Appendix~\ref{app:sum}
to evaluate the real and imaginary parts of the fluctuation propagator analytically continued to real
frequencies. The benefits of this rather complicated derivation are manifest in the increased 
computational efficiency of having analytic expressions as opposed to resorting to a Kramers-Kronig
relation.  Again, we start from
\begin{equation}
\Pi_T(q,\On,R) = \iske \frac{1}{(k^2+R+|\en|)[(k+q)^2+R+|\en+\On|]}
\end{equation}
and need to analytically continue to real frequencies $i\On \to \Omega + i\eta$.  
Thus, using Eq.~(\ref{eq:I2}) we perform the Matsubara summation to give
\begin{align}
& \Pi_T(q,\Omega+i\eta,R) = \frac{1}{2} \int \frac{dk}{2\pi} \int \frac{d\epsilon}{2\pi} \Bigg\{
\coth\left(\frac{\epsilon}{2T}\right)\Big[F_{\Pi}(q,k,\epsilon + i\eta ,\epsilon+\Omega + i\eta) 
\nonumber
\\
& \qquad -\;  F_{\Pi}(q,k,\epsilon-i\eta,\epsilon+\Omega+i\eta)\Big] 
+ \coth\left(\frac{\epsilon+\Omega}{2T}\right) \nonumber
\\
& \qquad \times \; \Bigl[F_{\Pi}(q,k,\epsilon-i\eta,\epsilon+\Omega+i\eta) 
- F_{\Pi}(q,k,\epsilon-i\eta,\epsilon+\Omega-i\eta\Big] \Bigg\}
\label{eq:PiTF}
\end{align}
where 
\begin{equation}
F_{\Pi}(q,k,\epsilon\pm i\eta,\nu\pm i\eta) = \frac{1}{(k^2+R \mp i\epsilon)[(k+q)^2 + R \mp i\nu]}.
\end{equation}
Considering each of the four terms in Eq.~(\ref{eq:PiTF}) separately, we will have to perform an
integral of the form
\begin{equation}
I_{\Pi} (q,a,b) = \int \frac{dk}{2\pi} \frac{1}{(k^2+a)[(k+q)^2+b]} = 
\frac{1}{2}\left(\frac{1}{\sqrt{a}} +
\frac{1}{\sqrt{b}}\right) \frac{1}{q^2+(\sqrt{a}+\sqrt{b})^2}
\label{eq:IPi}
\end{equation}
which was evaluated using Feynman parameters and in the particular case considered here 
$\Re a = \Re b = R > 0$. Using the relation
\begin{equation}
\frac{1}{\sqrt{a \mp i b}} = \frac{1}{\sqrt{2}\md{a}{{b}}}
\left( \sqrt{\md{a}{b}+a} \pm i \sgn{b}\sqrt{\md{a}{b}-a} \right)
\label{eq:1osq}
\end{equation}
and
\begin{align}
& \Re\frac{1}{q^2+(\sqrt{a-i\zeta_b b}+\sqrt{a-i\zeta_c c})^2} = 
\frac{1}{\Delta(q,a,b,\zeta_b,c,\zeta_c)} \nonumber 
\\ 
& \qquad \times \; \Bigg[ q^2 + 2 a + \sqrt{(\md{a}{b}+a)(\md{a}{c}+a)} \nonumber
\\
& \qquad \qquad -\; \zeta_b\zeta_c \sgn{b}\sgn{c} \sqrt{(\md{a}{b}-a)(\md{a}{c}-a)}\ \Bigg]
\\
& \Im\frac{1}{q^2+(\sqrt{a-i\zeta_b b}+\sqrt{a-i\zeta_c c})^2} = 
\frac{1}{\Delta(q,a,b,\zeta_b,c,\zeta_c)}\nonumber
\\
& \qquad \times \; \Bigg[ \zeta_b b + \zeta_c c + \zeta_c \sgn{c} \sqrt{(\md{a}{b}+a)(\md{a}{c}-a)} 
\nonumber
\\
& \qquad \quad +\; \zeta_b \sgn{b} \sqrt{(\md{a}{b}-a)(\md{a}{c}+a)}\ \Bigg]
\end{align}
where
\begin{align}
\Delta(q,a,b,\zeta_b,c,\zeta_c) &= 
\Bigg[ q^2 + 2 a + \sqrt{(\md{a}{b}+a)(\md{a}{c}+a)} \nonumber 
\\
&  \qquad -\; \zeta_b\zeta_c \sgn{b}\sgn{c} \sqrt{(\md{a}{b}-a)(\md{a}{c}-a)}\ \Bigg]^2
\nonumber 
\\
& \quad +\; \Bigg[ \zeta_b b + \zeta_c c + \zeta_c \sgn{c} \sqrt{(\md{a}{b}+a)(\md{a}{c}-a)} 
\nonumber 
\\
&  \qquad +\; \zeta_b \sgn{b} \sqrt{(\md{a}{b}-a)(\md{a}{c}+a)}\ \Bigg]^2
\label{eq:Delta}
\end{align}
with $a,b,c \in \mathbb{R}$ and $\zeta_{b,c} = \pm 1$.

Therefore, using Eq.~(\ref{eq:IPi}) to Eq.~(\ref{eq:Delta}) we can write the real and imaginary
parts of the analytically continued fluctuation propagator as (suppressing all $R$ dependence)
\begin{align}
& \Re \Pi_T(q,\Omega\pm i\eta) = \frac{1}{2} \int \frac{d\epsilon}{2\pi} \Bigg\{
	\coth\left(\frac{\epsilon}{2T}\right) \Im f_{\Pi}(q,\epsilon+i\eta,\epsilon+\Omega+i\eta)
	 \nonumber
\\
&  \qquad -\;  \coth\left(\frac{\epsilon+\Omega}{2T}\right) \Im
f_{\Pi}(q,\epsilon-i\eta,\epsilon+\Omega-i\eta)  \nonumber 
\\
&  \qquad + \left[\coth\left(\frac{\epsilon+\Omega}{2T}\right) -
\coth\left(\frac{\epsilon}{2T}\right)  \right]
\Im f_{\Pi}(q,\epsilon-i\eta,\epsilon+\Omega+i\eta) \Bigg\} 
\\
& \Im \Pi_T(q,\Omega\pm i\eta) = \mp \frac{1}{2} \int \frac{d\epsilon}{2\pi} \Bigg\{
	\coth\left(\frac{\epsilon}{2T}\right) \Re f_{\Pi}(q,\epsilon+i\eta,\epsilon+\Omega+i\eta) 
\nonumber
\\
& \qquad -\;  \coth\left(\frac{\epsilon+\Omega}{2T}\right) 
\Re f_{\Pi}(q,\epsilon-i\eta,\epsilon+\Omega-i\eta) \nonumber 
\\
&  \qquad + \left[\coth\left(\frac{\epsilon+\Omega}{2T}\right) -
\coth\left(\frac{\epsilon}{2T}\right)  \right]
\Re f_{\Pi}(q,\epsilon-i\eta,\epsilon+\Omega+i\eta) \Bigg\}
\end{align}
with
\begin{multline}
\Re f_{\Pi}(q,\epsilon+i\zeta_\epsilon \eta,\nu+i\zeta_\nu \eta) = 
\frac{1}{2 \sqrt{2} \Delta(q,R,\epsilon,\zeta_\epsilon,\nu,\zeta_\nu)} 
\\
\times \; \Biggl\{
\left(\frac{\sqrt{\md{R}{\epsilon}+R}}{\md{R}{\epsilon}} + 
      \frac{\sqrt{\md{R}{\nu}+R}}{\md{R}{\nu}}\right) 
\\
\times \; \left[ q^2 + 2 R + \sqrt{(\md{R}{\epsilon}+R)(\md{R}{\nu}+R)} \right. 
\\
\left. -\; \zeta_\epsilon\zeta_\nu \sgn{\epsilon}\sgn{\nu}
                \sqrt{(\md{R}{\epsilon}-R)(\md{R}{\nu}-R)}\ \right] 
\\
-\; \left(\zeta_\epsilon \sgn{\epsilon} \frac{\sqrt{\md{R}{\epsilon}-R}}{\md{R}{\epsilon}} + 
      \zeta_\nu \sgn{\nu}\frac{\sqrt{\md{R}{\nu}-R}}{\md{R}{\nu}}\right)  
\\
\times \; \left[ \zeta_\epsilon \epsilon + \zeta_\nu \nu  + \zeta_\nu \sgn{\nu}
\sqrt{(\md{R}{\epsilon}+R)(\md{R}{\nu}-R)} \right. 
\\
\left. +\; \zeta_\epsilon\sgn{\epsilon} \sqrt{(\md{R}{\epsilon}-R)(\md{R}{\nu}+R)}\ \right] \Biggr\} 
\end{multline}
and
\begin{multline}
\Im f_{\Pi}(q,\epsilon+i\zeta_\epsilon \eta,\nu+i\zeta_\nu \eta) = 
\frac{1}{2 \sqrt{2} \Delta(q,R,\epsilon,\zeta_\epsilon,\nu,\zeta_\nu)} 
\\
\times \; \Biggl\{
\left(\zeta_\epsilon \sgn{\epsilon} \frac{\sqrt{\md{R}{\epsilon}-R}}{\md{R}{\epsilon}} + 
      \zeta_\nu \sgn{\nu}\frac{\sqrt{\md{R}{\nu}-R}}{\md{R}{\nu}}\right)  
\\
\times\; \left[ q^2 + 2 R + \sqrt{(\md{R}{\epsilon}+R)(\md{R}{\nu}+R)} \right. 
\\
\left. - \; \zeta_\epsilon\zeta_\nu \sgn{\epsilon}\sgn{\nu}
                \sqrt{(\md{R}{\epsilon}-R)(\md{R}{\nu}-R)}\ \right] 
\\
+\; \left(\frac{\sqrt{\md{R}{\epsilon}+R}}{\md{R}{\epsilon}} + 
      \frac{\sqrt{\md{R}{\nu}+R}}{\md{R}{\nu}}\right) 
\\
\times\; \left[ \zeta_\epsilon \epsilon + \zeta_\nu \nu  + \zeta_\nu \sgn{\nu}
\sqrt{(\md{R}{\epsilon}+R)(\md{R}{\nu}-R)} \right. 
\\
\left.+ \; \zeta_\epsilon\sgn{\epsilon} \sqrt{(\md{R}{\epsilon}-R)(\md{R}{\nu}+R)}\ \right] \Biggr\},
\end{multline}
where $\zeta_{\epsilon,\nu} = \pm 1$. Such a formulation allows us to compute both the real and 
imaginary parts of $\Pi_T$ without having to resort to a Kramers-Kronig relation, leading to
\begin{equation}
\Re \left[ \frac{1}{\Pi_T(q,\Omega,R)} \right] = \frac{\Re\Pi_T(q,\Omega,R)}
{[\Re\Pi_T(q,\Omega,R)]^2 + [\Im\Pi_T(q,\Omega,R)]^2}.
\end{equation}

\section{Details on the Evaluation of Matsubara Sums}
\label{app:sum}

This appendix provides details on the evaluation of multiple Matsubara summations which appear in
the evaluation of various current-current correlation functions coming from the Kubo formula.
We begin with the basic identity \cite{agd}
\begin{equation}
T \sum_{\epsilon_n} \mathcal{F}(i \epsilon_n) = 
\frac{1}{2} \int_{-\infty}^{\infty}
\frac{ d \epsilon}{2 \pi i} \coth \left( \frac{\epsilon}{2T} \right) 
\left[ F( \epsilon + i \eta) - F(\epsilon - i \eta) \right],
\end{equation}
noting that if $F(i \epsilon_n ) = \mathcal{F} (|\epsilon_n|)$, then after analytic continuation
$F(\epsilon \pm i \eta) = \mathcal{F} ( \mp i \epsilon)$.  By a similar application of contour
integration we obtain
\begin{align}
I_2 (i \omega_n) &= T \sum_{\epsilon_n} \mathcal{F}(i \epsilon_n, i (\epsilon_n + \omega_n)) 
\nonumber 
\\
&= \frac{1}{2} \int_{-\infty}^{\infty}
\frac{ d \epsilon}{2 \pi i} \coth \left( \frac{\epsilon}{2T} \right) 
\Bigl[ F( \epsilon + i \eta, \epsilon + i \omega_n) - F(\epsilon - i \eta, \epsilon+ i \omega_n) \nonumber \\ 
& \qquad\qquad +\; F(\epsilon-i \omega_n, \epsilon+i\eta) - 
F(\epsilon- i \omega_n, \epsilon - i \eta) \Bigr]
\label{eq:I2pre}
\end{align}
and so 
\begin{align}
& I_2 ( \omega + i \eta) = \nonumber
\\
& \; \int_{-\infty}^{\infty} \frac{d \epsilon}{2 \pi i}
\Biggl\{  \coth \left( \frac{\epsilon}{2 T} \right)  \Bigl[
F (\epsilon + i \eta, \epsilon + \omega + i \eta) 
- F (\epsilon - i \eta, \epsilon + \omega + i \eta) \Bigr]  \nonumber 
\\
& \quad + \; \coth \left( \frac{\epsilon+ \omega}{2 T} \right) \Bigl[
F (\epsilon - i \eta, \epsilon + \omega + i \eta) - 
F (\epsilon - i \eta, \epsilon + \omega - i \eta) \Bigr] \Biggr\} \nonumber 
\\
& = \int_{-\infty}^{\infty}\! \frac{d \epsilon}{2 \pi i}
\Biggl\{ \coth \left( \frac{\epsilon}{2 T} \right) \Bigl[ F (++) - F (-+) \Bigr] \nonumber
\\
& \qquad +\; \coth \left( \frac{\epsilon+ \omega}{2 T} \right) \Bigl[ F (-+) - F (--) \Bigr] \Biggr\}
\label{eq:I2}
\end{align}
where  in the last expression we only denote the sign of the $i\eta$ term, because the frequency
arguments remain the same in {\em all\/} terms:
\begin{equation}
F(\pm\pm) \equiv F(\epsilon \pm i\eta, \epsilon + \omega \pm i\eta).
\end{equation}
Rearranging the terms to preserve the order of the frequency arguments will allow us to pull out
common factors in the numerator and lead to many simplifications.

Any corrections coming from the presence of a finite self energy at order $1/N$ require that we
perform a dual Matsubara summation over a function with four frequency arguments.  Through a further
generalization of the method of contour integration used to obtain Eq.~(\ref{eq:I2pre}) we find
\begin{align}
I_4 ( i \omega_n) &= T^2 \sum_{\epsilon_n, \Omega_n} 
\mathcal{F}( i \epsilon_n, i \Omega_n, i (\epsilon_n + \Omega_n) , i (\epsilon_n + \omega_n)) \nonumber 
\\
&=  \frac{1}{4} \int_{-\infty}^{\infty} \frac{d \Omega}{2 \pi i} \coth \left( \frac{\Omega}{2T} \right)
\int_{-\infty}^{\infty} \frac{d \epsilon}{2 \pi i} \coth \left( \frac{\epsilon}{2T} \right)
\nonumber 
\\
& \qquad \times \; \Bigl[ \Upsilon_4^+ (\epsilon,\Omega,i\wn; i\eta) 
- \Upsilon_4^- (\epsilon,\Omega,i\wn;i\eta) \Bigr] 
\end{align}
where
\begin{align}
\Upsilon_4^+ (\epsilon,\Omega,i\wn; i\eta) & = 
F( \epsilon+i \eta, \Omega+i \eta, \Omega +  \epsilon + i \eta , \epsilon + i \omega_n)  \nonumber
\\
& \qquad +\; F( \epsilon+ i \eta, \Omega- \epsilon- i \eta, \Omega + i \eta, \epsilon + i \omega_n)
\nonumber
\\
& \qquad +\; F(  \epsilon- i \eta, \Omega-i \eta, \Omega + \epsilon-i \eta, \epsilon + i \omega_n) 
\nonumber
\\
& \qquad +\; F( \epsilon- i \eta, \Omega-  \epsilon+ i \eta, \Omega - i \eta, \epsilon + i \omega_n) 
\nonumber
\\
& \qquad +\; F( \epsilon - i \omega_n , \Omega+i \eta, \Omega +  \epsilon - i \omega_n, \epsilon + i
\eta) \nonumber
\\
& \qquad +\; F( \epsilon - i \omega_n, \Omega- \epsilon + i \omega_n, \Omega + i \eta,  \epsilon + i \eta)
\nonumber
\\
& \qquad +\; F( \epsilon - i \omega_n, \Omega-i \eta, \Omega +\epsilon - i \omega_n, \epsilon - i \eta) 
\nonumber
\\
& \qquad + \;  F( \epsilon - i \omega_n, \Omega- \epsilon + i \omega_n, \Omega - i \eta, \epsilon - i \eta) 
\end{align}
and
\begin{align}
\Upsilon_4^- (\epsilon,\Omega,i\wn; i\eta) &= 
F(  \epsilon+ i \eta, \Omega-i \eta, \Omega + \epsilon+i \eta, \epsilon + i \omega_n) \nonumber
\\
& \qquad +\; F( \epsilon+ i \eta, \Omega-  \epsilon- i \eta, \Omega - i \eta, \epsilon + i \omega_n) \nonumber 
\nonumber
\\
& \qquad +\; F( \epsilon-i \eta, \Omega+i \eta, \Omega +  \epsilon - i \eta , \epsilon + i \omega_n)
\nonumber
\\
& \qquad +\; F( \epsilon- i \eta, \Omega- \epsilon+ i \eta, \Omega + i \eta, \epsilon + i \omega_n)
\nonumber
\\
& \qquad +\; F( \epsilon - i \omega_n, \Omega-i \eta, \Omega +\epsilon - i \omega_n, \epsilon + i \eta) \nonumber 
\nonumber
\\
& \qquad +\; F( \epsilon - i \omega_n, \Omega- \epsilon + i \omega_n, \Omega - i \eta, \epsilon + i \eta) \nonumber 
\nonumber
\\
& \qquad +\; F( \epsilon - i \omega_n , \Omega+i \eta, \Omega +  \epsilon - i \omega_n, \epsilon - i \eta)
\nonumber
\\
& \qquad +\; F( \epsilon - i \omega_n, \Omega- \epsilon + i \omega_n, \Omega + i \eta,  \epsilon - i \eta)
\end{align}
so that
\begin{align}
I_4 (\omega + i \eta) &= 
\frac{1}{4} \int_{-\infty}^{\infty} \frac{d \Omega}{2 \pi i} \int_{-\infty}^{\infty} 
\frac{d \epsilon}{2 \pi i} \nonumber 
\\
& \quad \times\; \Biggl\{\coth \left( \frac{\Omega}{2T} \right)
\coth \left( \frac{\epsilon}{2T} \right) \Bigl[ F(++++) - F(+-++) \nonumber
\\
& \qquad\qquad\qquad\qquad\qquad\qquad\qquad +\; F(---+) - F(-+-+) \Bigr] \nonumber 
\\
&\qquad +\; \coth \left( \frac{\Omega+\epsilon}{2T} \right)
\coth \left( \frac{\epsilon}{2T} \right) \Bigl[ F(+-++) - F(+--+) \nonumber
\\
& \qquad\qquad\qquad\qquad\qquad\qquad\qquad +\; F(-+-+) - F(-+++) \Bigr] \nonumber 
\\
&\qquad +\; \coth \left( \frac{\Omega}{2T} \right)
\coth \left( \frac{\epsilon+\omega}{2T} \right) \Bigl[ F(-+-+) - F(---+) \nonumber
\\
& \qquad\qquad\qquad\qquad\qquad\qquad\qquad +\; F(----) - F(-+--) \Bigr] \nonumber 
\\
& \qquad + \; \coth \left( \frac{\Omega+\epsilon}{2T} \right)
\coth \left( \frac{\epsilon+\omega}{2T} \right) \Bigl[ F(-+++) - F(-+-+) \nonumber
\\
& \qquad\qquad\qquad\qquad\qquad\qquad\qquad +\; F(-+--) - F(-++-) \Bigr] \Biggr\}.
\label{eq:I4}
\end{align}
Multiple change of variable transformations have been performed to ensure that the function $F$ 
in the terms above have the same arguments $\epsilon$, $\Omega$, $\epsilon+\Omega$, $\epsilon+\omega$, 
and thus only the signs of the $i\eta$ terms have been denoted,
\begin{equation}
F(\pm\pm\pm\pm) = F(\epsilon \pm i\eta, \Omega \pm i\eta, \epsilon+\Omega\pm i\eta, \epsilon+\omega
\pm i\eta).
\end{equation}

Finally, the vertex corrections are similar in that we still need to perform a dual Matsubara sum, 
but now the frequency arguments are more complicated, and there are five unique combinations
\begin{align}
I_5 (i \omega_n) &= T^2 \sum_{\epsilon_n, \Omega_n} \mathcal{F}( i \epsilon_n, 
i (\epsilon_n + \omega_n), i \Omega_n, i (\Omega_n+\omega_n) , i (\Omega_n - \epsilon_n)) \nonumber 
\\
&= \frac{1}{4} \int_{-\infty}^{\infty} \frac{d \Omega}{2 \pi i} \coth \left( \frac{\Omega}{2T} \right)
\int_{-\infty}^{\infty} \frac{d \epsilon}{2 \pi i} \coth \left( \frac{\epsilon}{2T} \right)
\\
& \qquad \times \; 
\Bigl[ \Upsilon_5^+ (\epsilon,\Omega,i\wn; i\eta) - \Upsilon_5^- (\epsilon,\Omega,i\wn;i\eta) \Bigr]
\end{align}
where
\begin{align}
\Upsilon_5^+ (\epsilon,\Omega,i\wn; i\eta) &= 
F( \epsilon+i \eta, \epsilon+i  \omega_n, \Omega+ i \eta, \Omega + i \omega_n, \Omega- \epsilon-i \eta) 
\nonumber \\
&\qquad +\; F( \epsilon+i \eta, \epsilon+i \omega_n, \Omega - i \omega_n, \Omega + i \eta , \Omega- \epsilon-i \omega_n)
\nonumber \\
&\qquad+\; F( \epsilon+i \eta, \epsilon+i  \omega_n, \Omega +  \epsilon+i \eta, \Omega + i \omega_n +  \epsilon , \Omega+i \eta)
\nonumber \\
&\qquad+\; F( \epsilon-i \eta, \epsilon+i \omega_n, \Omega- i \eta, \Omega + i \omega_n, \Omega- \epsilon+i \eta) 
\nonumber \\
&\qquad+\; F( \epsilon-i \eta, \epsilon+i  \omega_n, \Omega - i \omega_n, \Omega - i \eta , \Omega- \epsilon-i \omega_n)
\nonumber \\
&\qquad+\; F( \epsilon-i \eta, \epsilon+i  \omega_n, \Omega +  \epsilon-i \eta, \Omega + i \omega_n +  \epsilon , \Omega-i \eta) 
\nonumber \\
&\qquad+\; F(\epsilon - i \omega_n, \epsilon+ i \eta, \Omega+ i \eta, \Omega + i \omega_n, \Omega-\epsilon + i \omega_n) 
\nonumber \\
&\qquad+\; F(\epsilon - i \omega_n, \epsilon+ i \eta, \Omega - i \omega_n, \Omega + i \eta , \Omega-\epsilon- i \eta)
\nonumber \\
&\qquad+\; F(\epsilon - i \omega_n, \epsilon+ i \eta, \Omega +  \epsilon - i \omega_n, \Omega + \epsilon + i \eta , \Omega+i \eta)
\nonumber \\
&\qquad+\; F(\epsilon - i \omega_n, \epsilon- i \eta, \Omega- i \eta, \Omega + i \omega_n, \Omega- \epsilon + i \omega_n)
\nonumber \\
&\qquad+\; F(\epsilon - i \omega_n, \epsilon- i \eta, \Omega - i \omega_n, \Omega - i \eta , \Omega-\epsilon + i \eta)
\nonumber \\
&\qquad+\; F(\epsilon - i \omega_n, \epsilon- i \eta, \Omega +  \epsilon - i \omega_n, \Omega + \epsilon - i \eta , \Omega-i \eta) 
\end{align}
and
\begin{align}
\Upsilon_5^- (\epsilon,\Omega,i\wn; i\eta) &= 
F( \epsilon+i \eta, \epsilon+i \omega_n, \Omega- i \eta, \Omega + i \omega_n, \Omega- \epsilon-i \eta) 
\nonumber \\
&\qquad +\; F( \epsilon+i \eta, \epsilon+i  \omega_n, \Omega - i \omega_n, \Omega - i \eta , \Omega- \epsilon-i \omega_n)
\nonumber \\
&\qquad +\; F( \epsilon+i \eta, \epsilon+i  \omega_n, \Omega +  \epsilon+i \eta, \Omega + i \omega_n +  \epsilon , \Omega-i \eta) 
\nonumber \\
&\qquad +\; F( \epsilon-i \eta, \epsilon+i  \omega_n, \Omega+ i \eta, \Omega + i \omega_n, \Omega- \epsilon+i \eta)
\nonumber \\
&\qquad +\; F( \epsilon-i \eta, \epsilon+i \omega_n, \Omega - i \omega_n, \Omega + i \eta , \Omega- \epsilon-i \omega_n)
\nonumber \\
&\qquad +\; F( \epsilon-i \eta, \epsilon+i  \omega_n, \Omega +  \epsilon-i \eta, \Omega + i \omega_n +  \epsilon , \Omega+i \eta)
\nonumber \\
&\qquad +\; F(\epsilon - i \omega_n, \epsilon+ i \eta, \Omega- i \eta, \Omega + i \omega_n, \Omega- \epsilon + i \omega_n)
\nonumber \\
&\qquad +\; F(\epsilon - i \omega_n, \epsilon+ i \eta, \Omega - i \omega_n, \Omega - i \eta , \Omega-\epsilon - i \eta)
\nonumber \\
&\qquad +\; F(\epsilon - i \omega_n, \epsilon+ i \eta, \Omega + \epsilon - i \omega_n, \Omega + \epsilon + i \eta , \Omega-i \eta) 
\nonumber \\
&\qquad +\; F(\epsilon - i \omega_n, \epsilon- i \eta, \Omega+ i \eta, \Omega + i \omega_n, \Omega-\epsilon + i \omega_n) 
\nonumber \\
&\qquad +\; F(\epsilon - i \omega_n, \epsilon- i \eta, \Omega - i \omega_n, \Omega + i \eta , \Omega-\epsilon+ i \eta)
\nonumber \\
&\qquad +\; F(\epsilon - i \omega_n, \epsilon- i \eta, \Omega +  \epsilon - i \omega_n, \Omega + \epsilon - i
\eta , \Omega+i \eta) .
\end{align}
Again performing multiple variable shifts yields a much simpler expression where the frequency
arguments of each term are the same. Suppressing the frequency dependence of $I_5 = I_5(\omega +
i\eta)$ we finally arrive at
\begin{align}
I_5  &=
\frac{1}{4} \int_{-\infty}^{\infty} \frac{d \Omega}{2 \pi i} 
\int_{-\infty}^{\infty} \frac{d \epsilon}{2 \pi i} \nonumber
\\
& \quad \times \;\Biggl\{  
\coth \left( \frac{\Omega}{2T} \right) \coth \left( \frac{\epsilon}{2T} \right) 
\Bigl[ F(+++++) - F(++++-) \nonumber
\\
& \hspace{16em} +\; F(-+-+-) - F(-+-++) \Bigr] \nonumber 
\\
& \qquad +\; \coth \left( \frac{\Omega+\epsilon}{2T} \right)
\coth \left( \frac{\epsilon}{2T} \right) \Bigl[ F(++++-) - F(++-+-) \nonumber
\\
& \hspace{16em} +\; F(-+-++) - F(-++++) \Bigr] \nonumber 
\\
& \qquad +\; \coth \left( \frac{\Omega+\epsilon+\omega}{2T} \right)
\coth \left( \frac{\epsilon}{2T} \right) \Bigl[ F(++-+-) - F(++---) \nonumber
\\
& \hspace{16em} +\; F(-+---) - F(-+-+-) \Bigr] \nonumber 
\\
& \qquad +\; \coth \left( \frac{\Omega+\epsilon}{2T} \right)
\coth \left( \frac{\epsilon+\omega}{2T} \right) \Bigl[ F(-++++) - F(-+-++) \nonumber
\\
& \hspace{16em} +\; F(---++) - F(--+++) \Bigr] \nonumber 
\\
& \qquad +\; \coth \left( \frac{\Omega+\epsilon+\omega}{2T} \right)
\coth \left( \frac{\epsilon+\omega}{2T} \right) \Bigl[ F(-+-+-) - F(-+---) \nonumber
\\
& \hspace{16em} +\; F(----+) + F(---++) \Bigr] \nonumber 
\\
&\qquad +\; \coth \left( \frac{\Omega}{2T} \right)
\coth \left( \frac{\epsilon+\omega}{2T} \right) \Bigl[ F(-+-++) - F(-+-+-) \nonumber 
\\
& \hspace{16em} +\; F(-----) + F(----+) \Bigr] \Biggr\} 
\label{eq:I5}
\end{align}
where the arguments of $F$ have been shifted to be $\epsilon$, $\epsilon + \omega$, 
$\epsilon + \Omega$, $\epsilon + \Omega + \omega$, $\Omega$,
\begin{equation}
F(\pm,\pm,\pm,\pm,\pm) \equiv
F(\epsilon\pm i\eta,\epsilon+\omega\pm i\eta,\epsilon+\Omega \pm i\eta, \epsilon+\Omega+\omega \pm
i\eta, \Omega \pm i\eta).
\end{equation}

\bibliographystyle{elsart-num}
\bibliography{wires-long}

\end{document}